\documentclass[twocolumn,superscriptaddress,amsmath,amssymb,floats,notitlepage,nopacs,nofootinbib,prx]{revtex4-1}
\usepackage[latin9]{inputenc}
\usepackage{amsmath}
\usepackage{graphicx}
\usepackage{color}
\usepackage{tabu}
\usepackage{nicefrac}
\usepackage{mathtools}
\usepackage{ulem}
\usepackage[bookmarks = false, colorlinks = true, urlcolor = black, citecolor = magenta, linkcolor = magenta, urlcolor = magenta]{hyperref}
\usepackage{comment}
\usepackage{cancel}

\makeatletter

\@ifundefined{textcolor}{}
{%
 \definecolor{BLACK}{gray}{0}
 \definecolor{WHITE}{gray}{1}
 \definecolor{RED}{rgb}{1,0,0}
 \definecolor{GREEN}{rgb}{0,1,0}
 \definecolor{BLUE}{rgb}{0,0,1}
 \definecolor{CYAN}{cmyk}{1,0,0,0}
 \definecolor{MAGENTA}{cmyk}{0,1,0,0}
 \definecolor{YELLOW}{cmyk}{0,0,1,0}
}

\usepackage{epstopdf}
\usepackage{color}
\usepackage{array}
\usepackage{mathrsfs}
\usepackage{dcolumn}	
\usepackage{bm}			

\newcolumntype{L}[1]{>{\raggedright\let\newline\\\arraybackslash\hspace{0pt}}m{#1}}
\newcolumntype{C}[1]{>{\centering\let\newline\\\arraybackslash\hspace{0pt}}m{#1}}
\newcolumntype{R}[1]{>{\raggedleft\let\newline\\\arraybackslash\hspace{0pt}}m{#1}}

\newcommand{\bea}{\begin{eqnarray}}
\newcommand{\eea}{\end{eqnarray}}
\newcommand{\bt}{\textbf}

\newcommand{\phd}{\phantom{\dag}}
\newcommand{\ph}{\phantom{.}}

\newcommand{\noi}{\noindent}
\newcommand{\no}{\nonumber}

\usepackage{tikz}
\usepackage{booktabs}
\usepackage{dsfont}
\usepackage{multirow}
\usetikzlibrary{calc,intersections}
\usetikzlibrary{shadings}
\usepgflibrary{shadings}

\makeatother

\AtBeginDocument{
\heavyrulewidth=.08em
\lightrulewidth=.05em
\cmidrulewidth=.03em
\belowrulesep=.65ex
\belowbottomsep=0pt
\aboverulesep=.4ex
\abovetopsep=0pt
\cmidrulesep=\doublerulesep
\cmidrulekern=.5em
\defaultaddspace=.5em
}

\begin{document}
\title{Topological superconductivity induced by magnetic {\color{black}texture crystals}}

\author{Daniel Steffensen}
\affiliation{Niels Bohr Institute, University of Copenhagen, Jagtvej 128, DK-2200 Copenhagen, Denmark}

\author{Morten H. Christensen}
\affiliation{Niels Bohr Institute, University of Copenhagen, Jagtvej 128, DK-2200 Copenhagen, Denmark}
\affiliation{School of Physics and Astronomy, University of Minnesota, Minneapolis, Minnesota 55455, USA}

\author{Brian M. Andersen}
\affiliation{Niels Bohr Institute, University of Copenhagen, Jagtvej 128, DK-2200 Copenhagen, Denmark}

\author{Panagiotis Kotetes}
\email{kotetes@itp.ac.cn}
\affiliation{CAS Key Laboratory of Theoretical Physics, Institute of Theoretical Physics, Chinese Academy of Sciences, Beijing 100190, China}
\affiliation{Niels Bohr Institute, University of Copenhagen, Jagtvej 128, DK-2200 Copenhagen, Denmark}

\begin{abstract}
{\color{black}We present a detailed investigation of the topological phases and Majorana fermion (MF) excitations that arise from the bulk interplay between (un)conventional one/two-band spin-singlet superconductivity and a number of magnetic texture crystals. The latter define inhomogeneous magnetization profiles which consist of a periodically-repeating primitive cell. Here we focus on magnetic texture crystals with a primitive cell of the helix, whirl, and skyrmion types, which feature distinct symmetry properties. We identify a multitude of accessible topological phases which harbor flat, uni- or bi-directional, (quasi-)helical, or chiral MF edge modes. This rich variety ori\-gi\-na\-tes from the interplay between topological phases with gapped and nodal bulk energy spectra. The types of the emerging topological superconducting phases and the features of the ari\-sing MFs are solely determined by the properties and compatibility of the so-called magnetic and pairing point/space groups. Our analysis is general and does not rely on specific parameters of the models employed here to exemplify the topological scenarios which become accessible. Therefore, our results can be extended to systems with multiple bands, are relevant for a wide range of layered materials and hybrid devices, and provide predictions for strong, weak and crystalline topological phases.}
\end{abstract}

\maketitle

\section{Introduction}

Since its discovery, superconductivity has served as an inspiration for countless new concepts and applications. A recent deve\-lop\-ment in the field concerns the material discovery and synthetic engi\-nee\-ring of to\-po\-lo\-gi\-cal superconductors (TSCs)~\cite{HasanKane,QiZhang,Alicea,CarloRev,Leijnse,TanakaSatoNagaosa,AndoFuRev,Franz,Aguado,SatoAndo,LutchynNatRevMat,Pawlak,PradaRev}, which harbor charge-neutral Majorana fermion (MF) quasiparticles~\cite{Majorana,Wilczek,ReadGreen,Volovik99,Ivanov,KitaevUnpaired,VolovikBook,Altland,SchnyderClassi,KitaevClassi,Ryu,TeoKane}. Remarka\-bly, 0D defects can trap zero-energy MFs~\cite{Altland,SchnyderClassi,KitaevClassi,Ryu,TeoKane}, i.e., the so-called Majorana zero modes (MZMs), which adhere to non-Abelian exchange statistics~\cite{Ivanov,KitaevTQC,Nayak,AliceaTQC} and open perspectives for cutting-edge quantum manipulations~\cite{KitaevTQC,Nayak,AliceaTQC,Clarke,Bernard,Milestones,Karzig,Tommy,Freedman}. MZMs are sought after in a variety of systems, such as those containing singular defects, e.g. vortices~\cite{ReadGreen,Volovik99,Ivanov,VolovikBook,SatoStrings,
FuKane,SarmaSuperfuid,Fujimoto,SauPRL,FeTeSeTopo1,
FeTeSeTopo2,XJLiu,KunJiang,SteffensenMZM}, disclinations~\cite{TeoDisclinations,Rex}, hedgehogs~\cite{TeoKaneHedgehog}, and nonsingular defects unfolding in one direction, e.g., termination edges~\cite{KitaevUnpaired,TeoDisclinations,Rex,Wimmer}, domain walls~\cite{VolovikBook,FuKane}, and {\color{black}isolated magnetic skyrmions}~\cite{BalatskySkyrmion,KlinovajaSkyrmion,MikeSkyrmion,Mirlin,Garnier,Mesaros}.

\begin{figure*}[t!]
\centering
\includegraphics[width=1\textwidth]{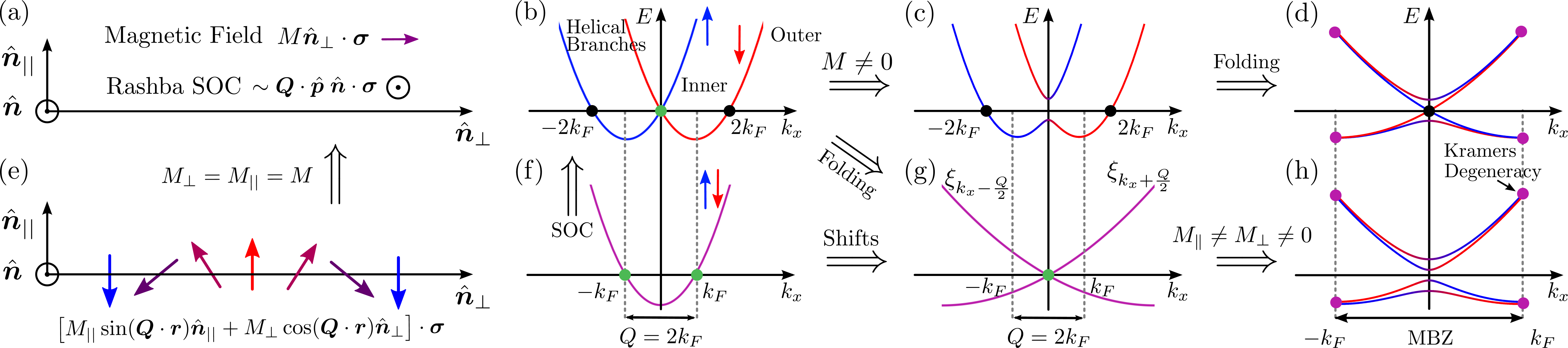}
\caption{(a)-(d) Standard mechanism to achieve 1D topological superconducting phases, which relies on Rashba-like spin-orbit coupling (SOC). (e)-(f) Reconstruction of the bulk spectrum of a spin-degenerate electron gas due to a magnetic helix crystal (MHC) which possesses a spatially-varying magnetic moment for $M_{||}\neq M_{\perp}$. (a) Schematic illustration of ISB SOC pointing in the $\hat{\bm{n}}=\hat{\bm{n}}_\perp\times\hat{\bm{n}}_{||}$ spin direction, in the presence of a homogeneous magnetic field pointing in the $\hat{\bm{n}}_\perp$ direction. (b) Spin-split bands of an electron gas in the absence of a Zeeman/exchange field, resulting in a degeneracy at the ISP $k_{\mathcal{I}}=0$. (c) The combination of SOC and a Zeeman/exchange field perpendicular to it lifts all spin degeneracies in (b). (d) Equivalent description of (c) after downfolding to the magnetic Brillouin zone (MBZ) of (e), where we depict the profile of a MHC with wave vector $\bm{Q}=Q\hat{\bm{n}}_\perp$, winding in the plane spanned by $\hat{\bm{n}}_{||}$ and $\hat{\bm{n}}_{\perp}$. Note that (a) and (e) map to each other for $M_{||}=M_\perp$. (f) shows the two Fermi points which become magnetically scattered in the presence of the MHC. Here, the magnetic wave number $Q$ coincides with the FS nesting wave number $Q_{\rm N}$. (g) Equivalent description of (f) in the MBZ. (h) The MHC opens a full gap at the intersecting point in (g) [green dot in (g)]. In this work, we focus on topological superconducting phases induced by various magnetic textures in 1D and 2D, by means of mechanisms similar to the one described in (f)-(h).}
\label{fig:Figure1}
\end{figure*}

In the majority of the most prominent engineered quasi-1D TSCs, where fingerprints of MZMs {\color{black}appear to have} been expe\-ri\-men\-tal\-ly recorded~\cite{Mourik,MTEarly,Das,Yazdani1,Yacoby,Ruby,Meyer,Molenkamp,MT,Sven,Fabrizio,Yazdani2,Gerbold,HaoZhang,Wiesendanger,Attila,Cren,Kontos,Moodera,Wiebe}, the pre\-sen\-ce of an inversion-symmetry-breaking (ISB) spin-orbit coupling (SOC) is crucial. Its role is to split the initially-degenerate spin bands, with the only re\-mai\-ning de\-ge\-ne\-ra\-cies surviving at inversion-symmetric points (ISPs) $k_{\cal I}$, which satisfy $k_{\cal I}\equiv-k_{\cal I}$. In superconductor-semiconductor nanowires~\cite{AliceaPRB,LutchynPRL,OregPRL} and collinear magnetic chains~\cite{Yazdani1,Brydon,Li,HeimesInter,Cadez}, the additional presence of a Zeeman/exchange field lifts the re\-si\-dual Kramers degeneracies, as sketched in Figs.~\ref{fig:Figure1}(a)-(d). The inclusion of spin-singlet pairing gaps out the remnant Fermi surface (FS) and compensates the magnetic gaps at $k_{\cal I}$, thus effecting the transition to a topological superconducting phase. 

However, there is still a large number of proposals for engineered quasi-1D TSCs which instead rely on a synthetic SOC, which is either induced by a magnetic texture~\cite{Choy,KarstenNoSOC,Ivar,KlinovajaGraphene,NadgPerge,Nakosai,KotetesClassi,Selftuned,Pientka,Ojanen,Sedlmayr,WeiChen,XiaoAn,Paaske,Zutic,Marra,Abiague,MorrSkyrmions,MorrTripleQ,FPTA,MohantaSkyrmion}, or alternatively, by antiferromagnetism~\cite{HeimesSynth} or ferromagnetism~\cite{Livanas} in the pre\-sen\-ce of currents and external or stray Zeeman fields. When it comes to magnetic textures, a magnetic helix crystal (MHC) is the mi\-ni\-mal profile that can induce topological superconducti\-vi\-ty, since it simultaneously ge\-ne\-ra\-tes the required ISB SOC and the per\-pen\-di\-cu\-lar exchange field mentioned above~\cite{BrauneckerSOC}. This is sketched in Figs.~\ref{fig:Figure1}(e)-(h). While a MHC is sufficient to gua\-ran\-tee the occurrence of topological superconductivity in 1D, engineering strong TSCs with a fully-gapped bulk energy spectrum in $d>1$ dimensions requires a magnetic profile which winds in all $d$ directions. Thus, while a MHC leads to spinless p-wave pai\-ring in 1D, a spin skyrmion crystal (SSC) phase is necessary to ge\-ne\-ra\-te an effective spinless chiral p+ip TSC in 2D~\cite{Nakosai}. Re\-mar\-ka\-bly, the key role of magnetic textures in TSCs has been recently highlighted by the expe\-ri\-men\-tal obser\-vations of Refs.~\onlinecite{Wiesendanger,Kontos,Wiesendanger2,Kubetzka,PanagopoulosSkyrmionVortex} where textures were shown to be pi\-vo\-tal for stabilizing to\-po\-lo\-gi\-cal super\-con\-ducti\-vi\-ty. This is also the case in Ref.~\onlinecite{Cren}, where the possible involvement of a skyrmion defect was invoked to explain the appearance of a pair of MZMs in topological magnetic-island heterostructures.

In this Article, we provide {\color{black}an in-depth exploration of the various 1D and 2D TSCs which emerge from the coe\-xi\-sten\-ce of (un)conventional one/two-band spin-singlet super\-con\-duc\-ti\-vi\-ty with a set of representative magnetic texture crystals. Our} star\-ting point is the synthetic SOC mechanism displayed in Fig.~\ref{fig:Figure1}(f)-(h), which opens the door to new and inte\-re\-sting topologically nontrivial phases. In fact, depending on the type of the texture and the strengths of the magnetic and superconducting gaps, we find either a fully-gapped or a nodal bulk ener\-gy spectrum, which give rise to a diversity of MF edge modes. We present a comprehensive classification for each type of to\-po\-lo\-gi\-cal band structure, and account for both strong and weak to\-po\-lo\-gi\-cal superconducting phases, as well as possible strong to\-po\-lo\-gi\-cal crystalline phases stabilized by additional magnetic {\color{black}point/space} group symmetries. 

\begin{table*}[t!]
\centering
\includegraphics[width=0.95\textwidth]{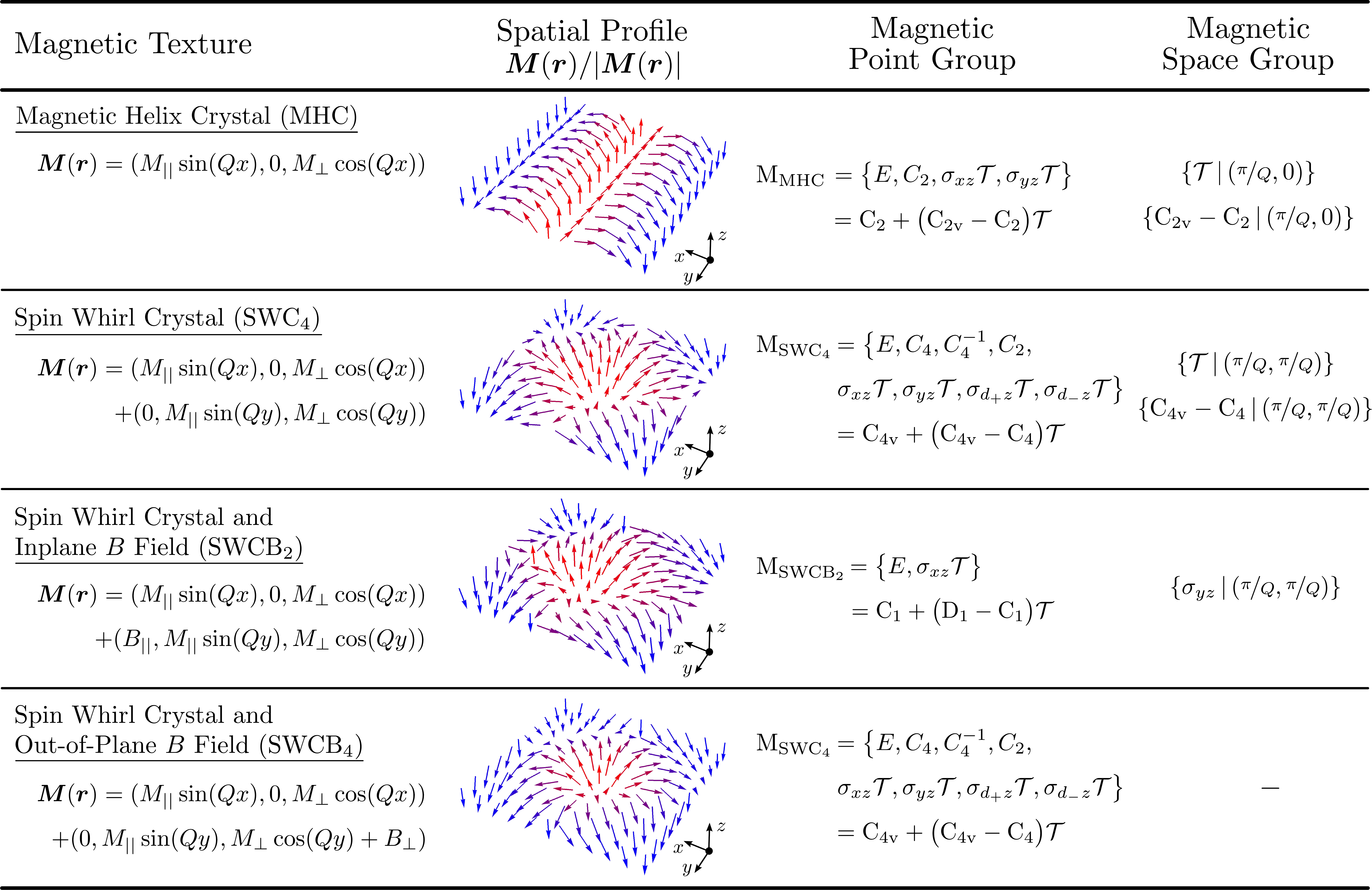}
\caption{Magnetic point and space group elements preserved for the different magnetic texture profiles studied in this work. These build upon the magnetic helix crystal (MHC) and the spin whirl crystal (SWC$_4$) textures. The SWC$_4$ profile is also investigated in the presence of an in (out-of) -plane Zeeman field $B_{||}$ ($B_\perp$). Note that the SWC$_4$ texture with a nonzero $B_\perp$ behaves as the spin skyrmion crystal (SSC$_4$) texture in certain parameter-value regimes. Here $\mathcal{T}$ represents the usual time-reversal operator, and $\sigma_{xz}$ ($\sigma_{yz}$) is the mirror operator in the $xz$ ($yz$) plane. The C$_{4{\rm v}}$ point group consists of the following five conjugacy classes $\{E\}$, $\{C_2\}$, $\{C_4,C_4^{-1}\}$, $\{\sigma_{\rm v}\}$ and $\{\sigma_{\rm d}\}$. Here, $E$ denotes the identity element, $C_n$ denotes a $2\pi/n$ counter-clockwise rotation of the system about the $z$ axis, and $\sigma_{\rm v}$ ($\sigma_{\rm d}$) contain the mirrors $\sigma_{xz,yz}$ ($\sigma_{d_\pm z}$) with vertical mirror planes extending in the $z$ direction and the main (dia\-gonal $d_\pm$ defined as $x=\pm y$) in-plane axis, as denoted by the respective index. Space group elements are denoted above using the Seitz notation, i.e., $\{g\,|\,\mathbf{t}\}$ where $g$ is a point group symmetry of D$_{\rm 4h}$, and $\mathbf{t}$ is a translation by a fraction of the Bravais lattice vectors. Note that deviations from the above generally appear for multiband implementations of the magnetic texture {\color{black}crystal} profiles. Examples of such situations are explored in detail in the main text.}
\label{table:TableI}
\end{table*}

The present work aims at setting a paradigmatic and general framework to study the topological properties arising from the interplay between {\color{black}magnetic texture cry\-stals and spin-singlet superconductivity. Since our analysis relies on the symmetry properties of the magnetic and pairing terms, and not on particular details of the models employed for the concrete demonstration, it} is applicable to a broad range of materials and hybrid devices, inclu\-ding platforms invol\-ving magnetic adatoms deposited on top of superconductors, alongside intrinsic TSCs that do not rely on ISB SOC. Our analysis naturally addresses topological superconducting phases in which magnetism and supercon\-ducti\-vi\-ty are assumed to originate from the same electronic degrees of freedom. Such a scenario {\color{black}may be of direct relevance} to iron-based superconductors (FeSCs), which feature coe\-xi\-sten\-ce of various magnetic phases and superconduc\-ti\-vi\-ty~\cite{ni08a,nandi10a,johrendt11,avci11,hassinger,avci14a,bohmer15a,wasser15,allred15a,allred16a,malletta,klauss15,mallettb,wang16a,zheng16a,meier17}. Among the experimentally discovered phases, {\color{black}one is of yet-unresolved nature}~\cite{wang16a}, and does not match with any of the three well-established commensurate magnetic phases known to exist in FeSCs~\cite{lorenzana08,eremin,brydon, giovannetti,gastiasoro15,wang15,kang15a,christensen15,christensen17}. This commensurate framework was recently extended in Ref.~\cite{Christensen_18} to include incommensurate {\color{black}magnetic texture crystals}. Given the currently incon\-clu\-sive status of the experimental observations, the phase disco\-ve\-red in Ref.~\onlinecite{wang16a} may as well be a magnetic texture {\color{black}crystal}. This opens up new possibilities for to\-po\-lo\-gi\-cal supercon\-duc\-ting phases in FeSCs, which are distinct to the ones that have so far been theoretically~\cite{FeTeSeTopo1,FeTeSeTopo2,KunJiang} and experimentally~\cite{JiaXinYin,hongding1,hongding2,Lingyuan,LingyuanZhu} explored. 

Motivated by the above, in the following we focus on the accessible TSCs in layered te\-tra\-go\-nal itinerant magnets, which possess a D$_{\rm 4h}$ point group symmetry in the nonmagnetic normal phase. We also consider that spin transforms according to the spatial symmetry operations, and we restrict to a single Kramers doublet of the double co\-ve\-ring D$_{\rm 4h}$ group. We additionally assume that ISB SOC and spin anisotropies (cf Ref.~\onlinecite{christensen15}) are negligible. Reference~\onlinecite{Christensen_18} has mapped out the types of single- and double-$\bm{Q}$ textures that such magnets support, and we here focus on the single-$\bm{Q}$ MHC and the fourfold-symmetric double-$\bm{Q}$ spin whirl crystal (SWC$_4$) profiles. We also consider the SWC$_4$ phase in the pre\-sen\-ce of an in- and out-of-plane Zeeman field, $B_{||}$ and $B_\perp$, leading to the here-termed SWCB$_2$ and SWCB$_4$ textures, respectively. Notably, for a range of $B_{\perp}$ values, the SWCB$_4$ texture is equivalent to a fourfold-symmetry pre\-serving SSC phase~\cite{Christensen_18}, which we here denote SSC$_4$. Table~\ref{table:TableI} provides an overview of these magnetic texture {\color{black}crystals}. 

\section{Summary of our Topological Classification Results}\label{sec:SectionII}

To perform the systematic topological classification of the various TSCs induced by the magnetic {\color{black}texture cry\-stals} presented in Table~\ref{table:TableI}, a number of aspects need to be taken into consi\-de\-ra\-tion in regards with the symmetry groups preserved by the magnetic and pairing terms.

A given magnetic texture {\color{black}crystal} preserves the so-called magnetic point group M, which is here a group related to the sub\-groups of the normal-phase symmetry group $\rm D_{4h}$. The elements of M are ge\-ne\-ral\-ly obtained from pro\-ducts of the ori\-gi\-nal double point group operations and time reversal ${\cal T}$. Products in\-vol\-ving ${\cal T}$ give rise to antiunitary mirror symmetries~\cite{Cano2019Oct,ClassiAFM,MagTQChem,Bouhon} which have non\-trivial implications on the topo\-lo\-gi\-cal classification in high-symmetry planes (HSPs), and open the door to novel types of cry\-stal\-line topological phases and MFs~\cite{FuCrystalline,SnTe,ChiuYaoRyu,Morimoto,Shiozaki,Chiu,Fang2014Mar,ZhangLiu,Mendler,ChiuRMP,Kruthoff,Elio,ZhidaSong}. Further information about the symmetry properties of the va\-rious textures considered in this work is listed in Table~\ref{table:TableI}.

\begin{table*}[t!]
\centering
\includegraphics[width=0.95\textwidth]{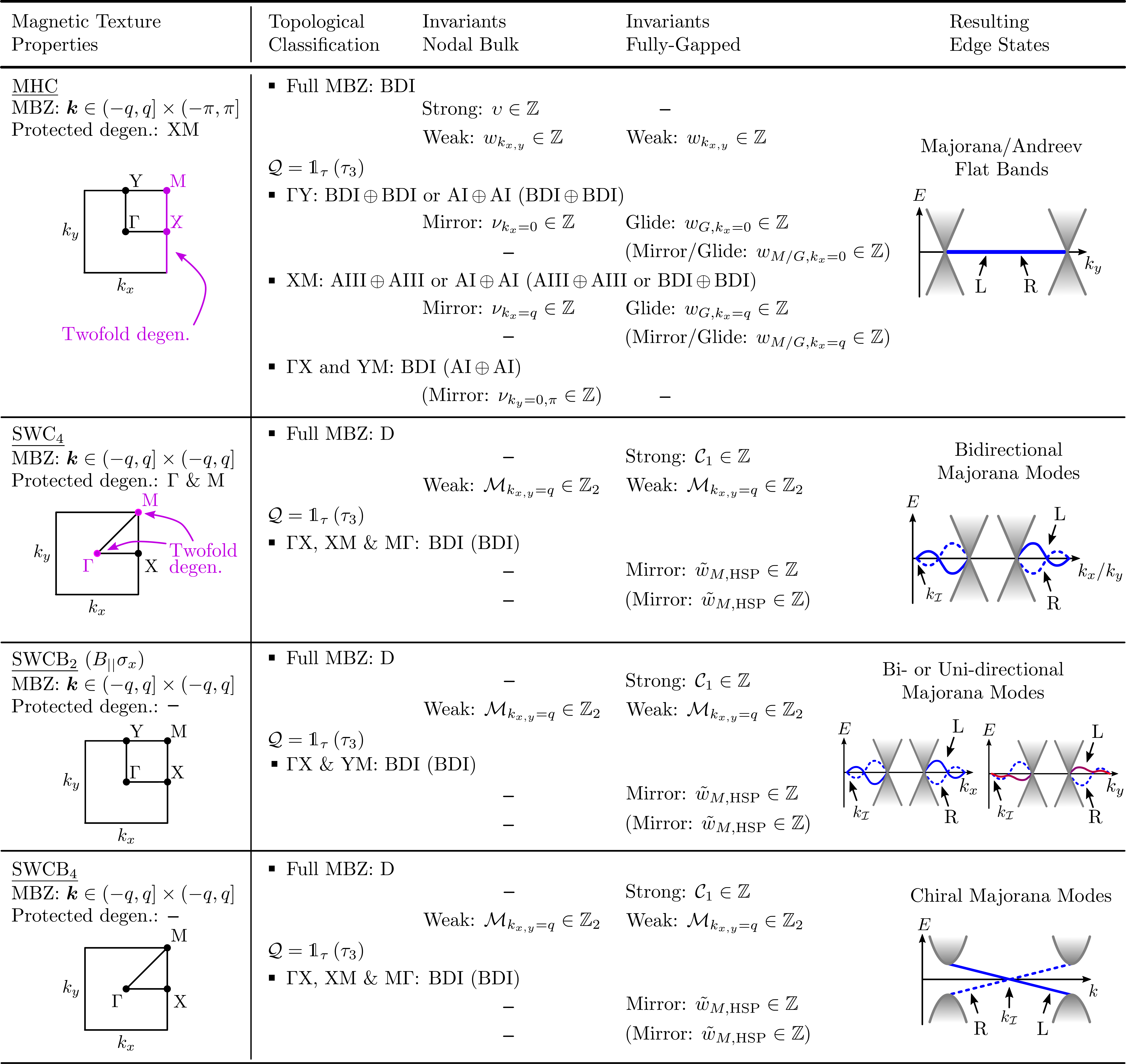}
\caption{Summarizing table of the broad variety of 2D topological superconducting phases induced by a magnetic helix crystal (MHC), a fourfold-symmetric spin whirl crystal (SWC$_4$), and a SWC$_4$ in the presence of in- and out-of-plane Zeeman field (SWCB$_2$ and SWCB$_4$). The present table holds for magnetic textures where both intra- and inter-band contributions are generally present. For each texture we display the respective magnetic Brillouin zone (MBZ), the relevant high symmetry planes (HSPs) and the arising symmetry-protected degeneracies induced by the given magnetic point group, see Table~\ref{table:TableI}. Furthermore, we also list the symmetry classification for the full MBZ and the HSPs, the relevant topological invariants, and the resulting type of Majorana/Andreev edges states. We arrive at three distinct types of invariants which become nontrivial. For a gapless energy spectrum these consist of the vorticities $\upsilon/\nu$ of the arising nodes, while for a fully-gapped energy spectrum, we find the winding number $w$, the Majorana number $\mathcal{M}$, and the 1st Chern number $\mathcal{C}_1$. Each invariant is in addition labeled as strong, weak, mirror or glide, depending on its type. Note that the classification in HSPs presents all the possible topological scenarios obtained by assuming the presence of only a single crystalline symmetry at a time. We further elaborate on these in the proceeding sections. The table also includes the HSP classification in the presence of a pairing function $\Delta_{\bm{k}}$ transforming as one of the irreducible representations (IRs) $\{\rm A_{\rm 1g},\rm B_{\rm 1g},B_{\rm 2g},A_{\rm 2g}\}$ of the group D$_{\rm 4h}$. Depending on the IR of $\Delta_{\bm{k}}$, the classification in HSPs splits into two branches, depending on whether $\Delta_{\bm{k}}$ is even or odd under the original D$_{\rm 4h}$ mirror operation defined for the HSP of interest. In the former case (${\cal Q}=\mathds{1}_{\tau}$), $\Delta_{\bm{k}}$ remains invariant under the original mirror operation, while in the latter (${\cal Q}=\tau_3$), $\Delta_{\bm{k}}$ is invariant under the mirror operation combined with a rotation in Nambu space, which is spanned by the unit $\mathds{1}_\tau$ and Pauli $\bm{\tau}$ matrices. Finally, the red/blue color coding is adopted throughout the text, and reflects the spin up/down orientation of the $z$ component of the electronic spin polarization stemming from the mode appearing on the corresponding left (L) and right (R) edge, when a termination is considered. Left (Right) edge modes are shown above with solid (dashed) lines.}
\label{table:TableII}
\end{table*}

\begin{table*}
\centering
\includegraphics[width=0.95\textwidth]{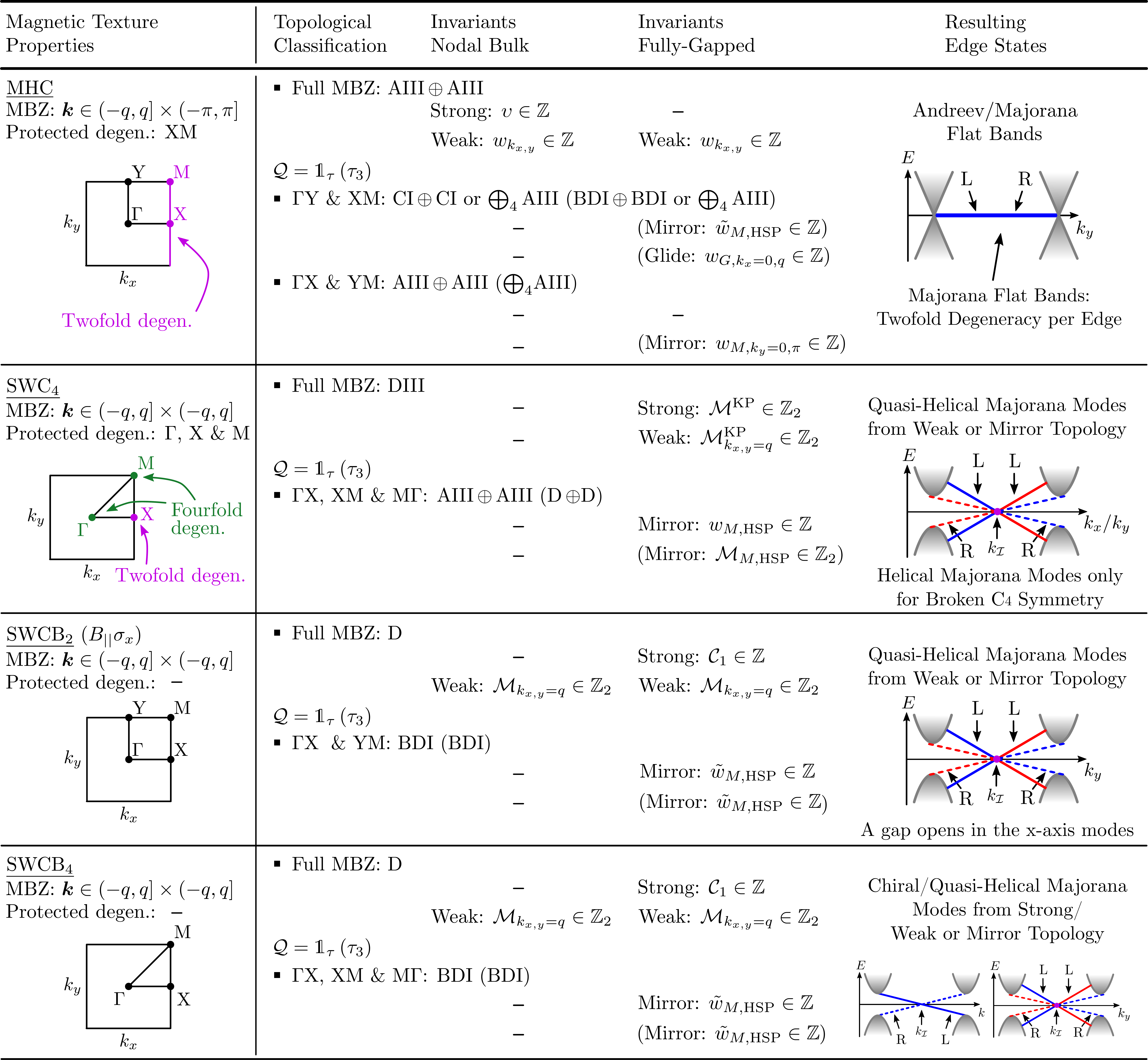}
\caption{Summarizing table of the accessible topological phases induced by interband-only magnetic textures. As in Table~\ref{table:TableII}, also here, we list the topological symmetry classification in the full MBZ and in HSPs, the relevant topological invariants and the resulting edges states, for all four magnetic textures of interest. As in the preceeding table, the above classification is performed for a pairing function $\Delta_{\bm{k}}$ transforming according to one of the $\{\rm A_{\rm 1g}, B_{\rm 1g}, B_{\rm 2g}, A_{\rm 2g}\}$ D$_{\rm 4h}$ IRs. Note that the topological invariants in the MHC case are identical for the two AIII blocks. Hence, we only enlist the invariants for a single AIII block.}
\label{table:TableIII}
\end{table*}

The classification in HSPs is also affected by magnetic space group symmetries~\cite{Ramazashvili,Mong,Slager,KotetesClassi,ZhangLiu,Fang2013Aug,Wang2016Jan,Shiozaki2016May,Yanase,WillXianxin}. Here, ${\cal T}$ or elements of the double D$_{\rm 4h}$ point group are combined with translations which make the texture ``slide'' in the plane~\cite{Sliding}. These constitute exact symmetries of the system only as long as the involved translation also constitutes a lattice translation, which takes place when the magnetic vector is commensurate. Nonetheless, in iti\-nerant magnetic systems the invariance under magnetic translations also emerges in an approximate man\-ner for low energies, since it is the Fermi wavelength rather than the lattice constant which sets the characteristic lengthscale that go\-verns the pro\-per\-ties of the system in that regime. In either situation, nonsymmorphic symmetries en\-rich the topological classification in bulk HSPs and at edges which preserve them.

The final crucial factor which influences the to\-po\-lo\-gi\-cal properties is the type {\color{black}of the pai\-ring point group associated with the pairing gap. In this work, we assume (un)conventional} {\color{black}one/two-band spin-singlet pairing with possible symmetry-imposed or accidental nodes. Going beyond a single band picture} allows us to capture salient features of realistic band structures of correlated magnets, such as the FeSCs~\cite{Raghu_08,TenBand1,Daghofer,ikeda10,graser10,smallAperis,cvetkovic13,TenBand2}. Moreover, depending on which irreducible representation (IR) of D$_{\rm 4h}$ enters the pairing term $\Delta_{\bm{k}}$, i.e., $\{\rm A_{1g},\rm B_{1g},B_{2g},A_{2g}\}$, we ge\-ne\-ral\-ly find a dif\-fe\-rent topological scenario in HSPs, since $\Delta_{\bm{k}}$ may possess symmetry-enforced zeros in these.

Our main findings regarding the rich diversity of TSCs are collected in Tables~\ref{table:TableII} and~\ref{table:TableIII}, for general and interband-only magnetic scattering, respectively. 2D systems in the presence of a MHC texture exhibit protected nodal points which lead to edge Majorana/Andreev flat bands (MFBs/AFBs). Bulk nodal points are also present when considering the SWC$_4$ texture, but are not protected. As a result, edge MFBs/AFBs are not accessible. Nonetheless, a new type of MF edge modes arises, which we denote bidirectional (see also Ref.~\onlinecite{Shiozaki2016May}), since they do not possess a fixed helicity or chirality, and depend strongly on the edge termination. These MF edge modes emerge due to mirror-sym\-metry pro\-tec\-ted de\-ge\-ne\-ra\-cies at ISPs, or alternatively due to weak topological superconducting phases. The addition of a magnetic field can either render the bidirectional MF edge modes unidirectional~\cite{Wong_13,Daido_17,Akbari}, or open a bulk gap and me\-dia\-te a transition to a chiral TSC, analogous to a p+ip superconductor. Remarkably, we find that the multiband character of the systems con\-si\-de\-red here not only allows for a more rea\-li\-stic description, but also results in novel topological superconducting phases and MF edge modes. In par\-ti\-cu\-lar, we show that two-band systems under the influence of multiband magnetic textures harbor Kramers (mirror-symmetry protected) pairs of (quasi-)helical MF edge modes, although time-reversal (TR) symmetry is broken. In fact, these pairs of MFs constitute topologically-protected Andreev zero modes (AZMs) in 1D~\cite{KlinovajaFF,SticletFF,NagaosaFF,MarraFF}. While AZMs have been poorly explored, their to\-po\-lo\-gi\-cal nature also renders them prominent candidates for quantum computing applications~\cite{Zazunov,HigginbothamParity}. 

{\color{black}At this point, it appears vital to underline the va\-rious novelties of the present work in comparison to a number of previous studies concerning TSCs from MHCs in 1D and 2D, cf Refs.~\onlinecite{KarstenNoSOC,Ivar,NadgPerge,KotetesClassi,Selftuned,Pientka,Ojanen,Sedlmayr,WeiChen,XiaoAn,Paaske,Zutic,Abiague,FPTA}, as well as TSCs stemming from the coupling of a conventional spin-singlet superconductor to a SSC$_4$ magnet~\cite{Nakosai,WeiChen,MorrSkyrmions,MorrTripleQ}. {\color{black}Regarding the investigation of topological pro\-per\-ties, the above studies proceeded along two different paths. Either by numerically evaluating a topological invariant or by considering special cases of} magnetic texture cry\-stals described by a magnetization of a spatially-constant mo\-du\-lus, in which cases, the spatial dependence can be fully or approximately removed by appropriate SU(2) rotations~\cite{BrauneckerSOC}. In stark contrast, here we adopt a metho\-do\-lo\-gy that addresses the general case where the magnetic textures possess a spatially-varying modulus. For {\color{black}the purposes of} demonstration, and with no loss of generality, we restrict to crystals which lead to a small number of bands after the reconstruction of the bandstructure. Importantly, this formalism also reveals how to construct low-energy models which facilitate the ana\-ly\-ti\-cal evaluation of the topological invariants{\color{black}. This aids in} the description of the to\-po\-lo\-gi\-cal properties of the system in spite of the added complexity introduced by the superlattice formation. Moreover, our work {\color{black}is not restricted} to the study of the MHC and SSC$_4$ textures, but further includes the investigation of the SWC$_4$, which together with the MHC have been shown to constitute global minima of a general Landau functional describing incommensurate magnetism in iti\-ne\-rant magnets~\cite{Christensen_18}. The topological properties of the SWC$_4$ texture is also analyzed in the presence of additional Zeeman/exchange fields, thus also connecting to the SSC$_4$ phase.

Our work provides a complete description of the topological properties of these systems, by means of a classification which accounts for strong, weak, and cry\-stalline phases. Notably, our analysis also considers the pos\-si\-bi\-li\-ty of unconventional pairing terms. {\color{black}We demonstrate how the interplay} of the magnetic and pai\-ring groups shapes the topological properties of these sy\-stems and gives rise to a multitude of MF excitations, with several of these not have been previously discussed in this context. Hence, our work naturally involves the construction of a number of topological invariants which have not been pre\-viously discussed for TSCs originating from magnetic texture cry\-stals. In fact, we reveal that a number of these topological invariants are responsible for the quantization of physical quantities, such as the bulk staggered magnetization, which can be harnessed to experimentally infer the topological phase of the sy\-stems in question. {\color{black}In contrast to previous works, that are primarily} focused on single-band systems, our work discusses two-band scenarios which open the door to engineering topologically protected {\color{black}Andreev zero/edge modes, as well as Kramers pairs of Majorana solutions. The latter is a notable} result, given the fact that the magnetic texture crystal vio\-lates the standard time-reversal symmetry.}

The remainder of this article is organized as follows. In Sec.~\ref{sec:Models} we describe the modeling assumptions that we use throughout this work. In Sec.~\ref{sec:1D_topological_superconductors}, we investigate TSCs in 1D systems induced by a MHC. In Sec.~\ref{sec:2D_topological_superconductors} we extend our study to 2D, and explore the full variety of possible TSCs and protected MF edge modes induced by the SWC$_4$ phase. {\color{black}The experimental implementation of the various topological scenarios of interest is examined in Section~\ref{sec:ExperimentalStuff}. There, we elaborate on the interplay of magnetic texture crystals and spin-singlet superconductivity, and enumerate prominent platforms that provide fertile ground for their viable coe\-xi\-stence.} Section~\ref{sec:conclusion_outlook} presents our conclusions and outlook. {\color{black}Lastly, Apps.~\ref{app:AppendixA}-\ref{app:Functions} contain various definitions, further technical details, and complementary numerical calculations.}

\section{Modeling Considerations}\label{sec:Models}

\begin{figure}[t!]
\centering
\includegraphics[width=0.95\columnwidth]{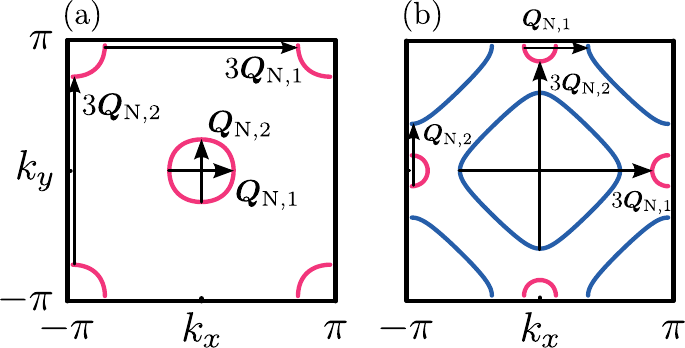}
\caption{One-band (a) and two-band (b) Fermi surfaces stu\-died here. In (a) the two pockets possess the same character, while in (b) the poc\-kets centered at the ${\rm \Gamma}$ and M (${\rm X}$ and ${\rm Y}$) points are of the hole (electron) type. The chosen electron- and hole-characters for the pockets in (a) and (b) allow a qualitative connection with the band structures of certain FeSCs~\cite{ikeda10,graser10}. {\color{black}Note that the results presented in Tables~\ref{table:TableII} and~\ref{table:TableIII} are not particular to the example bandstructures of (a) and (b), but hold generally for sy\-stems which share common magnetic and pairing point/space groups. The precise details of the bandstructures mainly influence the quantitative details of the topological phase diagrams, but not the accessible phases. Moreover,} our results for the two-band model and the ge\-ne\-ral mechanisms underlying the TSC phases can be directly generalized to system with more bands.}
\label{fig:Figure2}
\end{figure}

Before proceeding, we specify the modeling assumptions employed in the upcoming analysis. While our ana\-ly\-ti\-cal and numerical investigations also aim at predic\-ting possible topological phases relevant to unconventional superconductors, we here treat these cases only in a qualitative fashion. Correlated systems generally exhibit complex band structures, which is an aspect that hinders a transparent discussion of the topological pro\-per\-ties as pursued here. For example, an accurate description of the FeSCs typically requires five- or ten-band models~\cite{graser10,ikeda10}. Therefore, in order to ensure a ba\-lan\-ce between analytical tractability and faithful mo\-de\-ling, we focus on simplified one- and two-band mo\-dels\footnote{Note that each one of the band dispersions employed in the upcoming models is chosen to be independently in\-va\-riant under all the $\rm D_{4h}$ point group operations.\label{foot:Footnote1}} which exhibit hole and electron pockets, as well as FS ne\-sting. These features are similar to those exhibited by some FeSCs, see also Fig.~\ref{fig:Figure2}. The one-band (two-band) models are particu\-lar\-ly suitable for exploring to\-po\-lo\-gi\-cal pro\-per\-ties arising from intra-pocket (inter-pocket) nesting.

Another aspect of realistic systems that needs to be accounted for, is the fact that the magnetic wave vectors may be incommensurate or exhibit a high-order degree of commensurability. As a result, such cases require an infinite or very large number of bands for an accurate description after downfolding to the magnetic Brillouin zone (MBZ). To avoid such a complication, we consider commensurate magnetic wave vectors $\bm{Q}_{1,2}$, with the pro\-per\-ty $\bm{k}+4\bm{Q}_{1,2}\equiv\bm{k}$. More precisely $\bm{Q}_1=Q\hat{\bm{x}}$ with $Q=-3\pi/2\equiv\pi/2$, with the lattice constant set to unity throughout, and $\bm{Q}_2=Q\hat{\bm{y}}$, since the latter is obtained via a counter-clockwise fourfold rotation of $\bm{Q}_1$. In most cases we consider that the magnetic wave vectors coincide with the FS nesting wave vectors $\bm{Q}_{{\rm N},1}$ and $\bm{Q}_{{\rm N},2}$, see Fig.~\ref{fig:Figure2}, which is a realistic assumption within {\color{black}an itinerant picture of magnetism which is also of interest in this work}. 

We note that our choice of $Q$ does not affect the ge\-ne\-ra\-li\-ty of the qualitative conclusions regarding the to\-po\-lo\-gi\-cal features of the systems under discussion, as the above wave vectors can be adiabatically connected to incommensurate ones. This is achieved by only deforming the FS of the system without modifying its to\-po\-lo\-gy, i.e., {\color{black}assuming that neither Lifshitz~\cite{Lifshitz,LifshitzVolovik} nor metal-to-insulator transitions~\cite{AA}, occur during this process.} In fact, the analytical tracta\-bi\-li\-ty which is ensured in this manner, allows for a deeper and transparent un\-der\-stan\-ding of the under\-lying mechanisms. Our conclusions thus serve as a basis for the study of more realistic multiband magnetic superconductors.

Furthermore, we point out that the upcoming analysis is not self-consistent with respect to the magnetic and superconducting order parameters. While our starting point {\color{black}partly builds upon the results of Ref.~\onlinecite{Christensen_18}, which have been derived using microscopic models, we do not exa\-mi\-ne the fate of the textured magnetic order and its interplay with superconductivity in full detail. The present work has a more explorative character, and thus, we allow for our search to be unconstrained in order to identify the most prominent routes for achieving TSCs. Section~\ref{sec:ExperimentalStuff} contains a detailed discussion of prominent routes and systems to realize the various topological phenomena investigated in this work.} 

We also wish to stress that the crystalline topological properties induced by the magnetic point and space group symmetries, only give rise to topologically-protected modes at boundaries which preserve the symmetries in question. This is in stark contrast to strong topological phases, for which the bulk-edge correspondence enforces the edge states to appear for a boundary of an arbitrary direction. For this reason, in the up\-coming sections we first carry out the classification of the strong TSC phases, and subsequently discuss the possible crystalline ones which result from the magnetic point and space group symmetries. {\color{black}On the other hand, weak TSCs are distinct to both crystalline and strong phases. Weak TSCs can be viewed as a network of strong TSCs in one spatial dimension lower. For instance, 2D weak to\-po\-lo\-gi\-cal phases emerge as a collection of strong 1D TSCs orien\-ted along a given axis. Consequently, such weak 2D TSCs do not give rise to topologically-protected modes along edges which are parallel to the characteristic axis determined by the constituent strong 1D TSCs. However, topologically-protected modes appear for any other edge orientation, and the characteristics of the edge-mode di\-spersions vary de\-pen\-ding on the orientation of the edge.}

{\color{black}To this end, we clarify that the results of Tables~\ref{table:TableII} and~\ref{table:TableIII} are not only applicable to the specific magnetic texture crystals depicted in Table~\ref{table:TableI}. Instead, these also hold for any other magnetic texture crystal combined with a pairing term which belong to the point and space groups which are studied in this work. Even more, our results do not rely on any par\-ti\-cu\-lar details of the underlying band dispersions, but only on the general assumption that these preserve ${\cal T}$ and each band is individually invariant under the point and space group operations go\-ver\-ning the normal phase, see also Footnote~\ref{foot:Footnote1}.}

Concluding this section, we wish to emphasize that we have confirmed the validity of all the analytical results presented in the upcoming sections by means of nu\-me\-ri\-cal investigations of the respective lattice models in 1D and 2D, with open boundary conditions, and observed the predicted edge states. 

\section{1D Topological Superconductors}\label{sec:1D_topological_superconductors}

In the following analysis we first explore possible topological phases in 1D. Apart from being simple to inve\-sti\-gate, the 1D case sets the stage and the formalism employed in the study of 2D systems, which is the main goal of this work. In 1D we consider the topological effects arising from the coexistence of either a conventional or an unconventional superconducting pairing with a MHC phase with either a constant or a spatially-varying magnitude of the magnetic moment. 

\subsection{One-band models}\label{subsec:1BM_1D}

We begin our study with one-band models (1BMs), which are defined on a lattice and describe electrons with a band dispersion $\xi_{k_x}$ set by the hopping matrix elements $t_{nm}$. After including the chemical potential $\mu$, we have:
\bea
H_0=-\sum_{n,m}\bm{\psi}^{\dag}_n\big(t_{nm}+\mu\delta_{nm}\big)\mathds{1}_{\sigma}\bm{\psi}_m\,,\label{eq:free_ham_1D_1BM}
\eea

\noi where the integers $n,m$ label the positions $R_{n,m}$ on the direct lattice, and $\bm{\psi}_n=(\psi_{n \uparrow},\psi_{n \downarrow})^{\intercal}$, with $^{\intercal}$ denoting transposition. The operator $\psi_{n\sigma}$ ($\psi_{n\sigma}^{\dag}$) annihilates (crea\-tes) an electron with spin projection $\sigma=\uparrow,\downarrow$ at the lattice site $R_n$. The magnetic part of the Hamiltonian describes a MHC texture winding in the $xz$ spin plane:
\begin{align}
H_{\rm mag}=\sum_n\bm{\psi}^{\dag}_n\big[M_{||}\sin(QR_n)\sigma_x+M_\perp\cos(QR_n)\sigma_z\big]\bm{\psi}_n\,.
\label{eq:MH}
\end{align}
\noi Here, $Q=\pi/2$ denotes the magnetic ordering wave number, which in low-dimensional itinerant magnets it often happens to coin\-ci\-de with the nesting wave number $Q_{\rm N}$. The texture mediates scattering between two distinct pairs of points. In this work, one pair usually lies at high ener\-gies and the other at low. When the condition $Q_{\rm N}=Q$ is met, the latter pair is identified with the two nested Fermi points. See Fig.~\ref{fig:Figure3}(a) for a concrete example. We observe that when $|M_\perp|\neq |M_{||}|$ ($|M_\perp|=|M_{||}|$) the MHC leads to a spatially va\-rying (constant) magnetic moment. Below we examine each case separately.

\subsubsection{MHC with a spatially constant magnetic moment: $|M_\perp|=|M_{||}|$}\label{sec:MH_constant_moment}

In this case, we follow Ref.~\onlinecite{BrauneckerSOC} and gauge away the spatial dependence of the magnetization profile through a unitary transformation $\bm{\psi}_n\mapsto \hat{U}\bm{\psi}_n$ with $\hat{U}=\exp\left(iqR_n\sigma_y\right)$, where $q=Q/2$. By employing the plane-wave basis, the single-particle Hamiltonian reads:
\bea
\hat{h}_{k_x}=\xi_{k_x\mathds{1}_{\sigma}-q\sigma_y}+M\sigma_z=\xi_{k_x;q}^+\mathds{1}_{\sigma}+\xi_{k_x;q}^{-}\sigma_y+M\sigma_z,\quad\label{eq:single_particle_ham_1BM}
\eea

\noi with $M_\perp=M_{||}=M>0$. The dispersions read:
\bea 
\xi^{\pm}_{k_x;q}=\big(\xi_{k_x-q}\pm\xi_{k_x+q}\big)/2\,.\quad\label{eq:ShiftedDispersion}
\eea

\begin{figure}[t!]
\centering
\includegraphics[width=0.95\columnwidth]{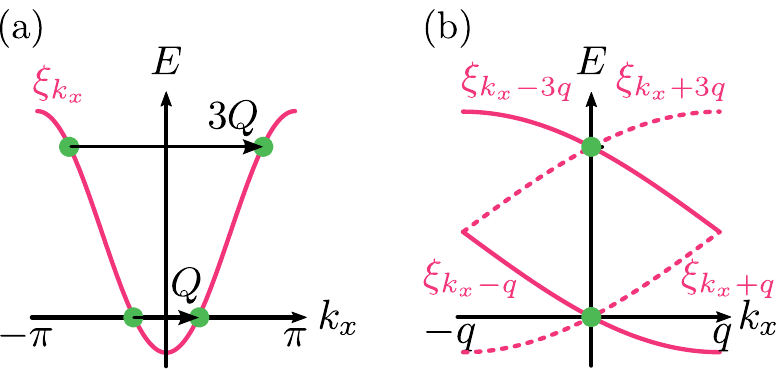}
\caption{Example of a 1BM in 1D, obtained by considering a nearest neighbor hopping with strength $t$ and a che\-mi\-cal po\-ten\-tial value $\mu=-\sqrt{2}t<0$. (a) The resulting dispersion $\xi_{k_x}=-2t\cos k_x-\mu$ contains two Fermi points at $k_x=\pm\pi/4$ of the 1st BZ, which are connected by $Q_{\rm N}=Q=\pi/2\equiv-3\pi/2=-3Q$. The texture also mediates the scattering between two points lying at $E=2|\mu|$. (b) The resulting four bands of the dispersion in the MBZ. The points connected by the magnetic wave vectors are depicted with green dots.}
\label{fig:Figure3}
\end{figure}

\noi One observes that the spin-dependent shift of the wave number splits the dispersion into an even and an odd function under inversion, i.e., $\xi^{\pm}_{-k_x;q}=\pm\xi^{\pm}_{k_x;q}$. The emergence of the odd function reflects the induction of a Rashba-type SOC. By further considering a generic spin-singlet pairing gap $\Delta_{k_x}$, we obtain the Bogoliubov - de Gennes (BdG) Hamiltonian:
\bea
\hat{\cal H}_{k_x}&=&\tau_3\otimes\big(\xi_{k_x;q}^+\mathds{1}_{\sigma}+\xi_{k_x;q}^-\sigma_y\big)+M\mathds{1}_{\tau}\otimes\sigma_z\no\\
&+&\tau_1\otimes\big(\Delta^+_{k_x;q}\mathds{1}_{\sigma}+\Delta^-_{k_x;q}\sigma_y\big),
\label{eq:isotropic_spiral_bdg}
\eea

\noi where we introduced the superconducting gaps $\Delta_{k_x;q}^{\pm}$ in a similar fashion to $\xi_{k_x;q}^{\pm}$ in Eq.~(\ref{eq:ShiftedDispersion}), as well as the spinor:
\bea
\bm{\Psi}_{k_x}^{\dag}=\left(\psi_{k_x\uparrow}^{\dag},\,\psi_{k_x\downarrow}^{\dag},\,\psi^{}_{-k_x\downarrow},\,-\psi^{}_{-k_x\uparrow}\right)\,.
\label{eq:BdGspinor}
\eea

\noi Hence, the many-body mean-field Hamiltonian reads $H=\frac{1}{2}\sum_{k_x}\bm{\Psi}_{k_x}^{\dag}\hat{\cal H}_{k_x}\bm{\Psi}_{k_x}$. In the above, we employed the $\tau_{1,2,3}$ Pauli matrices defined in Nambu electron-hole space. From now on, we adopt the shorthand notation $A\otimes B\equiv AB$ for Kronecker products, and we omit wri\-ting all identity matrices unless this is deemed necessary for reasons of clarity. 

Leaving aside for the moment the magnetic point and space group symmetries present, the BdG Hamiltonian in Eq.~(\ref{eq:isotropic_spiral_bdg}) resides in the BDI symmetry class with ge\-ne\-ra\-li\-zed TR, charge-conjugation and chiral symmetries effected by the ope\-ra\-tors $\Theta=\mathcal{K}$, $\Xi=\tau_2\sigma_y{\cal K}$ and $\Pi=\tau_2\sigma_y$, respectively. Here, ${\cal K}$ denotes complex conjugation. 

When $\Delta_{k_x}$ leads to a fully-gapped spectrum, the system harbors an integer number of topologically protected MZMs per edge, with the corresponding $\mathbb{Z}$ topological inva\-riant given by the win\-ding number $w$~\cite{Jay}. To define the win\-ding number, we rely on the chiral symmetry dictating the Hamiltonian and block off-diagonalize it via the unitary transformation ${\cal S}=(\Pi+\tau_3)/\sqrt{2}$:
\bea
{\cal S}^{\dag}\hat{\cal H}_{k_x}{\cal S}=\begin{pmatrix}0&\hat{A}_{k_x}\\\hat{A}^{\dag}_{k_x}&0\end{pmatrix}\,.\label{eq:BlockOffDiag}
\eea

\noi Given the above, we calculate $\det(\hat{A}_{k_x})$, which reads:
\bea
\det(\hat{A}_{k_x})&=&\left(\xi_{k_x;q}^+\right)^2+\left(\Delta^+_{k_x;q}\right)^2-\left(\xi_{k_x;q}^-\right)^2-\left(\Delta^-_{k_x;q}\right)^2\no\\
&&-M^2+2i\left(\xi_{k_x;q}^-\Delta^+_{k_x;q}-\xi_{k_x;q}^+\Delta^-_{k_x;q}\right)\label{eq:det_A_1BM}
\eea

\noi and allows us to define the normalized complex function:
\bea
z_{k_x}=\det(\hat{A}_{k_x})/|\det(\hat{A}_{k_x})|\,,\label{eq:zTopoInv}
\eea

\noi and the associated winding number in the complex plane:
\bea
w=\frac{1}{2\pi i}\int_{\text{BZ}}\frac{\text{d}z_{k_x}}{z_{k_x}}\,.\label{eq:winding_number}
\eea

To facilitate the evaluation of $w$, one relies on its invariance under smooth deformations of the Hamiltonian, i.e., deformations that do not lead to any gap closings of the bulk spectrum. Hence, one assumes that the pa\-ra\-me\-ters take such values, so that the system is close to topological phase transitions. In such cases the main contributions to $w$ arise from the gap-clo\-sing points $k_{\rm c}$ of the bulk energy spectrum, determined by $|\det(\hat{A}_{k_{\rm c}})|=0$. This equation yields the conditions ${\rm Im}\big[{\rm det}(\hat{A}_{k_{\rm c}})\big]=0$ and ${\rm Re}\big[{\rm det}(\hat{A}_{k_{\rm c}})\big]=0$, whereof the first one reads:
\bea
{\rm Im}\big[\det(\hat{A}_{k_{\rm c}})\big]=\xi_{k_{\rm c};q}^-\Delta_{k_{\rm c};q}^+-\xi_{k_{\rm c};q}^+\Delta_{k_{\rm c};q}^-=0\no\\
\Rightarrow\phd\xi_{k_{\rm c}-q}\Delta_{k_{\rm c}+q}=\xi_{k_{\rm c}+q}\Delta_{k_{\rm c}-q}\,.\quad\phd
\eea 

\noi The above is always satisfied at the ISPs ($k_{\cal I}\equiv-k_{\cal I}$). If we momentarily assume that the $k_x$ dependence of the pairing gap does not lead to any additional gap clo\-sings besides the ones at the ISPs, and take into account the remai\-ning gap-clo\-sing condition ${\rm Re}\big[\det(\hat{A}_{k_{\rm c}})\big]=0$, we obtain the topological phase transition criterion:
\bea
M=\sqrt{(\xi_{k_{\cal I};q}^+)^2+(\Delta_{k_{\cal I};q}^+)^2},\label{eq:nested_points_1db}
\eea

\noi because $\xi_{k_{\cal I};q}^-=\Delta_{k_{\cal I};q}^-=0$. Since the 1D BZ contains only the two ISPs $k_{\cal I}=\{0,\pi\}$, the winding num\-ber reads: 
\bea
&&w=\sum_{k_{\cal I}=0,\pi}{\rm sgn}\left.\left(\Delta_{k_x;q}^+\frac{\text{d}\xi_{k_x;q}^-}{\text{d}k_x}-\xi_{k_x;q}^{+}\frac{{\rm d}\Delta^-_{k_x;q}}{{\rm d}k_x}\right)
\right|_{k_x=k_{\cal I}}\no\\
&&\qquad\quad\phd\phd\ph\cdot\frac{{\rm sgn}\big[M^2-(\Delta_{k_{\cal I};q}^+)^2-(\xi^+_{k_{\cal I};q})^2\big]}{2}\,.
\label{eq:general_winding_1BM_iso}
\eea

\begin{figure}[t!]
\centering
\includegraphics[width=0.95\columnwidth]{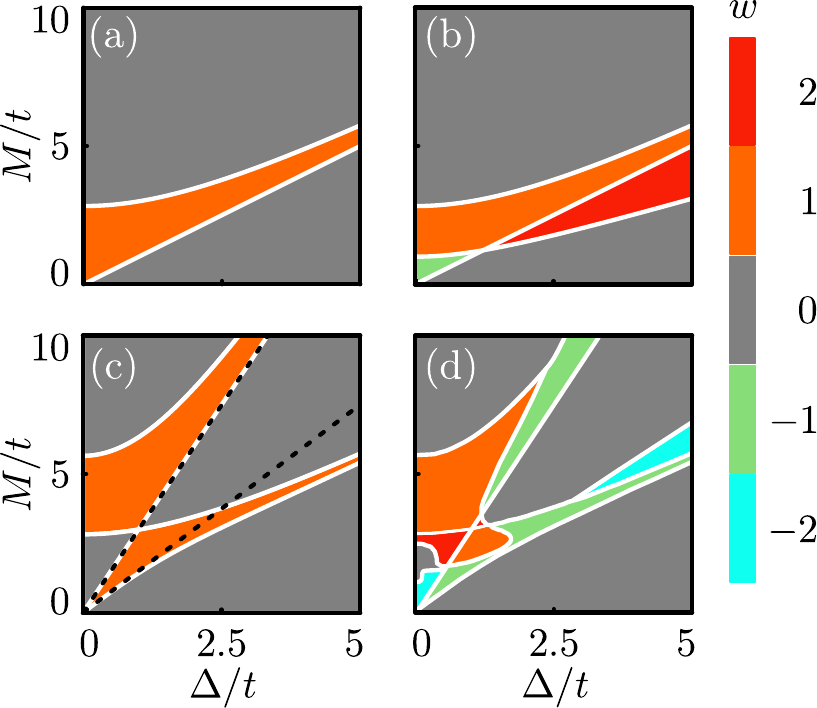}
\caption{Topological phase diagrams for the 1BM in Fig.~\ref{fig:Figure3} in the MHC phase for: (a) a constant pairing gap $\Delta_{k_x}=\Delta$ and (b) an unconventional pairing gap  $\Delta_{k_x}=\Delta\big[1+2\cos(2k_x)\big]$, which leads to additional gap-closing points. Both (a) and (b) were obtained for $Q=Q_{\rm N}$ and a spatially-constant magnetic moment, i.e., $M_\perp=M_{||}=M>0$. Panels (c) and (d) display the same as in (a) and (b), respectively, but for a spatially-varying magnetic moment with $M_\perp=M>0$ and $M_{||}=M/3$. The black dotted lines in (c) span the topologically nontrivial region in the weak-coupling limit, see Eq.~\eqref{eq:1BM_Low_Energy_Criterion}. As one observes, this limit is only valid when $0<M_{||,\perp},\Delta\ll2|\mu|$ are satisfied.}
\label{fig:Figure4}
\end{figure}

We note that when the magnetic wave vector $Q\hat{\bm{x}}$ coin\-ci\-des with the FS nesting vector $Q_{\rm N}\hat{\bm{x}}$, the expression for the topological invariant further simplifies since $\xi^+_{k_{\cal I};q}=0$ for at least one of the two ISPs. Remarkably, this special, but actually realistic situation, appears to be the sweet spot for entering the topologically nontrivial phase, since the minimum magnetic gap $M_{\rm c}=|\Delta_{k_{\cal I};q}^+|$ is required in this case. Away from this special point, the critical magnetic gap increases, and its value is controlled by the degree of the $|Q-Q_{\rm N}|$ detuning, which is reflected in the size of $|\xi_{k_{\cal I};q}^+|$. Therefore, a topological phase transition still occurs even if the special condition $Q=Q_{\rm N}$ is not met. We thus conclude that the system exhibits gap closings at wave numbers for which $\xi_{k_x+Q}=\xi_{k_x}$ and ${\rm d}\xi_{k_x+Q}/{\rm d}k_x=-{\rm d}\xi_{k_x}/{\rm d}k_x$. Notably both conditions hold trivially for two nested Fermi points given that $Q=Q_{\rm N}$. 

For a 1BM model with $\xi_{0;q}^+=0$ and $\xi_{\pi;q}^+=2|\mu|$, which happens to hold for the 1BM in Fig.~\ref{fig:Figure3}, Eq.~\eqref{eq:general_winding_1BM_iso} yields:
\bea
w=\frac{1}{2}\sum_{k_{\cal I}=0,\pi}e^{ik_{\cal I}}{\rm sgn}\left[M^2-(\xi^+_{k_{\cal I};q})^2-(\Delta_{k_{\cal I};q}^+)^2\right]\,,\quad
\label{eq:invariant_k_y_0}
\eea

\noi which implies that the topologically nontrivial regime is realized in the interval:
\bea
\sqrt{(\xi^+_{0;q})^2+(\Delta_{0;q}^+)^2}<M<\sqrt{(\xi^+_{\pi;q})^2+(\Delta_{\pi;q}^+)^2}\,.
\label{eq:nontrivial_interval}
\eea

\noi Note, however, that the upper boundary may not be reached in practice, since for this purpose, magnetic gap values larger than the Fermi energy are required. 

In Fig.~\ref{fig:Figure4}(a), we numerically determine the topological pha\-se dia\-gram for a con\-ven\-tio\-nal s-wave pairing gap $\Delta_{k_x}=\Delta>0$, given the dispersion in Fig.~\ref{fig:Figure3}. The orange region displays the parameter space, for which, the system is in a topologically nontrivial phase with $w=1$. For the parameters used in the figure, Eq.~\eqref{eq:nontrivial_interval} reduces to $\Delta<M<\sqrt{8t^2+\Delta^2}$, and coin\-ci\-des with the numerically-obtained upper and lower bounds of the nontrivial region.

We now continue with addressing the case of an unconventional superconducting order parameter which ge\-ne\-ra\-tes additional gap closings away from ISPs. As a result of chiral and charge-conjugation symmetries, the additional gap closings come in pairs $\pm k_*$~\cite{HeimesInter}, and thus each pair of gap-closing points of this type generally contributes with $\pm1$ units to $w$. For an illustration, we consider the gap function $\Delta_{k_x}=\Delta\big[1+2\cos(2k_x)\big]$ which features a single pair of such nodes. The latter nodes have an accidental origin, since they are not imposed by the presence of a symmetry, and further contribute to the winding number of Eq.~\eqref{eq:invariant_k_y_0}. 

A numerically-obtained example for this case is depicted in Fig.~\ref{fig:Figure4}(b). The regions with $w=-1$ and $w=+2$ appear due to the fact that the signs of the fractional contributions arising from the $k_{\cal I}=0,\pi$ points are no longer determined by Eq.~\eqref{eq:invariant_k_y_0}, as a result of the unconventional pairing. Therefore, the contributions of the $k_{\cal I}$ points for $w=-1$ ($w=+2$) cancel out (add up to $+1$), while the contribution from the $\pm k_*$ gap-closing points is $-1$ ($+1$). Thus, the inclusion of an unconventional pairing function which leads to additional gap clo\-sings at $\pm k_*$, does not significantly alter the nontrivial region here, but does increase the overall complexity of the phase diagram.

We conclude this section by discussing the impact of the various magnetic point group symmetries on the topological classification, given that the nonmagnetic part of the BdG Hamiltonian is invariant under the symmetry group of the normal phase. The addition of magnetism reduces the initial point group down to the magnetic point group ${\rm M}_{\rm MHC}={\rm C}_2+({\rm C_{2v}}-{\rm C_{2}})\mathcal{T}$, whose elements are presented in Table~\ref{table:TableI}. Any possible implications of the magnetic point group on the topological pro\-per\-ties of the system are asso\-cia\-ted with the emergence of the two \textit{antiunitary} mirror symmetries ${\rm(C_{2v}-C_2)}\mathcal{T}=\{\sigma_{xz}{\cal T},\sigma_{yz}{\cal T}\}$. Their presence implies that the symmetry classes for the $xz$ and $yz$ HSPs generally differ from the BDI class which was obtained by solely considering the ge\-ne\-ra\-li\-zed TR symmetry $\Theta={\cal K}$. This is because these anti\-uni\-tary mirror symmetries act as additional ge\-ne\-ra\-li\-zed TR symmetries in the HSPs. Furthermore, the pre\-sen\-ce of this set of three TR symmetries induces two \textit{unitary} symmetries $\big\{{\cal O}_{xz},{\cal O}_{yz}\big\}$, {\color{black}whose presence allows for} the block dia\-go\-na\-li\-zation of the Hamiltonian.

It is customary to describe these effects in terms of $\{\Theta,\Xi\}$ and their (anti)commutation relations with the generators of the induced unitary symmetries $\{{\cal O}_{xz},{\cal O}_{yz}\}$~\cite{ChiuYaoRyu,Morimoto,Shiozaki,ChiuRMP}, whose expressions are inferrable from the two \textit{unitary} mirror symmetries: 
\bea
\mathcal{R}={\rm(C_{2v}-C_2)}\mathcal{T}\Theta\equiv\{\mathcal{R}_{xz},\,\mathcal{R}_{yz}\}\,.
\eea 

\noi Hence, by restricting to the HSPs, we find the expressions ${\cal O}_{xz}\equiv{\cal R}_{xz}=\mathds{1}_{\sigma}$ and ${\cal O}_{yz}\equiv{\cal R}_{yz}=\sigma_z$. 

For the 1D system examined here ${\cal R}_{xz}$ is trivial, while ${\cal R}_{yz}$ is expected to only affect the bulk topological pro\-per\-ties of the system, since it is violated when boun\-da\-ries are introduced. Specifically, ${\cal R}_{yz}$ can influence the classification at the mirror-symmetry-invariant points $k_{\mathcal{R}_{yz}}=\{0,\pi\}$ satisfying ${\cal R}_{yz}k_{{\cal R}_{yz}}\equiv k_{{\cal R}_{yz}}$. For these points $\xi_{k_{{\cal R}_{yz}};q}^-=\Delta^-_{k_{{\cal R}_{yz}};q}=0$, and the BdG Hamiltonian in Eq.~\eqref{eq:isotropic_spiral_bdg} becomes block diagonal and reads:
\bea
\hat{\cal H}_{k_{{\cal R}_{yz}},\sigma}=\sigma M+\xi_{k_{{\cal R}_{yz}};q}^+\tau_3+\Delta^+_{k_{{\cal R}_{yz}};q}\tau_1,
\label{eq:isotropic_spiral_bdg_little}
\eea

\noi where we used the eigenvectors of $\sigma_z$, which are labelled by $\sigma=\pm1$. The above Hamiltonian belongs to symmetry class AI$\oplus$AI with $\Theta={\cal K}$, which in 0D yields the mirror topological invariant $n_M\in\mathbb{Z}$. See Refs.~\onlinecite{ChiuYaoRyu,VolovikBook}.

To calculate $n_M$, we choose an approach which keeps the va\-riety of the topological-inva\-riant constructions used in this work to a minimum. This is achieved by eva\-lua\-ting the mirror invariant in an augmented space, which is spanned by the spatial dimensions relevant for the classification, and the continuous frequency $\epsilon\in(-\infty,\infty)$ obtained as the zero-temperature limit of the Ma\-tsu\-ba\-ra frequencies. Adding $\epsilon$ compensates the reduction of the phy\-si\-cal dimensions by one, which occurs when re\-stric\-ting to a HSP. As a result, this allows us to describe both bulk and HSPs using a winding number defined in the natural and augmented 1D spaces, respectively. 

The topological inva\-riant for the AI class in 0D is thus given here by the winding number of the normalized complex number $Z_\epsilon$, obtained from the normalized determinant of the inverse single-particle Matsubara Green function $\hat{{\cal G}}_\epsilon^{-1}=i\epsilon-\hat{\cal H}_{\rm BdG}$. {\color{black}See Refs.~\onlinecite{VolovikBook,Ishikawa,QiTFT} for further details on topological invariants based on Green functions. In the present case}, we have for each $\sigma$ block: 
\bea
n_{k_{{\cal R}_{yz}},\sigma}=\frac{1}{4\pi i}\int_{-\infty}^{+\infty}\frac{{\rm d}Z_{\epsilon,\sigma}}{Z_{\epsilon,\sigma}}\,,\label{eq:Augmented_winding_number}
\eea

\noi with $Z_{\epsilon,\sigma}=-\det\big(\hat{{\cal G}}_{\epsilon,\sigma}^{-1}\big)/|\det\big(\hat{{\cal G}}_{\epsilon,\sigma}^{-1}\big)|$. We thus find:
\bea
-\det\big(\hat{{\cal G}}_{\epsilon,\sigma}^{-1}\big)&=&\big(\xi_{k_{{\cal R}_{yz}};q}^+\big)^2+\big(\Delta^+_{k_{{\cal R}_{yz}};q}\big)^2-(i\epsilon)^2\no\\&&-M^2+2i\sigma\epsilon M\,,
\eea

\noi which has a form similar to that of the determinant in Eq.~\eqref{eq:det_A_1BM}, that was employed to calculate the winding number $w$ in $k_x$ space.\footnote{Note that the Green function approach could have been also used to evaluate the win\-ding number $w$ in Eq.~\eqref{eq:winding_number}, by means of a 1st Chern number in the augmented $(\epsilon,k_x)$ space. This method has clear advantages if we wish to include possible self-energy effects.} To be in accordance with Ref.~\onlinecite{ChiuYaoRyu}, given the definition in Eq.~\eqref{eq:Augmented_winding_number}, we define the mirror invariant $n_M$ using a single $\sigma$ block of our choice as follows:
\begin{align}
n_M=2{\rm sgn}\big(n_{k_x=0,\sigma}-n_{k_x=\pi,\sigma}\big)\big(|n_{k_x=0,\sigma}|-|n_{k_x=\pi,\sigma}|\big).\label{eq:Ryu_mirror_inv}
\end{align}

$n_M$ counts the number of mirror-symmetry {\color{black}protected edge states for edges} preserving the respective mirror symmetry. However, as we announced, the above inva\-riant becomes obsolete in 1D systems, since mirror symmetry is expected to be broken when termination edges are present. Nevertheless, here we rely on the translational invariance of the system and instead introduce a set of bulk mirror invariants. In analogy to the spin Chern number construction~\cite{KaneMeleI,DNSheng,BHZ}, we define:
\bea
n_{M;k_{{\cal R}_{yz}}}=\sum_\sigma\sigma n_{k_{{\cal R}_{yz}},\sigma}\,.\label{eq:mirror_winding_number_def}
\eea

\noi After evaluating $n_{k_{{\cal R}_{yz}},\sigma}$, we find that $n_{M;k_{{\cal R}_{yz}}}$ becomes:
\begin{align}
n_{M;k_{{\cal R}_{yz}}}=\frac{{\rm sgn}\big[(\xi^+_{k_{{\cal R}_{yz}};q})^2+(\Delta_{k_{{\cal R}_{yz}};q}^+)^2-M^2\big]-1}{2}.
\label{eq:mirror_winding_number_res}
\end{align}

This bulk mirror invariant reflects the quantization of the $z$ axis magnetization\footnote{Recall that the above is expressed in the local spin frame. Ro\-ta\-ting back with $\hat{U}^\dag$, implies that $n_{M;k_{{\cal R}_{yz}}}$ leads to the quantization of the staggered magnetization in the $xz$ spin plane.\label{footnote:localspinframe}} in HSPs, since $\left<\sigma_z\right>_{k_x=k_{{\cal R}_{yz}}}=n_{M;k_{{\cal R}_{yz}}}$. One finds $|n_{M;k_{{\cal R}_{yz}}}|=1$ only after certain level cros\-sings occur at $k_{{\cal R}_{yz}}$, thus bea\-ring si\-mi\-la\-ri\-ties to parity-switching level crossings known for Yu-Shiba-Rusinov bound states~\cite{YSRS}, which are induced by magnetic impurities in spin-singlet superconductors~\cite{BalatskyRMP}. 

The measurement of the $k_x$-resolved magnetization appears experimentally feasible by means of spin-resolved angle resolved photoemission spectroscopy, which has already been successfully applied to map out the spin cha\-rac\-ter of the surface states of TR-invariant 3D topological insulators~\cite{Hsieh}. We thus find that, although the unitary mirror symmetry $\mathcal{R}_{yz}$ is generally broken when edges are introduced to the system, we can still use $n_{M;k_{\mathcal{R}_{yz}}}$ as a bulk experimental probe for topological super\-con\-ducti\-vi\-ty. Furthermore, we remark that the above calculations also serve as a simple example of similar derivations that we plan to carry out in the upcoming sections.

\subsubsection{MHC with a spatially varying magnetic moment: $|M_\perp|\neq|M_{||}|$}\label{sec:1BM_MH_ANI}

We now extend the study of the previous section to the more general si\-tua\-tion, in which $|M_\perp|\neq|M_{||}|$. In this case, using a spin-dependent unitary transformation to gauge away the spatial dependence of the MHC is no longer possible, and one has to approach the problem in the MBZ, i.e., $k_x\in(-q,q]$, with $q=Q/2=\pi/4$. To describe the downfolding to the MBZ, one can either consider a sublattice description which is briefly discussed in App.~\ref{app:AppendixA}, or choose to express the Hamiltonian in $\{k_x,k_x+Q\}$ and $\{k_x,k_x+2Q\}$ wave-number-transfer spaces. The former is advantageous for carrying out the topological classification, since the resulting BdG Hamiltonian is $2\pi$-periodic and thus suitably compactified. The results presented in Table~\ref{table:TableII} were obtained using this approach. However, throughout the main text we follow the second route, which is implemented by intro\-du\-cing the enlarged spinor:
\begin{align}
\bm{{\cal X}}_{k_x}^{\rm 1BM}=\mathds{1}_\eta\otimes\frac{\rho_2+\rho_3}{\sqrt{2}}
\left(\bm{\Psi}_{k_x-q},\bm{\Psi}_{k_x+q},\bm{\Psi}_{k_x+3q},\bm{\Psi}_{k_x-3q}\right)^{\intercal}
\label{eq:Enlarged_Spinor_1BM}
\end{align}

\noi where we introduced the Pauli matrices ($\rho_{1,2,3}$) $\eta_{1,2,3}$ defined in the ($Q$-) $2Q$-transfer space. The above basis reveals more transparently the mechanisms underlying the nontrivial topological properties induced by the magnetic textures, it highlights the emergent Dirac physics, and it provides a simpler and more convenient framework to evaluate the various topological invariants.

The wave-number shifts in the arguments of the above spinor were chosen to connect to the spinor obtained after performing the unitary transformation $\hat{U}$ in the case $|M_\perp|=|M_{||}|$. See also Fig.~\ref{fig:Figure3}(b) and note that, given our spinor choice, the periodic magnetization opens gaps at $k_x=0$ when $Q=Q_{\rm N}$. The extended BdG Hamiltonian reads $\hat{{\cal H}}_{k_x}=\hat{{\cal H}}_{k_x}^0+\hat{{\cal H}}_{\rm mag}$ with:
\bea
\hat{\cal H}_{k_x}^0&=&\left[h_{k_x}^{(0)}+h_{k_x}^{(1)}\rho_2+h_{k_x}^{(2)}\eta_3+h_{k_x}^{(3)}\eta_3\rho_2\right]\tau_3\no\\
&+&\left[\Delta_{k_x}^{(0)}+\Delta_{k_x}^{(1)}\rho_2+\Delta_{k_x}^{(2)}\eta_3+\Delta_{k_x}^{(3)}\eta_3\rho_2\right]\tau_1,\no\\
\hat{\cal H}_{\rm mag}&=&-\frac{M_\perp\big(\mathds{1}+\eta_1\big)\rho_1\sigma_z+M_{||}\big(\mathds{1}-\eta_1\big)\rho_3\sigma_x}{2}\,,\quad
\label{eq:bloch_1BM_ani_helix}
\eea

\noi where the functions $h_{k_x}^{(s)}$ and $\Delta_{k_x}^{(s)}$ appearing above with $s=0,1,2,3$, constitute linear combinations of $\xi_{k_x}$ and $\Delta_{k_x}$, which follow from the general definitions:
\bea
f_{k_x}^{(0)}&=&\big(f_{k_x-q}+f_{k_x+q}+f_{k_x+3q}+f_{k_x-3q}\big)/4\,,\quad\label{eq:DefsShiftedVars1}\\
f_{k_x}^{(1)}&=&\big(f_{k_x-q}-f_{k_x+q}+f_{k_x+3q}-f_{k_x-3q}\big)/4\,,\quad\label{eq:DefsShiftedVars2}\\
f_{k_x}^{(2)}&=&\big(f_{k_x-q}+f_{k_x+q}-f_{k_x+3q}-f_{k_x-3q}\big)/4\,,\quad\label{eq:DefsShiftedVars3}\\
f_{k_x}^{(3)}&=&\big(f_{k_x-q}-f_{k_x+q}-f_{k_x+3q}+f_{k_x-3q}\big)/4\,.\quad\label{eq:DefsShiftedVars4}
\eea

The above construction further implies that inversion $k_x\mapsto-k_x$, acts as $f_{-k_x}^{(s)}=(-1)^sf_{k_x}^{(s)}$. One observes that the four linear combinations resulting from the electron part $h_{k_x}^{(0)}+h_{k_x}^{(1)}\rho_2+h_{k_x}^{(2)}\eta_3+h_{k_x}^{(3)}\eta_3\rho_2$ of the nonmagnetic BdG Hamiltonian, give rise to the four spin-degenerate band segments in the MBZ shown in Fig.~\ref{fig:Figure3}(b).

We point out that the set of ISPs in the MBZ reads $k_{\cal I}=\{0,q\equiv-q\}$, while the $k_{\cal I}=\pi$ of the original BZ coincides now with $k_{\cal I}=0$ in the MBZ. Note that the inversion-symmetric nature of $k_x=\pm q$ is established via the equivalence relation $-q\equiv q+nQ$ with $n\in\mathbb{Z}$.

The magnetic point group dictating the above BdG Hamiltonian is identical to the one examined in the pre\-vious section for a MHC with a spatially-constant magnetic moment ($|M_\perp|=|M_{||}|$). However, before discussing its implications on the topological classification, we point out that given the enlarged basis in Eq.~\eqref{eq:Enlarged_Spinor_1BM}, which is indispensable here for the description of the MHC, one needs to account for possible space group symmetries. 

Spe\-ci\-fi\-cal\-ly, as also presented in Table~\ref{table:TableI}, the additional space group symmetries $\{\mathcal{T}\,|\,\nicefrac{\pi}{Q}\}$ and $\{\sigma_{xz,yz}\,|\,\nicefrac{\pi}{Q}\}$ become now relevant. Throughout this work, we adopt the Seitz notation $\{g\,|\,\mathbf{t}\}$, which combines the point group element $g$ with the translation $\mathbf{t}$. Given our choice of basis, $\{\mathds{1}\,|\,\nicefrac{\pi}{Q}\}=-ie^{ik_x\pi/Q}\rho_2$ and ${\cal T}=i\sigma_y{\cal K}$ as usual, while the mirror operations have the fol\-lowing $\rho\otimes\sigma$ space matrix structure, i.e., $\hat{\sigma}_{yz}=i\rho_1\sigma_x$ and $\hat{\sigma}_{xz}=i\sigma_y$. Hence, we find the unitary symmetries with $\{\hat{\sigma}_{yz}\,|\,\nicefrac{\pi}{Q}\}=ie^{ik_x\pi/Q}\rho_3\sigma_x$, $\{\hat{\sigma}_{xz}\,|\,\nicefrac{\pi}{Q}\}=e^{ik_x\pi/Q}\rho_2\sigma_y$, and the antiunitary symmetry $\tilde{\Theta}_{k_x}\equiv\{{\cal T}\,|\,\nicefrac{\pi}{Q}\}=e^{ik_x\pi/Q}\rho_2\sigma_y{\cal K}$. In contrast to the physical TR operation which satisfies ${\cal T}^2=-\mathds{1}$, here $\tilde{\Theta}_{k_x}^2=e^{i\pi k_x/q}\mathds{1}$ and leads to a Kramers degeneracy only at the $k_x=q$ ISP in the MBZ~\cite{Bradley1968Apr,Bible,Dresselhaus}. Notably, this is the mechanism underlying the persistent Kramers de\-ge\-ne\-ra\-cies at the purple-colored points of the MBZ shown in Fig.~\ref{fig:Figure1}(h). As also discussed in App.~\ref{app:AppendixB}, the above space-group symmetries do not lead to any further symmetry-protected degeneracies in the spectrum, and thus influence the topological classification only in HSPs. For this reason, their implications are discussed later, together with the magnetic point group symmetries.

The extended BdG Hamiltonian belongs to class BDI and is classified by a winding number $w\in\mathbb{Z}$. Applying the methods of Sec.~\ref{sec:MH_constant_moment}, and assu\-ming for simplicity that $Q=Q_{\rm N}$, and $\Delta_{k_x}=\Delta>0$ so that $\Delta_{k_x}^{(0)}=\Delta$ and $\Delta_{k_x}^{(s)}=0$ for $s=1,2,3$, lead to (see App.~\ref{app:AppendixC1} for details):
\begin{align}
w=\sum_\nu\frac{\nu}{2}{\rm sgn}\left\{\big(M_\nu^2-\Delta^2\big)-\frac{\big(M_\perp^2-\Delta^2\big)\big(M_{||}^2-\Delta^2\big)}{(2\mu)^2}\right\},
\label{eq:winding_number_1BM_ani}
\end{align}

\noi with $M_\nu=(M_{||}+\nu M_\perp)/2$ and $\nu=\pm1$. 

Figure~\ref{fig:Figure4}(c) depicts the topological phase diagram for the 1BM in Fig.~\ref{fig:Figure3} when $M_\perp=M$ and $M_{||}=M/3$. The orange regions are phases with a single MZM per edge. While the anisotropic nature of the MHC does not lead to the removal of the topologically nontrivial phase, it still si\-gni\-fi\-cantly modifies the phase dia\-gram. It is straightforward to verify that for $\Delta_{k_x}=\Delta$, the gap clo\-sings responsible for the topological phase transition take place only at $k_x=0$. Nonetheless, Eq.~\eqref{eq:winding_number_1BM_ani} also holds for an unconventional superconducting order parameter after replacing $\Delta\mapsto\Delta_{k_x=0}$, under the condition that additional gap-closing points do not emerge. Instead, for an unconventional pairing order parameter which leads to additional gap closings at $\pm k_*$, the topological phase diagram ends up to be quite complex. After obtaining $w$ for a generic $\Delta_{k_x}$ (see App.~\ref{app:AppendixC2}), we focus on a pairing gap $\Delta_{k_x}=\Delta\big[1+2\cos(2k_x)\big]$. The related topological phase diagram is depicted in Fig.~\ref{fig:Figure4}(d).

To gain deeper insight, we set $\Delta_{k_x}=\Delta>0$ and restrict to the weak-coupling limit $|\Delta|,|M_{||,\perp}|\ll2|\mu|$. Since gap closings now occur only near the FS, Eq.~\eqref{eq:winding_number_1BM_ani} becomes:
\bea
w=\frac{{\rm sgn}\big(M_+^2-\Delta^2\big)-{\rm sgn}\big(M_-^2-\Delta^2\big)}{2}\,.
\label{eq:winding_number_1BM_Low_Energy}
\eea

\noi Thus, the topological phases arising from gap closings occurring in the low-energy sector, are solely determined by the inequality:
\bea
M_-<\Delta<M_+\quad{\rm for}\quad M_{\perp,||}\geq0\,.\label{eq:1BM_Low_Energy_Criterion}
\eea

\noi The spatial variation of the magnetic moment, which is reflected in the size of the difference $|M_-|=|M_{||}-M_\perp|$, sets a maximum value for the magnetic anisotropy that can still allow for the system to enter the nontrivial phase. For the specific parameters used in Fig.~\ref{fig:Figure4}(c), the above ine\-qua\-li\-ty reduces to $3\Delta/2<M<3\Delta$. The low-energy nontrivial region is therefore spanned by the black dotted lines in Fig.~\ref{fig:Figure4}(c), which verifies that Eq.~\eqref{eq:winding_number_1BM_Low_Energy} indeed describes the exact model well in the weak-coupling limit.

The results in the weak-coupling limit can be alternatively obtained by directly restricting the multicomponent spinor of Eq.~\eqref{eq:Enlarged_Spinor_1BM} to the operators crea\-ting/annihilating electrons in the low-energy sector, i.e.:
\bea 
\bm{\chi}_{k_x}^{\rm 1BM}=\frac{\rho_2+\rho_3}{\sqrt{2}}\big(\bm{\Psi}_{k_x-q},\,\bm{\Psi}_{k_x+q}\big)^{\intercal}\,. 
\eea

\noi The projection of the Hamiltonian in Eq.~\eqref{eq:bloch_1BM_ani_helix} onto this subspace is achieved by setting $\eta_3=+1$, and dropping the term proportional to $\eta_1$ which connects the low- and high-energy sectors. These steps lead to the Hamiltonian:
\bea
\hat{{\cal H}}_{k_x}^{\rm low-en}&=&\left(\xi_{k_x;q}^++\xi_{k_x;q}^-\rho_2\right)\tau_3-\frac{M_\perp\rho_1\sigma_z+M_{||}\rho_3\sigma_x}{2}
\no\\&+&\left(\Delta_{k_x;q}^++\Delta_{k_x;q}^-\rho_2\right)\tau_1\,,
\label{eq:Low_Energy_1BM}
\eea

\noi with $\xi_{k_x;q}^\pm$ and $\Delta_{k_x;q}^\pm$ following once again from Eq.~\eqref{eq:ShiftedDispersion}. 

We now discuss the impact of the unitary mirror ${\cal R}_{xz,yz}$ and space group $\{\sigma_{xz,yz}\,|\,\nicefrac{\pi}{Q}\}$ symmetries on the to\-po\-lo\-gi\-cal clas\-si\-fi\-ca\-tion in HSPs. First of all, $\tilde{\Theta}_{k_x}$ imposes a twofold de\-ge\-ne\-racy at $k_x=q$, thus implying that the magnetic texture does not induce any new topologically nontrivial phases at $k_x=q$. See Table~\ref{table:TableII} and App.~\ref{app:AppendixA} for more details. Even more, as in Sec.~\ref{sec:MH_constant_moment}, also here, the effects of ${\cal R}_{xz}$ are trivial, since it leads to a unitary symmetry ${\cal O}_{xz}=\mathds{1}$. The remaining four symmetries modify the topological properties at $k_x=0$. The symmetries ${\cal R}_{yz}$ and $\{\sigma_{yz}\,|\,\nicefrac{\pi}{Q}\}$ ($\{{\cal T}\,|\,\nicefrac{\pi}{Q}\}$ and $\{\sigma_{xz}\,|\,\nicefrac{\pi}{Q}\}$) lead to a ${\rm AI}\oplus{\rm AI}$ (${\rm BDI}\oplus{\rm BDI}$) class, thus providing an additional $\mathbb{Z}$ ($\mathbb{Z}_2$) topological invariant to the winding number $w$. 

The ${\cal R}_{yz}$ symmetry allows defining two types of mirror invariants, in analogy to Eqs.~\eqref{eq:Ryu_mirror_inv} and~\eqref{eq:mirror_winding_number_def}. The block-diagonalization of the BdG Hamiltonian in Eq.~\eqref{eq:bloch_1BM_ani_helix}, that is effected by the unitary transformation $({\cal O}_{yz}+\sigma_x)/\sqrt{2}$, yields in the weak-coupling limit (see also App.~\ref{app:AppendixC3}):
\bea
2n_{M,k_x=0}=\sum_{\nu=\pm1}\nu\,{\rm sgn}\big[(\xi^+_{0;q})^2+(\Delta_{0;q}^+)^2-M_{\nu}^2\big].\quad
\label{eq:0Dinvariant_k_y_0_ani_sigma_blocks}
\eea

\noi It is straightforward to verify that $n_{M,k_x=0}=n_M$, since no topological gap closings can take place at $k_x=q$. In a similar fashion, the offcentered\footnote{The nonmenclature highlights that this symmetry is not nonsymmorphic, because we can choose a coordinate system in the direct lattice for which $\{\sigma_{yz}\,|\,\nicefrac{\pi}{Q}\}\mapsto\sigma_{yz}$. See also App.~\ref{app:AppendixB}.} space-group symmetry $\{\sigma_{yz}\,|\,\nicefrac{\pi}{Q}\}$ allows introducing the here-termed glide inva\-riant $n_{G,k_x=0}\in\mathbb{Z}$, which is de\-fi\-ned following Eqs.~\eqref{eq:Augmented_winding_number} and~\eqref{eq:mirror_winding_number_def}. As pointed out in Sec.~\ref{sec:MH_constant_moment} and Footnote~\ref{footnote:localspinframe}, also here, the staggered magnetization in the $z$ ($x$) spin axis becomes quantized when $n_{M,k_x=0}$ ($n_{G,k_x=0}$) is nonzero, since $2n_{M,k_x=0}=\left<\rho_1\sigma_z\right>_{k_x=0}$ ($2n_{G,k_x=0}=\left<\rho_3\sigma_x\right>_{k_x=0}$). For the given model $|n_{M,k_x=0}|=|n_{G,k_x=0}|$.

We now proceed with the $\mathbb{Z}_2$ topological invariants which emerge from the $\{{\cal T}\,|\,\nicefrac{\pi}{Q}\}$ and $\{\sigma_{xz}\,|\,\nicefrac{\pi}{Q}\}$ symmetries. In the presence of either one of these, the unitary symmetry $\tilde{\cal O}=\rho_2\sigma_y$ is induced, and allows us to block-diagonalize the Hamiltonian in Eq.~\eqref{eq:bloch_1BM_ani_helix} via the unitary trans\-for\-mation $(\tilde{\cal O}+\sigma_z)/\sqrt{2}$. This yields the BDI blocks:
\bea
\hat{\cal H}_{k_x=0,\sigma}=\hat{\cal H}_{k_x=0}^0-\big(M_\sigma-M_{-\sigma}\eta_1\big)\rho_1\,,\label{eq:GlideBlocks1D}
\eea

\noi with $\sigma=\pm1$ labelling the eigenstates of $\sigma_z$. The two blocks see a chiral symmetry $\Pi=\rho_2\tau_2$. Following the same approach that led to Eq.~\eqref{eq:BlockOffDiag}, we block-off diagona\-li\-ze each $\sigma$ block by means of the unitary transformation $(\Pi+\tau_3)/\sqrt{2}$, with the upper block denoted $\hat{A}_{k_x=0,\sigma}$. The $\mathbb{Z}_2$ invariant, that we here term glide Majorana parity $P_G$, is constructed in App.~\ref{app:AppendixC4} and is defined as:
\bea
P_{G,k_x=0}={\rm sgn}\prod_{\sigma=\pm1}\det\big(\hat{A}_{k_x=0,\sigma}\big)\,.\label{eq:GlideMajorana}
\eea

\noi Within the weak-coupling limit, we obtain the result:
\bea
P_{G,k_x=0}={\rm sgn}\prod_{\sigma=\pm1}\big[(\xi^+_{0;q})^2+(\Delta_{0;q}^+)^2-M_{\sigma}^2\big]\,.\quad\label{eq:GlideMajoranaCalculation}
\eea

\noi Hence, here we end up with $P_{G,k_x=0}=(-1)^{n_{M,k_x=0}}$, which implies that in the present model $P_{G,k_x=0}$ is nontrivial, i.e., equal to $-1$, when $n_{M,k_x=0}\in2\mathbb{Z}+1$.

Concluding this section, we remark once again that in strictly 1D systems the above point and space group symmetries affect only the bulk topological properties, since these are all broken when edges are introduced. When this takes place, it is only $w$ together with the weak $\mathbb{Z}_2$ invariant of class BDI,\footnote{We define the weak invariant $P_{\cal M}={\rm sgn}\prod_{k_{\cal I}}\det(\hat{A}_{k_{\cal I}})\equiv {\rm sgn}[\det(\hat{A}_{k_x=0})]$, with $\hat{A}_{k_{\cal I}}$ the upper off-diagonal block of the block off-diagonalized Hamiltonian in Eq.~\eqref{eq:bloch_1BM_ani_helix} evaluated at $k_{\cal I}$. The arising equivalence is a result of the Kramers degeneracy at $k_x=q$. When either $\{{\cal T}\,|\,\nicefrac{\pi}{Q}\}$ or $\{\sigma_{xz}\,|\,\nicefrac{\pi}{Q}\}$ is a symmetry, $P_{\cal M}$ coin\-ci\-des with the glide Majorana parity $P_{G,k_x=0}$ in Eq.~\eqref{eq:GlideMajorana}.\label{footnote:PM}} which are capable of predicting the number of the arising MZMs. This is in stark contrast to 2D systems, where certain edges also support crystalline and/or weak invariants, as discussed further in later sections.

\subsection{Two-band models}\label{sec:2BM_1D}

\begin{figure}[t!]
\centering
\includegraphics[width=0.95\columnwidth]{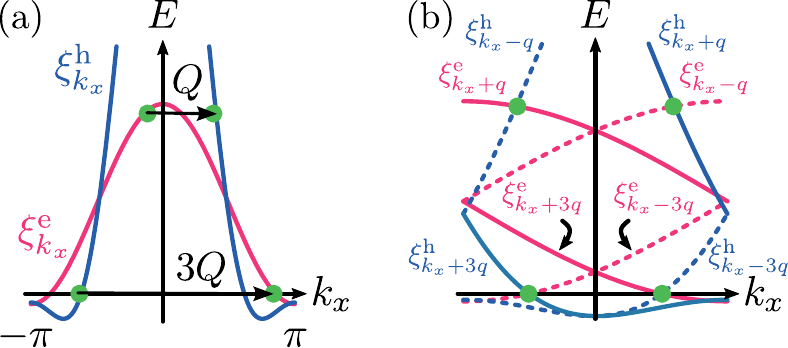}
\caption{Example of a 2BM in 1D. (a) The dispersions $\xi^{\rm e}_{k_x}=t_{{\rm e}}\cos k_x-\varepsilon_{\rm e}$ (fuchsia) and $\xi^{\rm h}_{k_x}=t_{\rm h}\cos k_x+t'_{\rm h}\big[1+\cos(2k_x)\big]-\varepsilon_{\rm h}$ (navy blue) in the first BZ. We set the parameters values $t_{\rm h}=2.86\,t_{\rm e}$, $t'_{\rm h}=t_{\rm e}$, $\varepsilon_{\rm e}\approx-0.92\,t_{\rm e}$ and $\varepsilon_{\rm h}=-0.80\,t_{\rm e}$. We also show the nesting wave vector $Q_{\rm N}\hat{\bm{x}}$ which, for the choice of parameters made here, coincides with the magnetic wave vector $Q\hat{\bm{x}}$. Thus, $Q\hat{\bm{x}}$ connects two pairs of points at the Fermi level and two more pairs at higher energies. (b) The resulting eight band-segments in the MBZ. The points connected by the magnetic wave number ($3Q=-Q$) come in pairs due to $C_2$ symmetry, and are depicted by green dots. For clarity, (a) depicts half of the ordering wave numbers.}
\label{fig:Figure5}
\end{figure}

In this section, we extend the previous analysis to the case of two-band models (2BMs) with dispersions $\xi_{k_x}^{\rm e,h}$, where the superscript ${\rm e/h}$ reflects the type of electron/hole pocket that ari\-ses from the respective band. An example of such a 2BM is shown in Fig.~\ref{fig:Figure5}. We employ the $\kappa_{1,2,3}$ Pauli matrices to represent Hamiltonian matrix elements in band space. This representation is also relevant for the magnetic and pairing terms, which now become matrices in this space. In particular, the direct-lattice profile of the magnetization generally reads:
\bea
\hat{\bm{M}}_{n}=\frac{\bm{M}_{n}^{\rm e}+\bm{M}_{n}^{\rm h}}{2}+\frac{\bm{M}_{n}^{\rm e}-\bm{M}_{n}^{\rm h}}{2}\kappa_3+\bm{M}_{n}^{\rm eh}\kappa_1\,.\quad
\label{eq:2BMMagnetization}
\eea

\noi One notes that a term proportional to $\kappa_2$ is not allowed, since this violates the requirement that the magnetization field of the texture is odd under ${\cal T}$. On the other hand, the spin-singlet pairing matrix reads:
\bea
\hat{\Delta}_{k_x}&=&\frac{\Delta_{k_x}^{\rm e}+\Delta_{k_x}^{\rm h}}{2}+\frac{\Delta_{k_x}^{\rm e}-\Delta_{k_x}^{\rm h}}{2}\kappa_3\no\\
&+&\frac{\Delta_{k_x}^{\rm eh}+\Delta_{k_x}^{\rm he}}{2}\kappa_1-\frac{\Delta_{k_x}^{\rm eh}-\Delta_{k_x}^{\rm he}}{2i}\kappa_2\,.
\eea

\noi The required antisymmetry of the superconducting matrix order parameter in the combined spin, band and $k_x$ spaces, implies that the terms proportional to $\mathds{1}_{\kappa}$, $\kappa_3$ and $\kappa_1$ are even under inversion, while the one proportional to $\kappa_2$ is odd. {\color{black} In the remainder, we focus on cases where the FSs associated with the various pockets neither overlap nor share similar shapes and, as a result, interband pai\-ring is expected to be substantially suppressed\footnote{{\color{black}Note also that interband pairing appears more possible to arise in 1D rather than 2D systems, since the FS in the former case consists of points. For nonoverlapping FSs, the appearance of interband pairing requires a suitable bosonic ``glue'', such as anti\-fer\-ro\-ma\-gne\-tic magnons, which can provide the wave-vector transfer that is required to connect the different bands~\cite{PDW}. Even more, the development of interband pairing in such cases either leads to a net momentum in the ground state {\color{black}(Fulde-Ferrell phase~\cite{FF}) or spontaneously violates translational invariance (Larkin-Ovchinikov phase~\cite{LO})}. This may in turn generate electric currents or elastic deformations which result in additional energy penalties for bandstructures which originally respect TR and translational symmetries. Finally, in the here more relevant case of 2D systems which harbor multiple non\-si\-mi\-lar and nonoverlapping FSs, interband pai\-ring is expected to be further disfavored since the establishment of a full gap on the FSs requires the development of a pairing {\color{black}order parameter} which consists of multiple wave vectors~\cite{PDW}.}}.}

In the present case we also consider that the magnetization possesses a MHC form, with the $M_{||,\perp}$ helix components of the previous section being upgraded to band-space matrices according to Eq.~\eqref{eq:2BMMagnetization}. In addition, we assume that the two intraband pairing order parameters $\Delta_{k_x}^{\rm e,h}$ do not contain any zeros, and thus, we set them to be constants. Nonetheless, our results qualitatively hold for more complex unconventional gap structures, as long as these do not contain any zeros. We further note that, here, the inclusion of an additional band does not lower the symmetry of the Hamiltonian, and the system is therefore left invariant under the magnetic point and space group symmetries discussed in the previous section.

Since the 2BMs bear similarities with the thoroughly-explored 1BMs, we confine the analysis to the novel features brought about by the additional band. We extend the spinor defined in Eq.~\eqref{eq:Enlarged_Spinor_1BM} as follows:
\bea
\bm{{\cal X}}_{k_x}^{\rm 2BM}=\left(\bm{{\cal X}}_{k_x}^{\rm 1BM;e}\,,\,\bm{{\cal X}}_{k_x}^{\rm 1BM;h}\right)^{\intercal}.
\label{eq:Enlarged_Spinor_2BM}
\eea

\noi and consider the Hamiltonian
\begin{align}
\hat{{\cal H}}_{k_x}=\sum_{s}^{{\rm e,h}}{\cal P}_s\hat{\cal H}_{k_x}^{0;s}
-\frac{\hat{M}_\perp\big(\mathds{1}+\eta_1\big)\rho_1\sigma_z+\hat{M}_{||}\big(\mathds{1}-\eta_1\big)\rho_3\sigma_x}{2}
\label{eq:Energy_2BM}
\end{align}

\noi where we introduced the electron-like [hole-like] band projectors ${\cal P}_{\rm e}=(\mathds{1}_{\kappa}+\kappa_3)/2$ [${\cal P}_{\rm h}=(\mathds{1}_{\kappa}-\kappa_3)/2$]. Depen\-ding on the precise matrix form of the magnetization, one can interpolate between intra- and inter-band scattering. 

For $\hat{M}_{||,\perp}\propto\mathds{1}_{\kappa},\,\kappa_3$, the magnetic scat\-te\-ring has only an intraband character, and the two bands are completely decoupled. Thus, the topological pro\-per\-ties of the system follow from applying the results of the previous paragraphs separately to each band, and the symmetry class is BDI$\oplus$BDI. In contrast, when $\hat{M}_\perp=M_\perp\kappa_1$ and $\hat{M}_{\parallel}=M_{\parallel}\kappa_1$, the Hamiltonian exhibits an additional unitary symmetry with the ge\-ne\-ra\-tor ${\cal O}=\kappa_3\sigma_y$. Note that this symmetry is due to the specific form of the magnetic texture in band space, and intraband magnetic scat\-te\-ring terms $\hat{M}_{\perp,\parallel}\propto\mathds{1}_{\kappa},\,\kappa_3$ generally violate it.

Assuming the pre\-sen\-ce of ${\cal O}$, we block diagonalize the Hamiltonian using the transformation ${\cal S}=({\cal O}+\sigma_z)/\sqrt{2}$, and find that the nonmagnetic part of Eq.~\eqref{eq:Energy_2BM} remains unaltered, while the magnetic part becomes:
\bea
\hat{{\cal H}}_{{\rm mag},\sigma}=-\frac{\sigma M_\perp\kappa_1\big(\mathds{1}+\eta_1\big)\rho_1+M_{\parallel}\kappa_2\big(\mathds{1}-\eta_1\big)\rho_3}{2},\phd\quad
\label{eq:Energy_2BM_Blocks}
\eea

\noi where $\sigma=\pm1$ correspond to the eigenvalues of $\sigma_z$. Each Hamiltonian block belongs to the symmetry class AIII with $\Pi=\kappa_3\tau_2$. The procedure for carrying out the 1D topological analysis is here identical to the one presented in Sec.~\ref{sec:1BM_MH_ANI} for systems in the BDI symmetry class, since also the AIII class supports a $\mathbb{Z}$ to\-po\-lo\-gi\-cal inva\-riant in 1D, which is identified with a winding number.

\begin{figure}[t!]
\centering
\includegraphics[width=0.95\columnwidth]{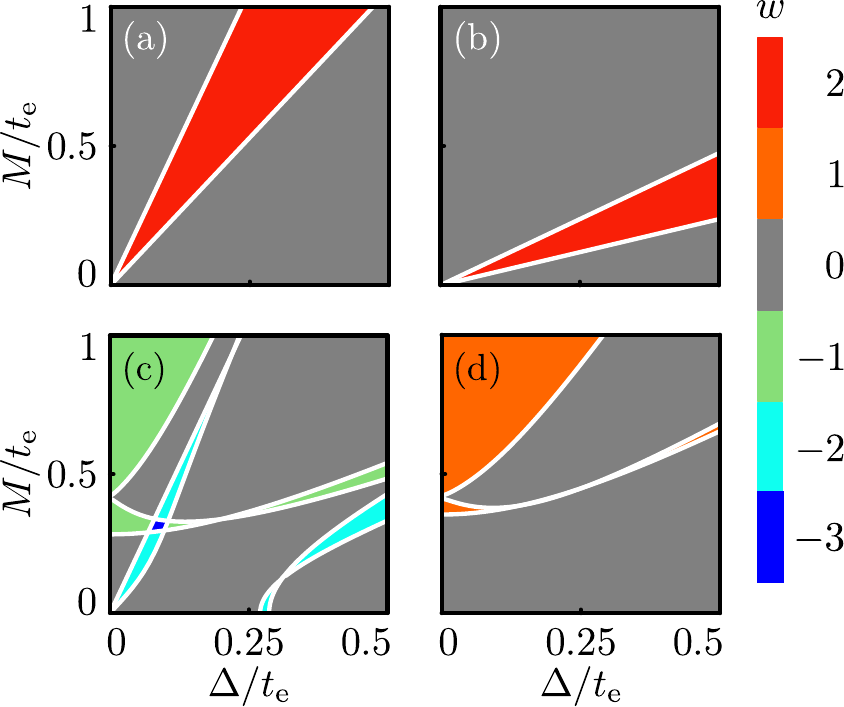}
\caption{Topological phase diagrams for the 2BM in Fig.~\ref{fig:Figure5} in the MHC phase, evaluated by pro\-jecting onto the low-energy sector. (a) Phase diagram when solely interband magnetic scat\-te\-ring is con\-si\-de\-red, and the order parameter values $\Delta^{\rm h}=\Delta$ and $\Delta^{\rm e}=\Delta/2$ are employed. (b) Same as in (a), but with $\Delta^{\rm{h}}/\Delta^{\rm{e}}=1/10$. When $\Delta^{\rm h}\Delta^{\rm e}<0$ the topologically nontrivial phase va\-ni\-shes. (c) Same as in (a), but with the inclusion of intraband magnetic scat\-te\-ring, with $M_{||,\perp}^{\rm{e}}=M_{||,\perp}^{\rm{h}}=M$. (d) Here, $\Delta^{\rm h}=-2\Delta^{\rm e}$ and $M_{||,\perp}^{\rm{e}}=M_{||,\perp}^{\rm{h}}=M$. Notably, the inclusion of intraband magnetic scat\-te\-ring induces a topologically-nontri\-vial phase in an otherwise trivial region. We considered $M^{\rm eh}_\perp=M$ and $M^{\rm eh}_{||}=M/3$ in all calculations.}
\label{fig:Figure6}
\end{figure}

Such an analysis, see App.~\ref{app:AppendixC5}, yields that the topological phase transition from the trivial to the nontrivial phase occurs when $\xi_{k_x\pm 3q}^{\rm e}\Delta^{\rm h}=\xi_{k_x\mp3q}^{\rm h}\Delta^{\rm e}$ and $\xi_{k_x\pm3 q}^{\rm e}\xi_{k_x\mp3q}^{\rm h}+\Delta^{\rm e}\Delta^{\rm h}=M_\sigma^2$, are simultaneously sa\-ti\-sfied, for a given $\sigma$. When the magnetic wave vectors connect two points at the Fermi level, we have $\xi_{k\pm3q}^{\rm e}=\xi_{k\mp3q}^{\rm h}=0$ and the topological criterion reads $\Delta^{\rm e}\Delta^{\rm h}=M_\sigma^2$. Remarkably, the latter condition can be satisfied only when the two magne\-ti\-cal\-ly connected points exhibit the \textit{same} sign for the pai\-ring term. Figures~\ref{fig:Figure6}(a)-(b) display the resul\-ting phase dia\-grams for $\Delta^{\rm h}/\Delta^{\rm e}=\{2,1/10\}$, respectively, with the topologically nontrivial regions marked in red. In agreement with the above criterion, we find that the nontri\-vial regime shrinks when $|\Delta^{\rm e}-\Delta^{\rm h}|$ increases.

We note the doubling of $w$ compared to the case of the 1BMs, cf Fig.~\ref{fig:Figure4}(a) and \ref{fig:Figure4}(c). This is due to the doubling of the number of gap closing points $k_{\rm c}$. Each one of the green dots in Fig.~\ref{fig:Figure5}(b) contributes with a single unit to $w$. The emergence of a number of $2\mathbb{Z}$ MZMs in conjunction with the {\color{black}AIII classification implies that each emergent} pair of MZMs should be seen as a single topologically protected AZM. Indeed, from the ana\-ly\-ses of Refs.~\onlinecite{Altland,SchnyderClassi}, it emerges that here a TSC ori\-gi\-na\-ting from an interband-only magnetic texture can be described with a spinor of halved dimensionality compared to the one defined in Eq.~\eqref{eq:Enlarged_Spinor_2BM}, since spin-up electrons of the electron-like pocket pair up only with spin-down electrons of the hole-like pocket, and vice versa. While so far there e\-xists only a little theoretical acti\-vi\-ty on AZMs compared to MZMs, cf Ref.~\onlinecite{KlinovajaFF,SticletFF,NagaosaFF,MarraFF,FPTA}, their experimental crea\-tion and ma\-ni\-pu\-la\-tion is a fasci\-na\-ting topic on its own. Indeed, when these AZMs are topologically protected for an extended region in the parameter space, as found here, they in principle enable quantum information processing with long-lived quasiparticles~\cite{Zazunov,HigginbothamParity}. 

In the more general case, intraband terms which break the ${\cal O}=\kappa_3\sigma_y$ symmetry may also be present. These can be divided into nonmagnetic and magnetic. When these are nonmagnetic, e.g., various types of ISB SOC, they stabilize a DIII symmetry class which supports MZM Kramers pairs. For a similar situation see Ref.~\onlinecite{SteffensenMZM}. However, throughout this work we consider that all types of ISB SOC have negligible strengths, thus implying that such a possibility is inaccessible here. Nonetheless, as we discuss later in this manuscript, DIII class Majorana fermions become generally accessible in 2D 2BMs for a SWC$_4$ texture. On the other hand, when additional magnetic terms are considered, these restore the BDI class found in the 1BMs, as well as the Majorana nature of the topologically-protected edge excitations. 

We proceed by investigating the effects of intraband magnetic texture terms. For this purpose, we recalculate the winding number (see App.~\ref{app:AppendixC5}), and obtain the phase diagrams in Fig.~\ref{fig:Figure6}(c) and \ref{fig:Figure6}(d) for $\Delta^{\rm h}/\Delta^{\rm e}=\pm2$, respectively. Strikin\-gly, as seen in Fig.~\ref{fig:Figure6}(d), the inclusion of intraband magnetic scat\-te\-ring induces a topologically nontrivial phase, even when the connected points exhibit different signs for the pairing term. Once again, assu\-ming that the transition occurs due to the gap closing at two magnetically-con\-nected Fermi points, and that the magnetic moments are spatially constant, the topological criterion reads:
\bea
\big(\Delta^{\rm{e}}\pm M^{\rm{e}}\big)\big(\Delta^{\rm{h}}\pm M^{\rm{h}}\big)=\left(M^{\rm{eh}}\right)^2\,.
\label{eq:GapClosingCondition}
\eea

\noi This expression imposes severe constraints on the unconventional superconducting order parameter, as well as the relative contributions of intra- and inter-band magnetism, which can lead to topologically nontrivial phases in 1D. Nonetheless, this condition is not as stringent in higher-dimensional systems, since the pairing term may 
lead to a gap closing for some of the BZ points, which is a sufficient condition to allow, but not necessarily gua\-ran\-tee, the transition to a TSC phase. 

To this end, we remark that crystal\-line symmetries general\-ly influence the bulk classification of multiband systems in a similar fashion to 1BMs when inter- and intra-band magnetic texture terms are simultaneously present. Instead, when only interband textures are considered, the presence of the ${\cal O}$ symmetry renders the effects of the crystalline symmetries trivial.

Before proceeding to the 2D cases, we here summarize what we learned from the 1D models, and how this will help us explore the 2D cases. First of all, we discussed that a system in the presence of a MHC with a spatially varying magnetic moment $|M_{||}|\neq|M_{\perp}|$ cannot be directly mapped onto the ISB SOC mechanism in Figs.~\ref{fig:Figure1}(a)-(d), i.e., one cannot gauge away the spatial dependence of the MHC. Hence one needs to adopt either a sublattice description (cf App.~\ref{app:AppendixA}), or perform a downfolding to the MBZ. The latter serves as a convenient basis for our calculations, and is adopted in the upcoming paragraphs. In this regard, the 1D cases additionally served as an introduction and motivation for our formalism in the more complicated 2D systems.

Concerning the topological classification, we esta\-blished that 1BMs and 2BMs in the MHC phase ge\-ne\-ral\-ly reside in the BDI symmetry class, regardless of the type of spin-singlet pairing gap. However, by con\-si\-de\-ring interband-only magnetic scattering, we found that the 2BMs display an emergent unitary symmetry ultimately resulting in the class AIII\,$\oplus$\,AIII. This class supports AZMs, which however, can be converted back to MZMs by including intraband terms. Lastly, we also performed the topological classification in the presence of magnetic point and space group symmetries, introduced the re\-le\-vant invariants, and discussed how these can lead to new topological phases. For 1D systems, the additional unitary symmetries proved to be obsolete when it comes to the prediction of edge modes, since edges generally break these. Nonetheless, their presence sets constraints on a number of bulk topological properties which can be harnessed to experimentally infer the TSC phase of the system. In fact, the methodology employed in the study of unitary symmetries sets the stage and introduces the concepts for the upcoming 2D cases, where magnetic point group symmetries play instead an essential role in determining the type of Majorana or Andreev edge modes.

\section{2D Topological Superconductors}\label{sec:2D_topological_superconductors}

We now extend our study to the case of 2D systems, which is the main topic on our agenda. We start with 1BMs and afterwards consider two-band extensions. We find that 2D systems exhibit a rich variety of MF-edge-mode types, i.e., flat, uni- and bi-directional modes when nodes are present in the bulk energy spectrum, or, quasi-helical, helical, and chiral modes when the bulk energy band structure is fully gapped. This MF diversity is obtained by con\-si\-de\-ring 1BMs and 2BMs in the presence of a MHC, a SWC$_4$ and, finally, a SWC$_4$ combined with an external in- and out-of-plane Zeeman field, where the latter situation also reproduces a SSC$_{4}$ phase.

\subsection{One-band models}\label{sec:2D1BM}

In this section we extend the 1BM dispersion $\xi_{k_x}$ to its 2D analog $\xi_{\bm{k}}$, with a focus on models leading to two hole pockets centered at ${\rm\Gamma}(0,0)$ and ${\rm M}(\pi,\pi)$. The two poc\-kets are assumed to feature intra-pocket FS ne\-sting at the mu\-tual\-ly ortho\-gonal wave vectors $\bm{Q}_{{\rm N},1}$ and $\bm{Q}_{{\rm N},2}$, thus generally supporting both single-$\bm{Q}$ and double-$\bm{Q}$ magnetic phases~\cite{Christensen_18}. The magnetic vectors $\bm{Q}_{1,2}$ may coincide with the nesting wave vectors. Such a type of band structure is shown in Fig.~\ref{fig:Figure7}, and bears qualitative similarities to substantially hole doped FeSCs. In principle, for highly symmetric FSs other ne\-sting vectors may play a substantial role in deciding the resulting magnetic phase. Nonetheless, for the explorative nature of this paper we simply restrict to the star $\pm\bm{Q}_{1,2}$.

\subsubsection{MHC texture - Majorana flat bands}\label{sec:1BM_2D_MH}

The construction of the Hamiltonian in 2D is straightforward, and is obtained by replacing the 1D dispersion in Eq.~\eqref{eq:bloch_1BM_ani_helix} by its 2D analog. The MBZ is now defined as the set $\bm{k}\in(-q,q]\times(-\pi,\pi]$ and the ISPs are $\bm{k}_{\cal I}=\{(0,0),(q,0),(0,\pi),(q,\pi)\}$, where $\bm{q}_{1,2}=\bm{Q}_{1,2}/2$ and $q=|\bm{q}_{1,2}|$. Out of these four, $(q,0)$ and $(q,\pi)$ observe a Kramers degeneracy imposed by the antiunitary magnetic space group symmetry $\tilde{\Theta}_{\bm{k}}=\{{\cal T}\,|\,(\nicefrac{\pi}{Q},0)\}$. 

\begin{figure}[t!]
\centering
\includegraphics[width=0.95\columnwidth]{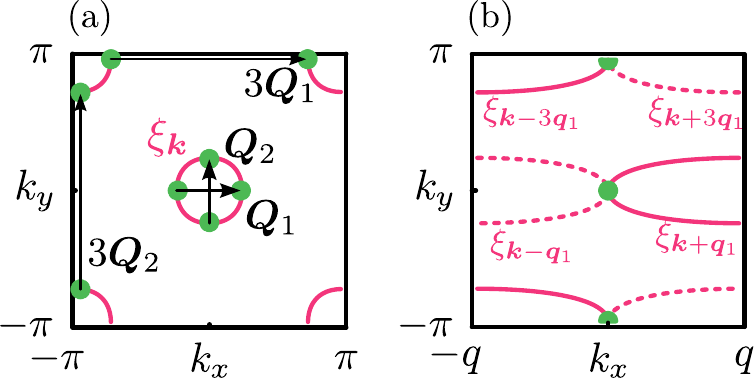}
\caption{Example of a 1BM in 2D. (a) Fermi surfaces (FSs) in the BZ obtained for a dispersion $\xi_{\bm{k}}=-2t\cos k_x\cos k_y-\mu$ with $\mu=-\sqrt{2}t<0$. We show the magnetic ordering wave vectors $\bm{Q}_{1,2}=Q\{\hat{\bm{x}},\hat{\bm{y}}\}$ and $3\bm{Q}_{1,2}\equiv-\bm{Q}_{1,2}$ connecting points of the FS. (b) depicts the resulting four FS segments after transferring to the MBZ for a MHC with the wave vector $\bm{Q}_1$. Similar to Fig.~\ref{fig:Figure3}, also here there exist points at higher energy which are connected by $\bm{Q}_{1,2}$.}
\label{fig:Figure7}
\end{figure}

The extension to 2D is complete after the addition of $k_y$ as a second argument to $h_{k_x}^{(s)}$ and $\Delta_{k_x}^{(s)}$ defined in Eqs.~\eqref{eq:DefsShiftedVars1}-\eqref{eq:DefsShiftedVars4}, which leads to $h_{\bm{k}}^{(s)}$ and $\Delta_{\bm{k}}^{(s)}$. Note, however, that this seemingly-trivial extension leads to a dichotomy in regards with the be\-ha\-vior of $\Delta_{\bm{k}}^{(s)}$ under mirror operations. Specifically, one can now distinguish two cases depending on whether $\Delta_{\bm{k}}$ transforms accor\-ding to the $\{{\rm A_{1g},B_{1g}}\}$ or the $\{{\rm B_{2g},A_{2g}}\}\equiv{\rm B_{2g}}\times\{{\rm A_{1g},B_{1g}}\}$ IRs of D$_{\rm 4h}$.\footnote{For example: $\{{\rm A_{1g},B_{1g},B_{2g}}\}\sim\{1,\cos k_x-\cos k_y,\sin k_x\sin k_y\}$.} Notably, pairing gaps transforming according to the former (latter) satisfy $\sigma_{xz,yz}\Delta_{\bm{k}}=+\Delta_{\bm{k}}$ ($\sigma_{xz,yz}\Delta_{\bm{k}}=-\Delta_{\bm{k}}$). While this difference does not diversify the BDI 2D symmetry classification for the two categories of pairing, it does lead to two distinct classifications in the HSPs depending on whether the magnetic and pairing point groups coincide or not. Below, we first focus on the 2D classification and study the influence of the magnetic point group at the end of this section.

In analogy to our previous analysis, we define the win\-ding number for each $k_y$ subsystem which, notably, for the BDI class in 2D defines a \textit{weak}, instead of a strong, to\-po\-lo\-gi\-cal invariant~\cite{FuKaneMele,Ringel}. Under the condition that $\Delta_{\bm{k}}$ does not induce additional gap closings in the MBZ other than the ones appearing for $\bm{k}=(0,k_y)$, we obtain:
\bea
w_{k_y}&=&\sum_{k_{\rm c}}{\rm sgn}
\left.\left(\Delta_{\bm{k};\bm{q}_1}^+\frac{\text{d}\xi_{\bm{k};\bm{q}_1}^-}{\text{d}k_x}-\xi_{\bm{k};\bm{q}_1}^{+}\frac{{\rm d}\Delta^-_{\bm{k};\bm{q}_1}}{{\rm d}k_x}\right)\right|_{k_x=k_{\rm c}}\no\\&&
\,\cdot\frac{{\rm sgn}\left\{M^2-\big[\Delta_{(k_{\rm c},k_y);\bm{q}_1}^+\big]^2-\big[\xi^+_{(k_{\rm c},k_y);\bm{q}_1}\big]^2\right\}}{2}\,,\quad
\label{eq:winding_ky}
\eea

\noi where we considered $M_{||,\perp}=M>0$, exploited the pro\-per\-ty $\xi_{-k_x,k_y}=\xi_{k_x,k_y}$, and made use of the constraint on $\Delta_{\bm{k}}$. Evi\-den\-tly, the 1D criterion for a gap closing at a point $k_{\rm c}$, for a given $k_y$, still holds, namely $M^2=[\Delta^+_{(k_{\rm c},k_y);\bm{q}_1}]^2+[\xi_{(k_{\rm c},k_y);\bm{q}_1}^+]^2$. Thus, gap closing points appear for $(k_x=0,k_y)$ and suitable values of $k_y$.\footnote{Note that the 1BM presented in Fig.~\ref{fig:Figure7} features additional gap closings at $\big(k_x,\pm\pi/2\big)$. However, these solely stem from the next-nearest-neighbor character of the hopping term considered, and do not constitute universal properties. In fact, this band peculiarity can be removed by consi\-de\-ring additional hopping matrix ele\-ments of a different range in the 2D version of Eq.~\eqref{eq:free_ham_1D_1BM}.} 

\begin{figure}[t!]
\centering
\includegraphics[width=0.95\columnwidth]{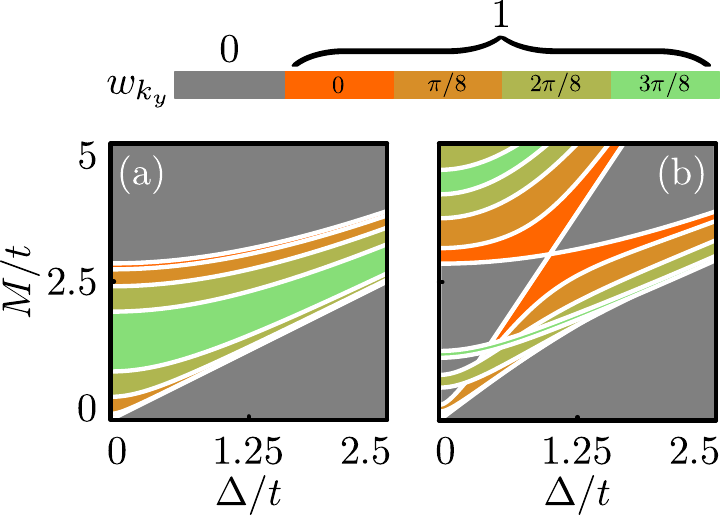}
\caption{(a)-(b) Topological phase diagrams for the 2D 1BM in Fig.~\ref{fig:Figure7}, in the presence of a MHC texture with spatially-constant and -varying ($M_\perp/M_{||}=3$) moment, respectively. For a given $k_y=0,\,\pi/8,\,2\pi/8,\,3\pi/8$ value, we find one (no) MZMs per edge in the correspondingly-colored (gray) regions.}
\label{fig:Figure8}
\end{figure}

In Fig.~\ref{fig:Figure8}(a) we display topological phase diagrams for the model in Fig.~\ref{fig:Figure7}, for various $k_y$ va\-lues which are depicted using a $k_y$-dependent color scale. In the weak-coupling limit, and for generally different $M_{||,\perp}$, each $k_y$ subsystem is dictated by the familiar criterion in Eq.~\eqref{eq:1BM_Low_Energy_Criterion}:
\bea
M_-<\sqrt{\big(\xi_{\bm{k}_{\rm c};\bm{q}_1}^+\big)^2+(\Delta^+_{\bm{k}_{\rm c};\bm{q}_1})^2}<M_+\,.\label{eq:single_Q_ani_criterion}
\eea

\noi However, in contrast to Eq.~\eqref{eq:1BM_Low_Energy_Criterion}, here, not all gap-closing points $\bm{k}_{\rm c}$ are at the Fermi level, and this de\-tu\-ning introduces an effective chemical potential $\xi_{\bm{k}_{\rm c};\bm{q}_1}^+$ in the above criterion. As expected, for $k_y=0$ we reproduce Eq.~\eqref{eq:1BM_Low_Energy_Criterion} after setting $\Delta_{\bm{k}}=\Delta$. Fig.~\ref{fig:Figure8}(b) presents the arising topological phase diagram for a system in the MHC phase with $|M_\perp|\neq|M_{||}|$. Clearly, the phase diagram becomes significantly modified as $k_y$ varies, and the various topologically nontrivial regions generally overlap. 

Based on the criteria for a gap closing at the various $k_y$ values, we infer that the nodes in the bulk spectrum move along the $k_x=0$ line in the MBZ when varying the superconducting and magnetic gaps. This resembles a gapless-gapful transition in topological insulators which leads to a Weyl semimetallic phase~\cite{Murakami_07}. In fact, such transitions have also been studied previously in: (i) p$\pm$ip TSCs, where an in-plane magnetic field drives the transition~\cite{Wong_13}, (ii) in nodal d-wave SCs~\cite{Daido_17}, and (iii) in nodal superconducting phases of FeSCs~\cite{Chubukov_nodes,Agterberg_nodes}. 

\begin{figure}[t!]
\centering
\includegraphics[width=0.95\columnwidth]{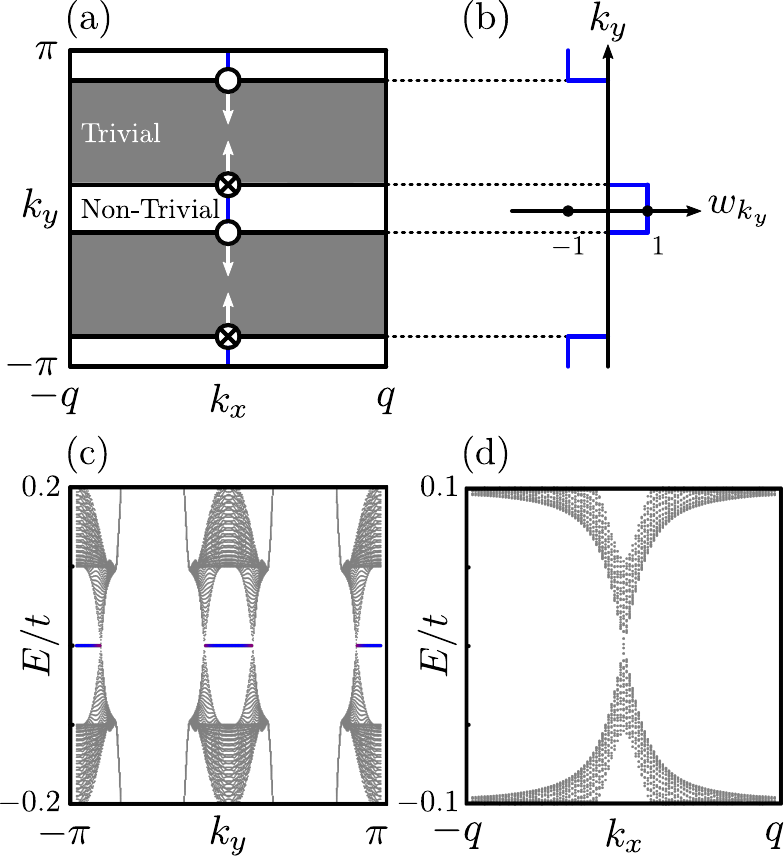}
\caption{Properties of the topologically-stable nodes obtained in the gapless bulk ener\-gy spectrum of the 1BM in Fig.~\ref{fig:Figure7} in the presence of the MHC texture. Here, nodes with vorticity $\upsilon =+1$ ($\upsilon =-1$) are discerned by a dot (cross). (a) Sketch of the path followed by the nodes when varying the superconducting and/or magnetic gaps. Nodes of opposite vor\-ti\-ci\-ties are connected by MFBs, in agreement with the win\-ding number values in (b). (c) and (d) Numerically-obtained dispersions with open boundary conditions along the $x$ and $y$ direction, respectively. Parameter values used: $\Delta=0.1\,t$, $M_\perp=M_{||}=0.2\,t$, and $L_{x,y}=401$ sites in the $x,y$ direction.}
\label{fig:Figure9}
\end{figure}

In Fig.~\ref{fig:Figure9}(a) we sketch the path followed by the nodes in the MBZ when varying the superconducting and magnetic gaps. Given the structure of the weak topological invariant $w_{k_y}$, we expect to find topologically-protected MF modes at the edges parallel to the $y$ direction, but no modes at the edges perpendicular to it. This is indeed verified in Figs.~\ref{fig:Figure9}(c) and~\ref{fig:Figure9}(d), where we plot the spectrum with open boundary conditions along the $x$ and $y$ direction, respectively. Here, only the former edge spectrum displays topologically-protected MF modes. This behavior is remi\-ni\-scent of graphene and the appearance of flat bands only when the termination is of the zig-zag type~\cite{GrapheneRev}. We observe that nodes related to each other by the discrete symmetries $\Theta$ and $\Xi$, are connected by Majorana flat bands (MFBs), in agreement with the va\-lues of $w_{k_y}$ shown in Fig.~\ref{fig:Figure9}(b). In the direct-lattice re\-pre\-sen\-ta\-tion, these MFBs manifest as standing MF waves only at edges parallel to the $y$ direction, i.e., they possess wave functions with a spatial part proportional to $\sin(n\pi R_y/N_y)$ where $n\in\mathbb{N}^+$ and $N_y$ being the number of lattice sites in the $y$ direction, cf Ref.~\onlinecite{PKBook}.

So far, we studied the emergence of the MFBs by vie\-wing $k_y$ as a mere parameter which controls the topological properties of each 1D $k_y$ subsystem. However, accounting for the correspondence of $k_y$ to the spatial coordinate $y$, and considering the stable character of the nodes in the bulk energy spectrum, allows us to cha\-racte\-ri\-ze the 2D nodal TSC using \textit{local} strong to\-po\-lo\-gi\-cal inva\-riants~\cite{TanakaFlatBands,Zhao_13_14,Matsuura_13}. In fact, the BDI symmetry class ensures that the MFBs enjoy a topological protection, which is in\-he\-ri\-ted from the respective robustness of the bulk nodes in the energy spectrum. Each node at $\bm{k}_{\rm c}$ possesses a $\mathbb{Z}$ topological charge, i.e., its vorticity:
\bea
\upsilon=\frac{1}{2\pi i}\oint_{\mathcal{C}}\frac{\text{d}z_{\bm{k}}}{z_{\bm{k}}},\label{eq:vorticity}
\eea

\noi where $\mathcal{C}$ is a contour encircling the node. Here, $z_{\bm{k}}$ corresponds to the 2D extension of Eq.~\eqref{eq:zTopoInv}.\footnote{{\color{black} The apparent similarity of the invariants in Eqs.~\eqref{eq:winding_number} and~\eqref{eq:vorticity} is not accidental. In the 2D case, the $k_y$ dependence of the dis\-persion can be viewed as a parameter that effectively controls the chemical potential of the 1D system. By deforming the contour ${\cal C}$ to two parallel lines at $k_y=k_c+0^\pm$, which ``close'' at infinity $|k_x|\rightarrow\infty$, the vorticity is given by the difference of the winding numbers $w(\mu_c^\pm)$. Here $\mu_c$ corresponds to the $k_c$ which tunes the system to the gap closing. Therefore, the vorticity defined in 2D, yields the difference between two winding number va\-lues which characterize two topologically distinct phases of the 1D TSC across the gap closing which appears at the location $(0,\mu_c)$ of the respective Berry singularity in $(k_x,\mu)$ space~\cite{VolovikBook}.}} 

For the present model, linearizing the Hamiltonian about a node yields $\upsilon={\rm sgn}\big(\alpha\beta\Delta \xi^{+}_{\bm{k}_{\rm c};\bm{q}_1}\big)$, where we set $\Delta_{\bm{k}}=\Delta$, and expanded the shifted dispersions about $\bm{k}_{\rm c}$ as follows: $\xi^+_{\bm{k};\bm{q}_1}\approx \xi^{+}_{\bm{k}_{\rm c};\bm{q}_1}+\alpha k_y$ and $\xi_{\bm{k};\bm{q}_1}^-\approx\beta k_x$. The above result reveals that the vorticity is ill-defined at the ISPs $(0,0)$ and $(0,\pi)$ since, there, $\alpha=0$.\footnote{For the model in Fig.~\ref{fig:Figure7}, $\upsilon$ is also ill-defined for $k_y=\pm\pi/2$, since at these points $\xi^+_{\bm{k}_{\rm c};\bm{q}_1}=0$ which, however, is only an artifact of the next-nearest-neighbor nature of the hopping considered.}

\begin{figure}[t!]
\centering
\includegraphics[width=0.95\columnwidth]{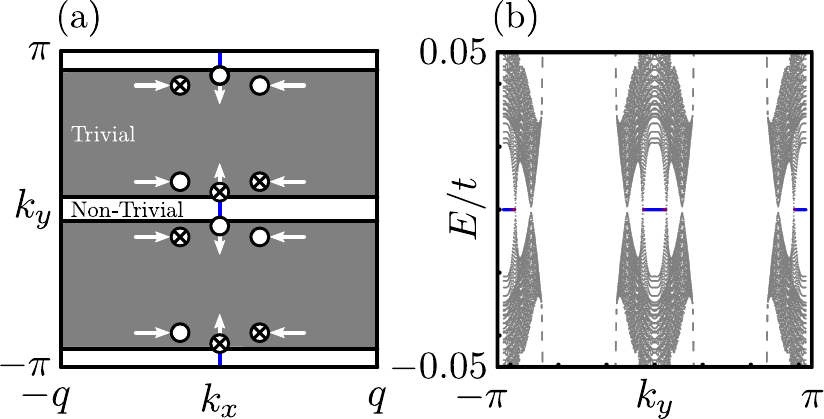}
\caption{Modification of the nodal spectrum in Fig.~\ref{fig:Figure9} due to the unconventional pairing function $\Delta_{\bm{k}}=\Delta(\cos k_x-\cos k_y)$. (a) The unconventional gap function induces 8 additional nodes in the bulk spectrum. In (b) we present the energy spectrum for open boundary conditions along the $x$ direction and $L_x = L_y = 1001$. We do not find any new MF bands since, for the given orientation of the termination edges, the contributions of the $8$ additional nodes cancel out. Parameter values used: $\Delta=0.1\,t$ and $M_\perp=M_{||}=0.05\,t$.}
\label{fig:Figure10}
\end{figure}

Let us now proceed and consider a superconducting order parameter which \textit{does} generate new nodal points. In order to determine the location of these new nodes, we employ the 2D analog of Eq.~(\ref{eq:Low_Energy_1BM}), which describes the low-energy features about the $\Gamma(0,0)$ point in the MBZ. Alternatively, we can perform the shift $\bm{k}\mapsto\bm{k}+4\bm{q}_1$ to obtain a description about ${\rm Y}(0,\pi)$. By repeating the steps detailed in Sec.~\ref{subsec:1BM_1D}, the gap closing points $\bm{k}_{\rm c}$ are given by ${\rm Im}\big[{\rm det}(\hat{A}_{\bm{k}_c,\sigma}^{\rm low-en})\big]=0$ which yields the equation:
\bea
\xi_{\bm{k}_c-\bm{q}_{1}}\Delta_{\bm{k}_c+\bm{q}_1}=\xi_{\bm{k}_c+\bm{q}_1}\Delta_{\bm{k}_c-\bm{q}_1}\,.\label{eq:Condition}
\eea

As an example, we consider the 1BM of Fig.~\ref{fig:Figure7} and the d-wave gap function $\Delta_{\bm{k}}=\Delta(\cos k_x-\cos k_y)$. We find the gap-closing points $\bm{k}_{\rm c}:\big\{k_x=0\phd{\rm or}\phd\cos(2k_y)=-1-\mu/t\big\}$. For the given values of $q$ and $\mu$, the additional nodes move on the lines $k_y\simeq\pm0.18\pi,\pm1.18\pi$. The complete information regarding the location of the above gap-closing points in the MBZ is found by further employing the re\-mai\-ning gap-closing condition ${\rm Re}\big[{\rm det}(\hat{A}_{\bm{k}_c}^{\rm low-en})\big]=0$, which is equivalent to the criterion in Eq.~\eqref{eq:single_Q_ani_criterion}. Ana\-ly\-zing the above yields nodes moving along different lines in the MBZ, as shown in Fig.~\ref{fig:Figure10}(a). Contrasting this to Fig.~\ref{fig:Figure9}, we clearly see the introduction of eight new nodes in addition to the four nodes for $k_x=0$. Here, however, we do not find any additional MF modes compared to Fig.~\ref{fig:Figure9}(c) when we consider open boun\-da\-ry conditions along the main $x,y$ axes, as is evident in Fig.~\ref{fig:Figure10}(b) for open boun\-da\-ry conditions in the $x$ direction. This is because, the contributions of the additional nodes cancel out by virtue of mirror symmetries when projected onto the edge where translational invariance persists. However, if the system were to be terminated along the, e.g., $(11)$ surface, that would indeed allow for the presence of new MF modes, cf Ref.~\onlinecite{Agterberg_nodes}. 

We now consider the additional presence of the magnetic point/space group symmetries and the modifications that these bring to the topological classification. A common feature of the crystalline classifications in 1D and 2D for the MHC, is that the effects of $\mathcal{R}_{xz}$ are tri\-vial in both. On the other hand, we find crucial dif\-fe\-ren\-ces which mainly relate to (i) the enhancement of the dimensiona\-li\-ty which generally leads to different to\-po\-lo\-gi\-cal inva\-riants even within the same symmetry class, and (ii) the structure of the pairing gap. Specifically, we find two distinct cases depending on whether $\Delta_{\bm{k}}$ is invariant under the action of the magnetic point group operations, or not. When $\Delta_{\bm{k}}\in\{{\rm A_{1g},B_{1g}}\}$ ($\Delta_{\bm{k}}\in\{{\rm B_{2g},A_{2g}}\}$), the magnetic point group M$_{\rm MHC}$ is (not) conserved. Notably, the pair of D$_{\rm 4h}$ IRs bunched together in a given set are equi\-va\-lent under the action of the magnetic point group. 

Another key aspect of the crystalline classification in 2D, is that the Kra\-mers de\-ge\-ne\-ra\-cy imposed by $\tilde{\Theta}_{\bm{k}}$ at $(q,0)$ and $(q,\pi)$, extends to the entire $k_x=q$ HSP. As discussed in App.~\ref{app:AppendixB} this is imposed by the pair of offcentered symmetries $\mathcal{R}_{yz}$ and $\{\sigma_{yz}\,|\,(\nicefrac{\pi}{Q},0)\}$. Hence, MHC-driven gap closings cannot take place anywhere in the $k_x=q$ HSP. Therefore, any possible crystalline topological features arising in this HSP originate from the structure of the pairing gap. For this reason, the remainder of this section focuses only on the $k_x=0$ HSP. The classification in the $k_x=q$ HSP appears in Table~\ref{table:TableII} and the topological invariants connect to the analysis below.

For $\Delta_{\bm{k}}\sim\{{\rm A_{1g},B_{1g}}\}$, ${\cal R}_{yz}$ and $\{\sigma_{yz}\,|\,(\nicefrac{\pi}{Q},0)\}$ lead to a AI$\oplus$AI class in the $k_x=0$ HSP. While a similar result was also encountered in Sec.~\ref{sec:1BM_MH_ANI}, here the topological consequences stemming from these symmetries differ because of the increased spatial dimen\-sio\-na\-li\-ty. Specifically, new topological features emerge only when the spectrum contains point nodes. In this case, the class AI allows defining an additional $\mathbb{Z}$ cry\-stal\-li\-ne topological invariant, which we denote $\nu_{k_{\mathcal{I}}}$, and associate with the fol\-lowing vorticity in $(\epsilon,k_y)$ space: 
\bea
\nu_{k_x=0}&=&\frac{1}{4\pi i}\sum_{\sigma=\pm1}\sigma\oint_{\mathcal{C}}\frac{\text{d}Z_{\epsilon,k_y,\sigma}}{Z_{\epsilon,k_y,\sigma}}\,,
\label{eq:vorticity_mirror}
\eea

\noi with ${\cal C}$ a path enclosing the node, and $\sigma=\pm1$ labelling the respective AI block. $Z_{\epsilon,k_y,\sigma}$ is obtained in a similar fashion to Eq.~\eqref{eq:Augmented_winding_number}. We remark that for the nodes shown in Fig.~\ref{fig:Figure9}, we find $|\nu_{k_x=0}|=|\upsilon|=1$.

The remaining two space group symmetries, i.e., $\{\sigma_{xz}\,|\,(\nicefrac{\pi}{Q},0)\}$ and $\{{\cal T}\,|\,(\nicefrac{\pi}{Q},0)\}$, need to be treated more carefully, since the former only leaves ISPs invariant, and thus does not introduce any changes to the topological classification in HSPs. The latter symmetry instead, combined with $\Theta$, leads to a class BDI$\oplus$BDI in the $k_x=0$ HSP. Hence, it only influences the topological properties of the system for a fully-gapped spectrum, since the BDI class cannot protect nodes in 1D~\cite{Matsuura_13}. Therefore, in the case of a full gap, we define the glide winding number: 
\bea 
w_{G,k_x=0}=\sum_{\sigma=\pm1}\sigma w_{k_x=0,\sigma}\equiv\frac{1}{4\pi i}\sum_{\sigma=\pm1}\sigma\int_{\text{BZ}}\frac{\text{d}z_{k_y,\sigma}}{z_{k_y,\sigma}},\qquad
\label{eq:glide_winding_number_def}
\eea 

\noi with $w_{k_x=0,\sigma}$ corresponding to the winding number of each BDI class Hamiltonian block of Eq.~\eqref{eq:GlideBlocks1D}, after the suitable $k_y$ dependence is accounted for, and the respective normalized complex function $z_{k_y,\sigma}=\det(\hat{A}_{k_x=0,k_y,\sigma})/|\det(\hat{A}_{k_x=0,k_y,\sigma})|$ is constructed. 

In contrast, when $\Delta_{\bm{k}}$ transforms according to the IRs $\{{\rm B_{2g},A_{2g}}\}$, we find deviations from the above behaviors. Notably, the magnetic and pairing point groups differ, since ${\rm M}_{\rm MHC}$ acts on $\Delta_{\bm{k}}\sim\{{\rm B_{2g},A_{2g}}\}$ as:
\bea
\big\{E,C_2,\sigma_{xz}{\cal T},\sigma_{yz}{\cal T}\big\}\Delta_{\bm{k}}=\big\{1,1,-1,-1\big\}\Delta_{\bm{k}}\,.
\eea

\noi Nevertheless, the sign-changing behavior of $\Delta_{\bm{k}}$ under mirror operations still allows us to define a point group ${\rm G}_{\rm MHC}$, which is preserved by the total Hamiltonian and is isomorphic to ${\rm M}_{\rm MHC}$. This group consists of the elements:
\bea
{\rm G}_{\rm MHC}=\big\{E,C_2,\sigma_{xz}^{\cal Q}{\cal T},\sigma_{yz}^{\cal Q}{\cal T}\big\}\,,
\eea

\noi with $\sigma_{xz,yz}^{\cal Q}={\cal Q}\sigma_{xz,yz}$, where we introduced the dimensionless electric charge operator ${\cal Q}=\tau_3$. Such symmetries have also been previously discussed in connection to p-wave superconductors~\cite{VolovikBook,SatoPwave}. There, these ensure that the energy spectrum is inversion symmetric, despite the fact that inversion itself is not preserved. Similarly here, ${\rm G}_{\rm MHC}$ ensures that the ener\-gy spectrum is inva\-riant under the original ${\rm M}_{\rm MHC}$ magnetic point group. Note also that the space group symmetries $\{\sigma_{xz,yz}\,|\,(\nicefrac{\pi}{Q},0)\}$, are correspondingly replaced by $\{\sigma_{xz,yz}^{\cal Q}\,|\,(\nicefrac{\pi}{Q},0)\}$.

To infer the arising topological modifications, we block diagonalize the 2D extension of Eq.~\eqref{eq:bloch_1BM_ani_helix} according to the symmetry of interest. We immediately observe that the effects of $\{{\cal T}\,|\,(\nicefrac{\pi}{Q},0)\}$ and $\{\sigma_{xz}^{{\cal Q}}\,|\,(\nicefrac{\pi}{Q},0)\}$ in HSPs remain the same, with the latter only affecting ISPs once again. The presence of ${\cal R}_{xz}^{\cal Q}$ establishes the class AI$\oplus$AI in the $k_y=\{0,\pi\}$ HSPs, where however the pairing gap is zero. In these HSPs ${\cal R}_{xz}^{\cal Q}$ can only protect nodes, with a $\mathbb{Z}$ invariant given by Eq.~\eqref{eq:vorticity_mirror} after interchanging $k_x$ and $k_y$, and con\-si\-de\-ring $k_y=\{0,\pi\}$. The remai\-ning two symmetries, i.e., $\{\sigma_{yz}^{\cal Q}\,|\,(\nicefrac{\pi}{Q},0)\}$ and ${\cal R}_{yz}^{\cal Q}$ lead to a BDI$\oplus$BDI class in the $k_x=\{0,q\}$ HSPs. When the spectrum is fully-gapped in these HSPs, one finds the glide $w_{G,{\rm HSP}}$ and mirror $w_{M,{\rm HSP}}$ winding numbers. These are correspondingly defined by, and in ana\-lo\-gy to, Eq.~\eqref{eq:glide_winding_number_def}. For details and concrete examples of the topological properties and the arising spectrum for a system with $\Delta_{\bm{k}}\sim\{{\rm B}_{\rm 2g},{\rm A}_{\rm 2g}\}$, see App.~\ref{app:AppendixC6}.

Closing this section, we point out that one can also introduce the BDI class weak $\mathbb{Z}$ invariants for $k_x$ points with a fully-gapped spectrum. These are given by the winding numbers $w_{k_x}$, which are defined in an analogous manner to $w_{k_y}$ in Eq.~\eqref{eq:winding_ky}. The $w_{k_x}$ invariants are expected to be particularly relevant when $\Delta_{\bm{k}}\sim\{{\rm B_{2g},A_{2g}}\}$. This is because, for a fixed $k_x$, the resulting pairing gap becomes effectively of the $p_y$-wave type, cf App.~\ref{app:AppendixC6}.

\subsubsection{SWC$_4$ phase - Majorana bidirectional edge modes}\label{subsec:SWC4_phase_majorana_dispersive_edge_modes}

\begin{figure}[t!]
\centering
\includegraphics[width=0.95\columnwidth]{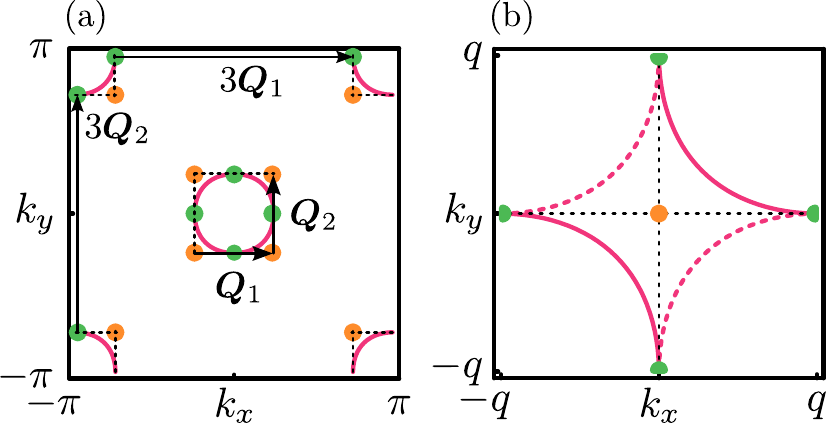}
\caption{2D 1BM of Fig.~\ref{fig:Figure7} with a double-$\bm{Q}$ magnetic texture. (a) FSs of the 1BM in the 1st BZ. We show the magnetic ordering wave vectors $\bm{Q}_1$ ($3\bm{Q}_1\equiv-\bm{Q}_1$) and $\bm{Q}_2$ ($3\bm{Q}_2\equiv-\bm{Q}_2$), connecting points at the Fermi level (green dots). As in the 1D case, cf Fig.~\ref{fig:Figure3}, points at higher energies are also connected by $\bm{Q}_{1,2}$, which upon increasing the magnetic energy scale give rise to nodes whose locations trace the dotted black lines. (b) Resul\-ting FS segments in the MBZ, where the points simultaneously experiencing magnetic scattering by both $\bm{Q}_1$ and $\bm{Q}_2$ (orange dots) are now centered at the $\Gamma(0,0)$ point.}
\label{fig:Figure11}
\end{figure}

In this section, we consider the case of a double-$\bm{Q}$ magnetic texture, with the ordering wave vectors depicted in Fig.~\ref{fig:Figure11}(a). Here, we focus on the SWC$_4$ profile which couples to the electrons through the exchange term:
\bea
&&H_{\rm mag}=\sum_n\bm{\psi}^{\dag}_n\big[M_\perp\cos(\bm{Q}_1\cdot\bm{R}_n)\sigma_z+M_{||}\sin(\bm{Q}_1\cdot\bm{R}_n)\sigma_x\no\\
&&\quad+M_\perp\cos(\bm{Q}_2\cdot\bm{R}_n)\sigma_z+M_{||}\sin(\bm{Q}_2\cdot\bm{R}_n)\sigma_y\big]\bm{\psi}_n\,.\label{eq:ham_swc4}
\eea

\noi The double-$\bm{Q}$ structure of the magnetic texture implies that the MBZ is obtained by folding in both $k_x$ and $k_y$ directions of the original BZ, and is defined as $\bm{k}\in(-q,q]\times(-q,q]$. The ISPs span the set $\bm{k}_{\cal I}=\{{\rm \Gamma}(0,0),{\rm X}(q,0),{\rm Y}(0,q),{\rm M}(q,q)\}$. To proceed, we employ the wave-vector-transfer Pauli matrices $\bm{\eta}$ and $\bm{\rho}$ related to foldings in the $k_x$ direction, as in Sec.~\ref{sec:1BM_MH_ANI}, and the Pauli matrices $\lambda_{1,2,3}$ and $\zeta_{1,2,3}$ related to foldings in the $k_y$ direction, acting in $\left\{\bm{k},\bm{k}+\bm{Q}_2\right\}$ and $\left\{\bm{k},\bm{k}+2\bm{Q}_2\right\}$ spaces, respectively. The re\-sul\-ting enlarged spinor reads:
\bea
&&\bm{\mathcal{X}}^{\rm 1BM,\,2D}_{\bm{k}}=\no\\
&&\mathds{1}_\zeta\otimes\frac{\lambda_2+\lambda_3}{\sqrt{2}}\left(\bm{\mathcal{X}}^{\rm 1BM}_{\bm{k}-\bm{q}_2},\,\bm{\mathcal{X}}^{\rm 1BM}_{\bm{k}+\bm{q}_2},\,
\bm{\mathcal{X}}^{\rm 1BM}_{\bm{k}+3\bm{q}_2},\,\bm{\mathcal{X}}^{\rm 1BM}_{\bm{k}-3\bm{q}_2}\right)^{\intercal}\quad
\label{eq:enlarged_spinor_SWC}
\eea

\noi where $\bm{\mathcal{X}}^{\rm 1BM}_{\bm{k}}$ is the 2D analog of Eq.~(\ref{eq:Enlarged_Spinor_1BM}). This yields the following class D ($\Xi=\tau_2\sigma_y{\cal K}$) bulk 2D Hamiltonian:
\bea
\hat{\mathcal{H}}_{\bm{k}}&=&\hat{F}\big(h_{\bm{k}}\big)\tau_3+\hat{F}\big(\Delta_{\bm{k}}\big)\tau_1\no\\
&-&\frac{M_\perp\big(\mathds{1}+\eta_1\big)\rho_1\sigma_z+M_{||}\big(\mathds{1}-\eta_1\big)\rho_3\sigma_x}{2}\no\\
&-&\frac{M_\perp\big(\mathds{1}+\zeta_1\big)\lambda_1\sigma_z+M_{||}\big(\mathds{1}-\zeta_1\big)\lambda_3\sigma_y}{2}\,,\quad
\label{eq:ham_SWC4}
\eea

\noi where $\hat{F}(h_{\bm{k}})$ and $\hat{F}(\Delta_{\bm{k}})$ are defined in Appendix~\ref{app:Functions}.

The band dispersions and the magnetic part of the BdG Hamiltonian of Eq.~\eqref{eq:ham_SWC4} are invariant under the magnetic point group $\rm M_{SWC_4}=C_{4}+(C_{4v}-C_{4})\mathcal{T}$, as well as the magnetic space group operations: $\{\mathcal{T},\,{\rm C_{4v}-C_{4}}\,|\,(\nicefrac{\pi}{Q},\nicefrac{\pi}{Q})\}$. See also Table~\ref{table:TableI}. Out of these five space group symmetries, the antiunitary $\tilde{\Theta}_{\bm{k}}=\{{\cal T}\,|\,(\nicefrac{\pi}{Q},\nicefrac{\pi}{Q})\}=ie^{i\pi(k_x+k_y)/Q}\lambda_2\rho_2\sigma_y{\cal K}$ defines a TR symmetry with $\tilde{\Theta}_{\bm{k}}^2=-e^{i\pi(k_x+k_y)/q}\mathds{1}$, and yields a Kramers degeneracy at $\bm{k}_{\tilde{\Theta}_{\bm{k}}}=\{{\rm \Gamma}(0,0),{\rm M}(q,q)\}$. The remaining space group symmetries do not lead to any additional symmetry-protected degeneracies in the spectrum. See App.~\ref{app:AppendixB} for further clarifications.

\begin{figure}[t!]
\centering
\includegraphics[width=0.95\columnwidth]{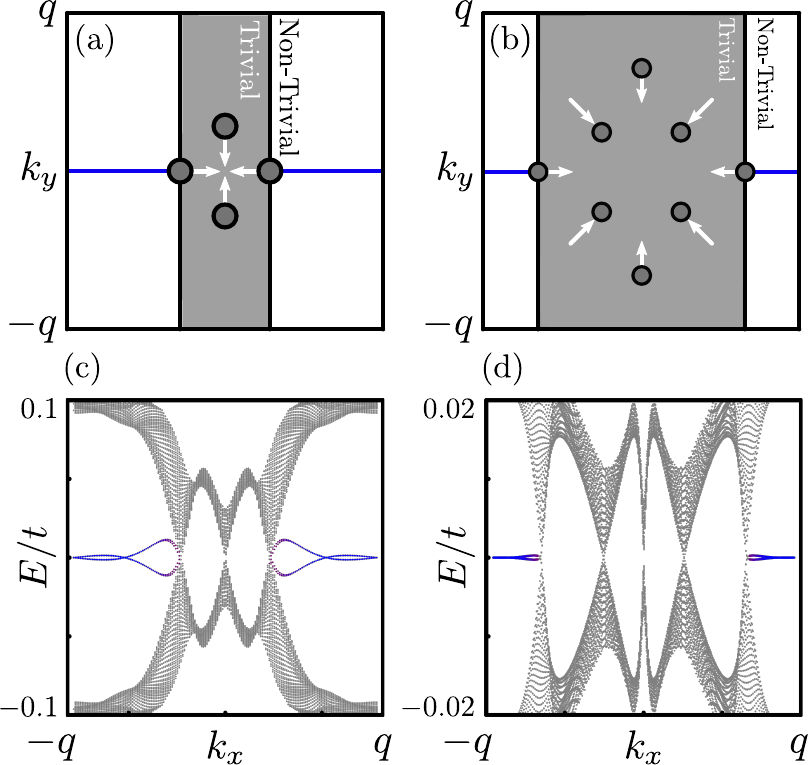}
\caption{(a) and (b) depict the paths swept by the pairs of nodes emer\-ging in the bulk ener\-gy spectrum for the 1BM in Fig.~\ref{fig:Figure7} in the SWC$_4$ phase. In contrast to Sec.~\ref{sec:1BM_2D_MH}, the nodes here are not topologically protected as reflected by the grey sha\-ding of the dots. (c) and (d) show the related dispersions for open boun\-dary conditions in the $y$ direction. (a) and (c) were obtained with $\Delta_{\bm{k}}=\Delta$ while for (b) and (d) we used the unconventional pairing gap $\Delta_{\bm{k}}=\Delta(\cos k_x - \cos k_y)$. For the latter we have four additional nodes in the spectrum compared to case (a). Note also that the resulting MF modes in (d) are also lifted from zero energy away from ISPs, but with a much flatter dispersion compared to the surface bands in (c). All the figures were obtained for $\Delta=0.1\,t$ and $L_x=L_y=701$, while in (c) and (d) we used $M_\perp=M_{||}=0.2\,t$ and $M_\perp=M_{||}=0.05\,t$, respectively.}
\label{fig:Figure12}
\end{figure}

\begin{figure*}[t!]
\centering
\includegraphics[width=\textwidth]{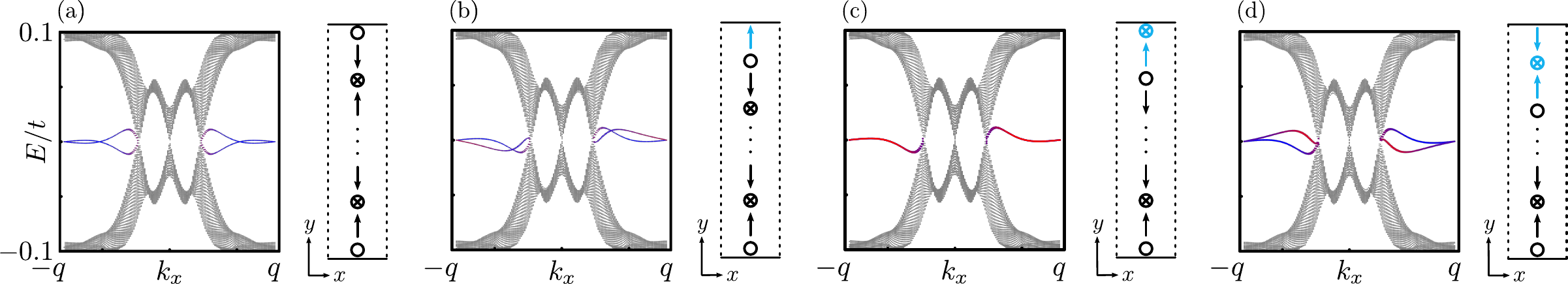}
\caption{Influence of crystal termination on the dispersion of the bidirectional MF edge modes in the 1BM in the SWC$_4$ phase. All spectra were obtained with open boundary conditions along the $y$ direction. The insets show the termination of the magnetic texture for: (a) $L_y=701$, (b) $L_y=702$, (c) $L_y=703$ and (d) $L_y=704$. We see that only the termination in (a) leads to a symmetric spectrum, since in (b)-(d) a net magnetization is accumulated at one edge (see cyan colored spin symbols). We used the parameter values $\Delta=0.1\,t$ and $M_\perp=M_{||}=0.2\,t$. The red/blue color coding is defined as in Table~\ref{table:TableII}.}
\label{fig:Figure13}
\end{figure*}

Similar to the previous section, also here, the point group ${\rm G}_{\rm SWC_4}$ preserved by the BdG Hamiltonian is decided by which \textit{one} out of the possible four IRs $\{\rm A_{1g},B_{1g},B_{2g},A_{2g}\}$, is stabilized for $\Delta_{\bm{k}}$. In a one-to-one correspondence to these four IRs, we find the scenarios:
\bea
{\rm G}_{\rm SWC_4}^{\rm A_{1g}}&=&\big\{E,C_2,2C_4,2\sigma_{\rm v}{\cal T},2\sigma_{\rm d}{\cal T}\big\}\,,\label{eq:GSWC4A1}\\
{\rm G}_{\rm SWC_4}^{\rm B_{1g}}&=&\big\{E,C_2,2C_4^{\cal Q},2\sigma_{\rm v}{\cal T},2\sigma_{\rm d}^{\cal Q}{\cal T}\big\}\,,\label{eq:GSWC4B1}\\
{\rm G}_{\rm SWC_4}^{\rm B_{2g}}&=&\big\{E,C_2,2C_4^{\cal Q},2\sigma_{\rm v}^{\cal Q}{\cal T},2\sigma_{\rm d}{\cal T}\big\}\,,\label{eq:GSWC4B2}\\
{\rm G}_{\rm SWC_4}^{\rm A_{2g}}&=&\big\{E,C_2,2C_4,2\sigma_{\rm v}^{\cal Q}{\cal T},2\sigma_{\rm d}^{\cal Q}{\cal T}\big\}\,.\label{eq:GSWC4A2}
\eea

\noi In a similar fashion, depending on the IR of $\Delta_{\bm{k}}$, we obtain four space group symmetries generated by products of $\{\mathds{1}\,|\,(\nicefrac{\pi}{Q},\nicefrac{\pi}{Q})\}$ and the mirror operations preserved by $\Delta_{\bm{k}}$. Note that the topological class remains D, irrespectively of the given point group $\rm G_{\rm SWC_4}$. As it is customary in this work, the effects of the point and space groups are presented at the end of the section.

For the 1BM in Fig.~\ref{fig:Figure11}, we find two pairs of nodes upon modifying the various parameters, as sketched in Fig.~\ref{fig:Figure12}(a). These pairs move along mutually-orthogonal HSPs in the MBZ as indicated by the white arrows in the figure. Spe\-ci\-fi\-cal\-ly, as the magnetic gap increases, the nodes first emerge at the ${\rm X}(q,0)$ and ${\rm Y}(0,q)$ points, and then move towards the ${\rm \Gamma}(0,0)$ point of the MBZ.

Similar nodes emerged for a MHC texture in 2D, cf. Fig.~\ref{fig:Figure9}. Hence, we expect that in this nodal regime, the topological pro\-per\-ties stemming from a SWC$_4$ texture are de\-scri\-ba\-ble by superim\-po\-sing the results originating from two MHC textures which wind in perpendicular spatial directions and dif\-fe\-rent spin planes. In this sense, the underlying to\-po\-lo\-gi\-cal mechanism is essentially 1D and, as long as these nodes are present, we do not expect to obtain any ge\-nui\-ne 2D topological superconducting phases. The latter become accessible only after the nodes meet at the ${\rm \Gamma}(0,0)$ point and annihilate. However, the Kramers de\-ge\-ne\-ra\-cy enforced by $\tilde{\Theta}_{\bm{k}}$ prohibits that, thus impo\-sing that \textit{only nodal} TSC phases become stabilized by a SWC$_4$ texture in such 1BMs. Nonetheless, as we show in the next section, the con\-si\-de\-ra\-tion of additional perturbations which violate $\tilde{\Theta}_{\bm{k}}$ unlock the possibility of gapped 2D topological superconducting phases.

We anticipate that the gapping of these nodes becomes possible by considering suitable perturbations of even infinitesimally-weak strength. This is because class D does not protect nodes in 2D~\cite{Matsuura_13}. In fact, one would expect that the nodes could be protected by some crystalline symmetry, but as we find, this is also not the case. Let us further elaborate on this, through exa\-mi\-ning the impact of the crystalline symmetries on the topological classification. Each one of the ${\rm (C_{4v}-C_{4})}\mathcal{T}$ symmetries acts as an effective TR symmetry in the HSPs that they leave inva\-riant. Each TR symmetry operator in the HSPs squares to $+\mathds{1}$, thus establishing the BDI symmetry class in these high-symmetry lines. However, neither the BDI class is capable of providing protection to nodes appea\-ring in these HSPs. Lastly, as explained in App.~\ref{app:AppendixB} and Ref.~\onlinecite{Shiozaki2016May}, nonsymmorphic symmetries in 2D systems can only affect the classification at ISPs, and not in HSPs.

The absence of a topological protection for the nodes is reflected in the lack of MFBs in the energy spectrum obtained when open boundary conditions are imposed in one of the two main axes. Related nu\-me\-ri\-cal results for $\Delta_{\bm{k}}\sim\{1,\cos k_x-\cos k_y\}$ are discussed in Fig.~\ref{fig:Figure12}, where we assume open boundary conditions in the $y$ direction\footnote{\label{footnote:deg}The particular choice of energy di\-sper\-sion and pairing order parameter $\Delta_{\bm{k}}$, leads to an additional unitary symmetry and renders the spectra twofold degenerate. A weak violation of this symmetry gets the degeneracy lifted away from ISPs, but preserves the number of MF edge modes.}. Re\-mar\-ka\-bly, instead of MFBs we find MF edge modes with the di\-stinctive feature that they do not have a fixed helicity or chirality. Even more remar\-ka\-bly, their spin-character and group velocity are $k_x$ dependent and become strongly affected by the type of crystal termination. See Ref.~\cite{Shiozaki2016May} for related findings, and Fig.~\ref{fig:Figure13} where we display the edge spectrum in Fig.~\ref{fig:Figure12} for various edge terminations. Clearly we see that the local spin content on a given edge modifies the MF dispersion on that same edge. On these grounds, we here term this type of less familiar MF edge modes as \textit{bidirectional}.

The properties of the bidirectional MF edge modes can be understood by viewing their presence as the outcome of the two coe\-xi\-sting MHCs. The MHC which winds spatially in the $y$ direction gives rise to MFBs in the conserved $k_x$ space, as long as the other MHC is completely neglected. In this ideal situation one obtains a spectrum similar to the one of Fig.~\ref{fig:Figure9}(c) after fol\-ding down to the MBZ. From this point of view, the secondary MHC me\-dia\-tes a BDI$\rightarrow$D symmetry-class transition for the 1D edge and, thus, lifts the protection of the MFBs. However, the pre\-sen\-ce of bidirectional MF edge modes is ensured by topologically-protected degeneracies at ISPs.

The emerging 1D physics implies that there should be suitable topological invariants that encode the presence of a persistent degeneracy at $k_x=q$, thus enforcing the presence of the bidirectional MF edge modes. These are no other than the $\mathbb{Z}_2$ weak invariants of class D, which correspond to the Majorana numbers ${\cal M}_{k_{x,y}=q}$~\cite{KitaevUnpaired,Chiu}:
\bea
{\cal M}_{k_x=q(k_y=q)}={\rm sgn}\prod_s^{\rm X,\,M\,(Y,\,M)}{\rm Pf}\big(\hat{B}_{\bm{k}_s}\big)\,,\label{eq:WeakDZ2}
\eea

\noi where ${\rm Pf}(\hat{B}_{\bm{k}_s})$ is the pfaffian of the skew-symmetric matrix $\hat{B}_{\bm{k}_{\cal I}}=\hat{U}_{\Xi}\hat{\cal H}_{\bm{k}_{\cal I}}$, where $\Xi=\hat{U}_{\Xi}{\cal K}$, with $\hat{U}_{\Xi}=\tau_2\sigma_y$. The pre\-sen\-ce of a C$_4$-sym\-me\-tric energy spectrum further renders the two inva\-riants equal. Within the weak-coupling limit, these are nontrivial for $M_-<\big|\Delta_{\bm{k}=\bm{0};\bm{q}_{1,2}}^+\big|<M_+$, which is sa\-ti\-sfied only after a simultaneous gap closing takes place at ${\rm X}(q,0)$ and ${\rm Y}(0,q)$. This me\-cha\-nism stabilizes the degeneracies at the edge ISPs.

Alternatively, as a consequence of the antiunitary magnetic point group elements $(\rm C_{4v}-C_4)\mathcal{T}$, each HSP resides in the 1D BDI class, for which, one can calculate the ensuing mirror winding number in 1D, $\tilde{w}_{M,{\rm HSP}}$, si\-mi\-lar to the weak invariant $w_{k_{x,y}}$ for the MHC models in 2D. $\tilde{w}_{M,{\rm HSP}}$ is a crystalline topological invariant which is distinct from previously discussed mirror invariants in this paper, in the sense that the symmetries $(\rm C_{4v}-C_4)\mathcal{T}$ do not induce any block diagonal structure of the Hamiltonian in their respective HSPs, but rather an emergent TR symmetry. Note lastly that the invariants for the $k_{x,y}=q$ HSPs fulfill the relation $\mathcal{M}_{k_{x,y}=q}=(-1)^{\tilde{w}_{M,{\rm HSP}}}$.

\subsubsection{SWC$_4$ texture - Genuine 2D TSCs}

As we pointed out in the previous section, the Kramers degeneracy that the $\tilde{\Theta}_{\bm{k}}$ symmetry imposes at the ${\rm \Gamma}(0,0)$ point of the MBZ, does not allow the nodes moving along the ${\rm \Gamma X}$ and ${\rm \Gamma Y}$ lines to annihilate, therefore prohibiting the emergence of a fully-gapped bulk energy spectrum and genuinely 2D topological superconducting phases. Nonetheless, a fully-gapped bulk energy spectrum is obtainable in the pre\-sen\-ce of additional Hamiltonian terms which achieve at least one of the following two possibilities: (i) either preserve $\tilde{\Theta}_{\bm{k}}$ but enforce the nodes {\color{black}to meet and annihilate away from the Kramers degenerate points of the MBZ}, i.e., away from $\bm{k}_{\Theta}=\{{\rm \Gamma}(0,0),{\rm M}(q,q)\}$, or (ii) violate $\tilde{\Theta}_{\bm{k}}$. 

\begin{figure}[t!]
\centering
\includegraphics[width=0.95\columnwidth]{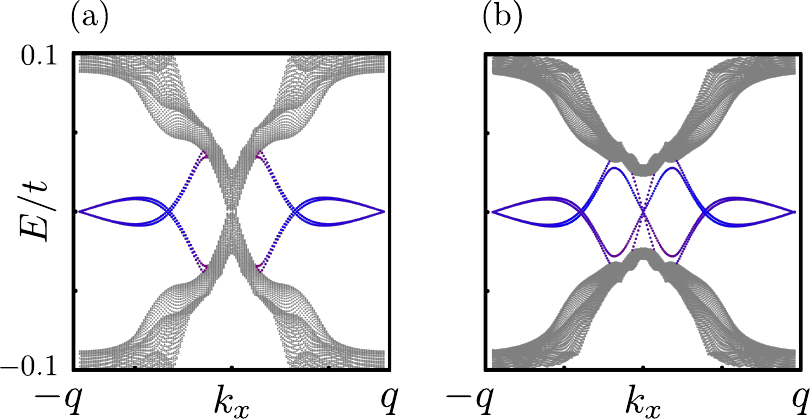}
\caption{(a) The nodal spectrum at $\Gamma(0,0)$ is protected by $\tilde{\Theta}_{\bm{k}}$ for the 1BM in Fig.~\ref{fig:Figure7} in the SWC$_4$ phase. (b) Resulting fully-gapped bulk spectrum for a broken C$_4$ symmetry due to a nematic dispersion $\xi^{\rm nem}_{\bm{k}}=\xi_{\bm{k}}+t_{\rm nem}\sin k_x\sin k_y$, where $\xi_{\bm{k}}$ is the dispersion used in (a). In the fully-gapped phase the preexisting bidirectional MF modes in (a) get accompanied by chiral MF modes. The figures were obtained using $L_x=L_y=1001$, $\Delta_{\bm{k}}=\Delta=0.1\,t$, $M_{||}=M_{\perp}=0.3\,t$ and $t_{\rm nem}=0.2\,t$.}
\label{fig:Figure14}
\end{figure}

\begin{figure*}[t!]
\centering
\includegraphics[width=\textwidth]{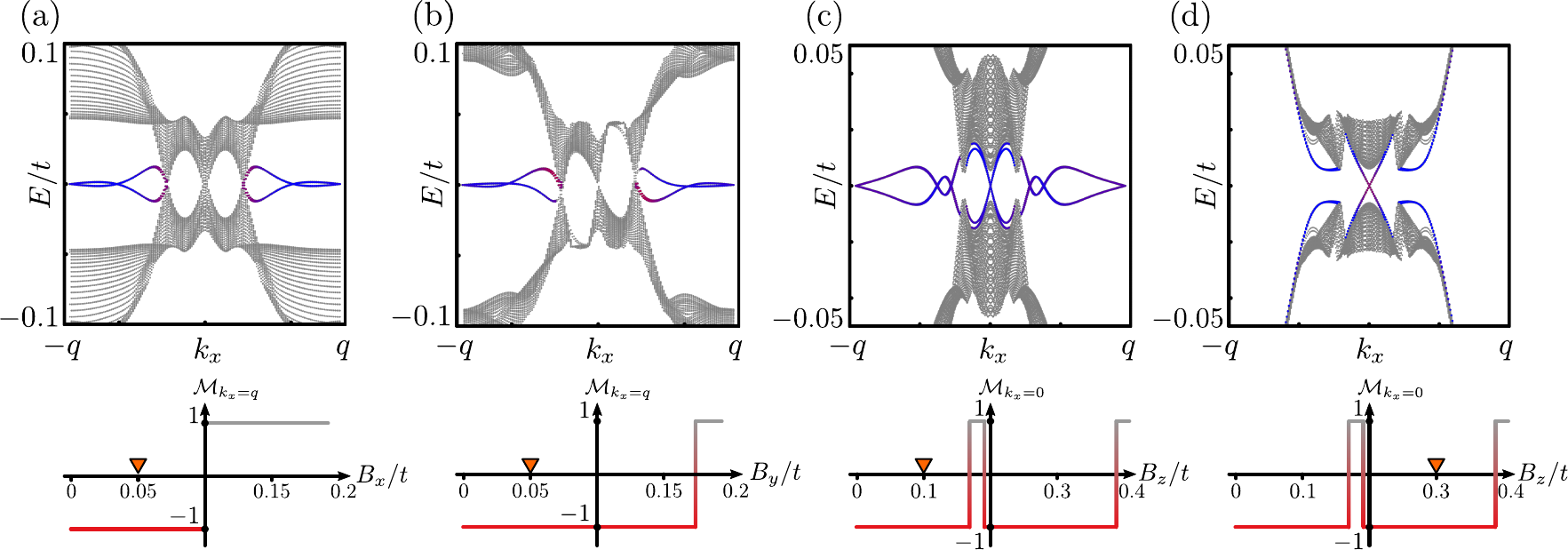}
\caption{The effects of Zeeman/exchange fields on the 1BM in Fig.~\ref{fig:Figure7} in the SWC$_4$ phase. (a)[(b)] Spectrum with the field in the $x$ [$y$] direction with strength $0.05\,t$. (c) displays the effect of a field oriented in the $z$ direction, thus, giving rise to a SSC$_4$ phase in the sufficiently-weak $B_z$ regime. Here the field does not lower the magnetic point group symmetry of the system, but it does lift the unprotected nodes of the bulk spectrum. Here we used $B_z=0.1\,t$. (d) Chiral edge modes for $B_z=0.3\,t$, which become accessible only after a band inversion at $\Gamma(0,0)$ takes place. For all figures we display the numerically calculated Majorana number Eq.~\eqref{eq:WeakDZ2} as a function of the magnetic field for the different cases. The invariant $\mathcal{M}_{k_x=0}$ in (c) and (d) is obtained by replacing (X, M) with (Y, $\Gamma$) in Eq.~\eqref{eq:WeakDZ2}, and allows us to infer the transition to the chiral TSC phase without residing to the calculation of the related 1st Chern number ${\cal C}_1$. The red/blue color coding is defined as in Table~\ref{table:TableII}. All figures were obtained with open boundary conditions in the $y$ direction and $\Delta_{\bm{k}}=\Delta=0.1\,t$, $M_{||}=M_\perp=0.2\,t$ and $L_x=L_y=1001$.}
\label{fig:Figure15}
\end{figure*}

In Fig.~\ref{fig:Figure14}, we present a situation in which the former scenario takes place. In this case, the addition of a term proportional to $\sin k_x\sin k_y$ to the dispersion preserves $\tilde{\Theta}_{\bm{k}}$ but violates $C_4$ symmetry. As a result, the nodes intersect away from ${\rm \Gamma}(0,0)$ and annihilate, therefore allowing for a chiral TSC. The second possibility is examined in the following section and is implemented by con\-si\-de\-ring the presence of a constant Zeeman field $\bm{B}$, which is added to the Hamiltonian via the term $\bm{B}\cdot\bm{\sigma}$.

Depending on the orientation of the Zeeman field, the magnetic point/space group symmetries can be fully or partially violated, thus, also affecting the type of the accessible dispersive MF edge modes. Spe\-ci\-fi\-cal\-ly, we find that an inplane Zeeman field leads to unidirectional (bidirectional) MF edge modes when its direction is pa\-ral\-lel (orthogonal) to the translationally-invariant termination edge. In contrast, an out-of-plane field preserves the bi\-di\-rec\-tio\-nal character of the edge modes. We insist that such edge modes and ISP degeneracies are still accessible even when the crystalline symmetries are all broken, since these are protected by the weak invariants defined in Eq.~\eqref{eq:WeakDZ2}, which still remain valid.

Apart from the abovementioned topological superconducting phases which have an underlying 1D character, the application of an out-of-plane field converts the SWC$_4$ phase into a SSC$_4$ for appropriate parameter regimes, and enables fully-gapped chiral topological superconducting phases. These are topologically equivalent to a p+ip TSC, and are classified according to the 1st Chern number ${\cal C}_1$ of the occupied bands~\cite{VolovikBook}.

Concluding this section, we remark that the introduction of the above perturbations is expected to influence the structure of the considered magnetic texture when the latter is treated in a self-consistent manner. However, sticking to the spirit of the explorative nature of this work, we neglect these modifications {\color{black}as they do not qualitatively modify the topological properties.}

{\color{black}\subsubsection{SWCB$_2$ texture - Majorana uni- and bi-directional\\ edge modes}}

An inplane Zeeman field with a direction which is not aligned with the main or diagonal axes defined by the HSPs $\{xz,yz,d_\pm z\}$ leads to the complete violation of the magnetic point and space group symmetries. In this case, it is the weak class D $\mathbb{Z}_2$ invariants which predict the appea\-rance of protected MZM crossings at edge ISPs independently of the orien\-ta\-tion of the termination edge. However, considering a magnetic field which is aligned with one of these axes, still allows for a nontrivial magnetic point group. For a Zeeman field in the $x$ ($y$) direction, the re\-sul\-ting magnetic point group becomes ${\rm M}_{\rm SWCB_2}=\{E,\sigma_{xz}\mathcal{T}\}$ (${\rm M}_{\rm SWCB_2}=\{E,\sigma_{yz}\mathcal{T}\}$), while the symmetry $\{\sigma_{yz(xz)}\,|\,(\nicefrac{\pi}{Q},\nicefrac{\pi}{Q})\}$ also remains intact. See also Table~\ref{table:TableI}. Hence, now, by virtue of the TR-symmetry $\sigma_{xz,yz}\mathcal{T}$ ac\-ting in the respective HSP, one can also define the BDI class mirror winding number $\tilde{w}_{M,{\rm HSP}}$.

In Figs.~\ref{fig:Figure15}(a)-(b) we present the edge spectra for a $B_x$ and a $B_y$ Zeeman field, respectively, with the system being open in the $y$ direction in both cases. By eva\-lua\-ting the respective weak invariant, we find protected de\-ge\-ne\-ra\-cies at the edge ISPs $k_x=\{0,q\}$. These persist until a gap closing takes place, which occurs for a Zeeman field value which depends on its orientation. Moreover, we observe the appearance of dispersive MF edge modes. In the open-system geometry of Fig.~\ref{fig:Figure15}(a), the antiunitary mirror symmetry implies that every state vector $\bm{\phi}_{n,k_x}$ corresponding to energy $E_{n,k_x}$ possesses a mirror par\-tner $\sigma_{xz}{\cal T}\bm{\phi}_{n,k_x}$ with energy $E_{n,-k_x}$, therefore resulting in a mirror symmetric spectrum. In contrast, the emergent antiunitary mirror symmetry $\sigma_{yz}{\cal T}$ in Fig.~\ref{fig:Figure15}(b) relates a state vector $\bm{\phi}_{n,k_x}$ with itself, thus, allowing for a mirror asymmetric spectrum and the emergence of unidirectional modes, see Fig.~\ref{fig:Figure15}(b).\\

{\color{black}\subsubsection{SWCB$_4$ texture - Majorana bidirectional/chiral edge modes}}

In the case of an out-of-plane $B_z$ field, the resulting SWCB$_4$ texture possesses nontrivial topological properties itself. Indeed, it has been shown~\cite{Christensen_18} that SWCB$_4$ is equivalent to a SSC$_4$ texture for $|B_z|<2|M_\perp|$. This allows us to esta\-blish a connection to prior works~\cite{Nakosai,TanakaYokoyamaNagaosa,LinderTanakaNagaosa,Ojanen2D,Jian2D,LadoSigrist} which have focu\-sed on the emergence of chiral topological superconducting phases in other magnetic platforms. The above criterion also implies that, for $|B_z|>2|M_\perp|$, SWCB$_4$ transforms into a ferromagnetic profile, which is expected to render the system trivial.

In connection to the detailed topological classification presented in the previous section for the SWC$_4$ phase, we observe that the addition of the $B_z$ field leaves the magnetic point group ${\rm M}_{\rm SWC_4}$ intact, but lifts the $\tilde{\Theta}_{\bm{k}}$ and the space group symmetries $\{{\rm C_{4v}-C_4}\,|\,(\nicefrac{\pi}{Q},\nicefrac{\pi}{Q})\}$. As a consequence, the classification of the SWCB$_4$ texture follows from the classification performed for the TSCs induced by a SWC$_4$ magnetic texture. Indeed, we find that the presence of the $B_z$ field still allows for MF edge modes crossings at ISPs, as seen in Fig.~\ref{fig:Figure15}(c). By evaluating the respective weak Majorana number, we find that for higher values of the external magnetic field, $B_z\sim0.2\,t$, the edge mode crossings at ISPs get lifted, while a band inversion at $\Gamma(0,0)$ takes place for slightly higher values of the field strength. Remarkably, the latter gives rise to two chiral MF edge mode branches as di\-splayed in Fig.~\ref{fig:Figure15}(d). The emergence of this chiral topological superconducting phase is also described by the Chern number $|{\cal C}_1|=2$. Note that the Chern-number value $|{\cal C}_1|=1$ is also generally accessible, as long as the accidental symmetry discussed in Footnote~\ref{footnote:deg} becomes lifted. 

We now summarize the key results for 1BMs in 2D. For a MHC the energy spectrum is nodal and leads to MFBs. Moreover, the classification in HSPs strongly depends on the IR of the pairing term. Nodes emerge also for a SWC$_4$ but they are not topologically stable. Nonetheless, degeneracies at ISPs persist and give rise to weak and crystalline TSC phases, which result into bidirectional MF edge modes. In the SWC$_4$ case, a fully-gapped spectrum is accessible only by violating C$_4$ or TR symmetries. Indeed, including a Zeeman field leaves the bidirectional modes intact, converts them into unidirectional modes, or, opens a gap in the spectrum and stabilizes chiral Majorana edge modes.\\

\subsection{Two-band models}\label{sec:2BM}

\begin{figure}[t!]
\centering
\includegraphics[width=0.95\columnwidth]{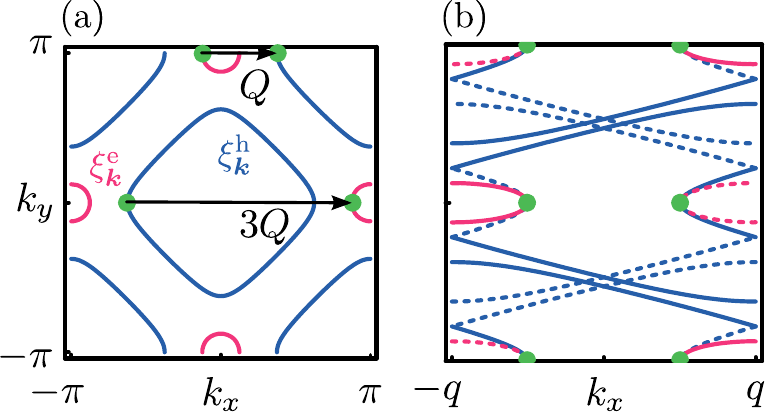}
\caption{Example of a 2BM in 2D, described by the dispersions $\xi_{\bm{k}}^{\rm h}=t_{\rm h}\cos k_x\cos k_y+t'_{\rm h}\big[\cos(2k_x)+\cos(2k_y)\big]-\varepsilon_{\rm h}$ and $\xi_{\bm{k}}^{\rm e}=t_{{\rm e}}\cos k_x \cos k_y-\varepsilon_{\rm e}$. We consider the parameters $t_{\rm h}=2.86\,t_{\rm e}$, $t'_{\rm h}=t_{\rm e}$, $\varepsilon_{\rm e}\approx-0.92\,t_{\rm e}$ and $\varepsilon_{\rm h}=-0.80\,t_{\rm e}$. (a) FSs of the 2BM in the first BZ. For clarity we only show half of the magnetic ordering wave vectors $Q$ and $3Q=-Q$ connecting bands at the Fermi level. (b) FS segments for the 2BM in the MBZ for a MHC, where the nested points at the Fermi level are marked by green dots. Note that nested points at finite energy, away from the Fermi level, are also present.}
\label{fig:Figure16}
\end{figure}

We now apply the classification methods discussed in the previous sections to 2D 2BMs. For an example of such a model see Fig.~\ref{fig:Figure16} which, for the chosen pa\-ra\-me\-ters, yields the FSs shown in Fig.~\ref{fig:Figure16}(a). Interband FS ne\-sting, with the two ordering wave vectors $\bm{Q}_{1,2}$, takes place between the hole (fuchsia) and electron (navy blue) pockets. Hence, the system has the pos\-si\-bi\-li\-ty to develop either a single-$\bm{Q}$ or a double-$\bm{Q}$ magnetic phase~\cite{Christensen_18}. 

Recall from our previous discussion in Sec.~\ref{sec:2BM_1D}, that a multiband system allows for an interplay of interband and intraband magnetic scattering, as well as for a here-assumed intraband pairing gap which is a matrix in band space $\hat{\Delta}_{\bm{k}}=(\Delta^{\rm e}_{\bm{k}}+\Delta^{\rm h}_{\bm{k}})/2+\kappa_3(\Delta^{\rm e}_{\bm{k}}-\Delta^{\rm h}_{\bm{k}})/2$. Even more importantly, we show here that the inclusion of the additional band, may in many realistic situations enrich the symmetry of the system. As we discuss below, a number of features that become unlocked for 2BMs open perspectives for new phe\-no\-me\-na and TSC phases.

\subsubsection{MHC texture - Majorana and Andreev flat bands}\label{subsubsec:MH_phase_majorana_flat_bands}

\begin{figure}[t!]
\centering
\includegraphics[width=0.95\columnwidth]{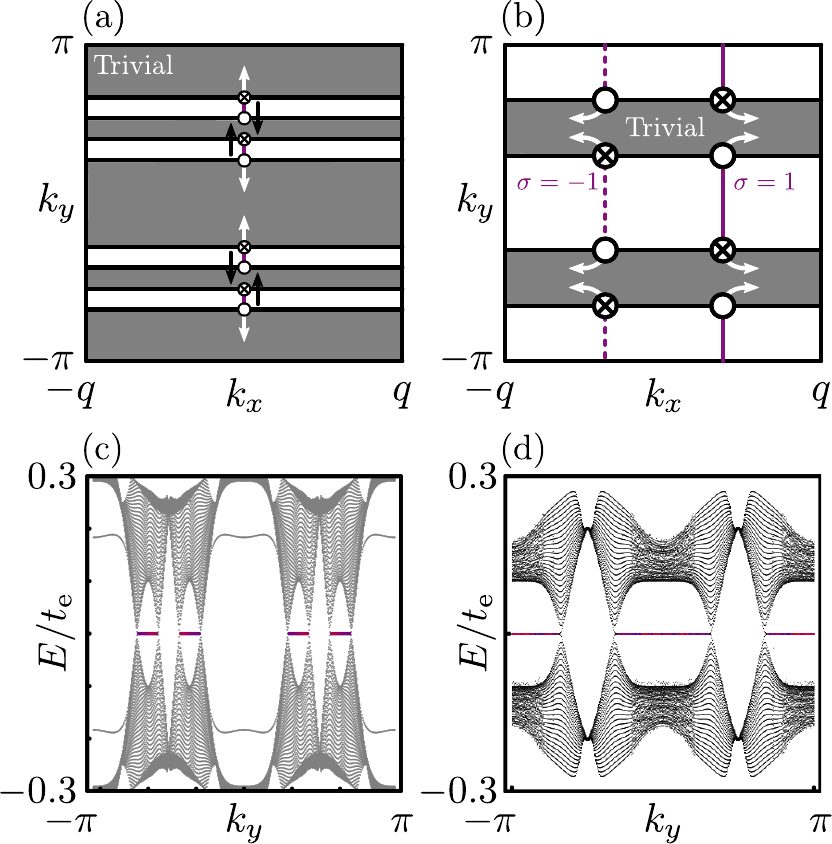}
\caption{Sketches of nodes and the numerically-obtained dispersions for the 2BM of Fig.~\ref{fig:Figure16} in the MHC phase. (a) and (c) $\Delta^{\rm e}=t_{\rm e}$, $\Delta^{\rm h}=-t_{\rm e}$ and $M^{\rm eh}_{||,\perp}=M^{\rm e}_{||,\perp}=M^{\rm h}_{||,\perp}=1.1\,t_{\rm e}$. The nodes are moving on straight lines similar to the 1BM in 2D, cf. Fig.~\ref{fig:Figure9}. The number of nodes has increased with the number of nested points, see green points in Fig.~\ref{fig:Figure16}. (b) and (d) $\Delta^{\rm e}=\Delta^{\rm h}=t_{\rm e}$, $M^{\rm eh}_{||}=M^{\rm eh}_{\perp}=1.1t_{\rm e}$ and $M_{||,\perp}^{\rm e}=M_{||,\perp}^{\rm h}=0$. The nodes are now moving on arcs, due to the interband-only scat\-te\-ring me\-dia\-ted by the magnetic texture. Both dispersions are obtained with open boundary conditions along the $x$ direction, and with $L_x=L_y=401$. In (b) $\sigma=\pm1$ labels the two blocks of the Hamiltonian after performing the unitary transformation with the operator $\mathcal{S}$ in Sec.~\ref{sec:2BM_1D}.}
\label{fig:Figure17}
\end{figure}

The present section builds upon the analyses of the 1D 2BMs and the 2D 1BMs under the influence of a MHC. In the general case, in which intra- and inter-band magnetic scatterings are present, the system is dictated by the same magnetic point and space group symmetries discussed in Sec.~\ref{sec:1BM_2D_MH}. The nodes in the bulk energy spectrum therefore possess a topological charge reflected in their vorticity $\upsilon$. Moreover, in HSPs one can also define the respective mirror vor\-ti\-ci\-ty $\nu_{{\rm HSP}}\in\mathbb{Z}$ following the definition in Eq.~\eqref{eq:vorticity_mirror}. By further assu\-ming spatially-constant pairing gaps $\Delta^{\rm e,h}$ for the two po\-ckets, we find that the edge spectrum contains MFBs, see Fig.~\ref{fig:Figure17}, whose topological protection can be genuinely described either by the strong local invariants mentioned above, or, by the weak invariant $w_{k_y}$.

As found previously for the 2BMs in 1D, the topologically-stable bulk nodes and edge MFBs become accessible even for $\Delta^{\rm e}\Delta^{\rm h}<0$, when intraband magnetic scattering is assumed. In Fig.~\ref{fig:Figure17}(a) and~\ref{fig:Figure17}(c) we display the re\-sul\-ting path of the nodes in the MBZ and spectrum, respectively, for $\Delta^{\rm e}=-\Delta^{\rm h}$. For the chosen values of inter- and intraband scattering, the to\-po\-lo\-gi\-cal pro\-per\-ties are essentially determined only by the hole pocket, thus, exhibiting a similar phe\-no\-me\-no\-lo\-gy to Fig.~\ref{fig:Figure9}(a), with the nodes moving on straight lines.

If instead we  restrict to an interband-only magnetic texture, we find that almost all the features of the 2BMs in 2D are directly inherited from the 1D interband versions, namely: (i) the Hamiltonian is block-diagonalizable into an AIII$\oplus$AIII fashion [see Eq.~\eqref{eq:Energy_2BM_Blocks} and Fig.~\ref{fig:Figure17}(b)], (ii) the number of gap clo\-sing points $\bm{k}_{\rm c}$ and edge modes double compared to the 1BMs, (iii) the gap clo\-sing points are found through the relation $\xi_{\bm{k}\pm3\bm{q}_1}^{\rm e}\Delta^{\rm h}=\xi_{\bm{k}\mp3\bm{q}_1}^{\rm h}\Delta^{\rm e}$, and (iv) nodal topological superconducting phases are accessible only for $\Delta^{\rm e}\Delta^{\rm h}>0$. 

All the above features are reflected in Fig.~\ref{fig:Figure17}(b) and (d) where we display the path taken by the nodes upon va\-ria\-tion of the magnetic and superconducting gaps, and the edge spectrum, respectively. Notably, here one obtains in most cases Andreev flat bands (AFBs), which extend the Andreev zero modes (AZMs) discussed in Sec.~\ref{sec:2BM_1D} to 2D. AFBs protected by the symmetry $\sigma^{\cal Q}_{xz}{\cal T}\Theta$ ($\{{\cal T}\Theta\,|\,(\nicefrac{\pi}{Q},0)\}$) are also accessible in $k_y=\{0,\,\pi\}$ for $\Delta_{\bm{k}}\sim\{{\rm B_{2g},A_{2g}}\}$ (for $\Delta_{\bm{k}}$ in any of the four D$_{\rm 4h}$ IRs), in which case the symmetry class is $\bigoplus_4{\rm AIII}$, and the topological invariant is a mirror (glide) win\-ding number $w_{M,{\rm HSP}}$ ($w_{G,{\rm HSP}}$). In contrast, MF excitations become possible only in crystalline TSC phases obtained for $\Delta_{\bm{k}}\sim\{{\rm B_{2g},A_{2g}}\}$, where the symmetry $\sigma^{\cal Q}_{yz}{\cal T}$ or $\{\sigma^{\cal Q}_{yz}\Theta\,|\,(\nicefrac{\pi}{Q},0)\}$ drives the symmetry-class transition AIII\,$\oplus$\,AIII\,$\rightarrow$\,BDI\,$\oplus$\,BDI, which in turn allows for MFBs. These are protected by a mirror winding number $\tilde{w}_{M,{\rm HSP}}$, which is similar to the weak inva\-riant $w_{k_{x,y}}$ for the 1BMs in 2D in the presence of a MHC. See Table~\ref{table:TableIII}.

Notably, a very crucial difference compared to the 2D 1BM is that, here, the paths along which the bulk nodes move in $\bm{k}$ space do not coincide with the main axes of the MBZ. Remar\-ka\-bly, here the nodes generally move on arcs, as indicated by the white arrows in Fig.~\ref{fig:Figure17}(b). This enables the bulk nodes to meet and annihilate away from Kramers degenerate points, thus, opening the perspective for fully-gapped spectra for class D or DIII topological superconducting phases in the SWC$_4$ phase. This implies that here strong 2D TSC phases seem to become accessible without the requirement of external perturbations, e.g., Zeeman fields, which was the case for 1BMs. 

\subsubsection{SWC$_4$ phase - Quasi-helical Majorana edge modes}\label{subsubsec:SWC4_phase_majorana_helical_edge_modes}

\begin{figure}[t!]
\centering
\includegraphics[width=\columnwidth]{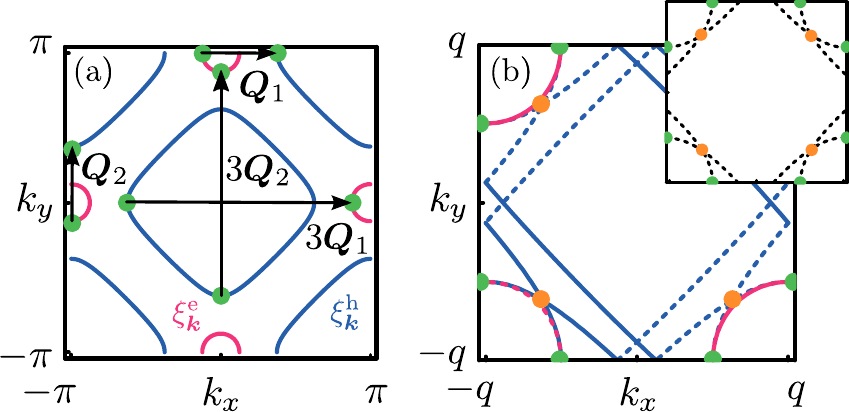}
\caption{2BM of Fig.~\ref{fig:Figure16} under the influence of a double-$\bm{Q}$ magnetic texture. (a) FSs of the 2BM in the first BZ. We sketch the magnetic ordering wave vectors $\bm{Q}_{1,2}$ ($3\bm{Q}_{1,2} \equiv-\bm{Q}_{1,2}$), which connect points at the Fermi level (green dots). For clarity we only show half of the ordering wave vectors. (b) The resulting FS segments in the MBZ. As in Fig.~\ref{fig:Figure11}, we also display the points connected by both $\bm{Q}_1$ and $\bm{Q}_2$ (orange dots) at energies away from the Fermi level. Inset: The dotted black lines show the gap closing points $\bm{k}_{\rm c}$ in the MBZ.}
\label{fig:Figure18}
\end{figure}

We now proceed by studying 2BMs with an \textit{interband-only} double-$\bm{Q}$ SWC$_4$ texture. As in previous sections, we employ the usual set of wave-vector-transfer Pauli matrices $\bm{\zeta}$, $\bm{\lambda}$, $\bm{\eta}$ and $\bm{\rho}$, in order to account for the magnetic scattering taking place in the two orthogonal directions, as displayed in Fig.~\ref{fig:Figure18}(a). The MBZ is displayed in Fig.~\ref{fig:Figure18}(b), where points connected by a single $\bm{Q}$-vector at the FS and points connected by both $\bm{Q}$-vectors, are marked by green and orange dots, respectively. 

Due to the interband nature of the magnetic scat\-te\-ring the BdG Hamiltonian now enjoys a TR symmetry $\Theta=\kappa_3\mathcal{T}$. This satisfies $\Theta^2=-\mathds{1}$ and leads to Kramers pairs (KP) at all the ISPs of the MBZ, thus enlisting the BdG Hamiltonian in the DIII symmetry class. Nodes in the bulk spectrum of a DIII Hamiltonian are topologically stable only at ISPs, and are classified by a vor\-ti\-ci\-ty akin to the one in Eq.~\eqref{eq:vorticity}. For a fully-gapped bulk spectrum, class DIII supports one strong and two weak $\mathbb{Z}_2$ to\-po\-lo\-gi\-cal invariants~\cite{QiTFT,QiTopoInv,Ardonne,LawDIII}, that we here construct as:
\bea
{\cal M}^{\rm KP}&=&\prod_s^{\rm \Gamma,\,X\,,M\,,Y}{\rm Pf}\big(\hat{\cal W}_{\bm{k}_s}\big)/\sqrt{\det\big(\hat{\cal W}_{\bm{k}_s}\big)},
\label{eq:StrongDIIIZ2}\\
{\cal M}^{\rm KP}_{k_x=q(k_y=q)}&=&\prod_s^{\rm X,\,M\,(Y,\,M)}{\rm Pf}\big(\hat{\cal W}_{\bm{k}_s}\big)/\sqrt{\det\big(\hat{\cal W}_{\bm{k}_s}\big)}.\quad\label{eq:WeakDIIIZ2}
\eea

\noi In the above, we defined the skew-symmetric ``sewing'' matrix $\hat{\cal W}_{\bm{k}_{\cal I}}\equiv\hat{\cal U}_\Theta\hat{A}_{\bm{k}_{\cal I}}$ at ISPs $\bm{k}_{\cal I}$ only. The $\hat{\cal W}$ matrix is the DIII analog of the AII class sewing matrix introduced by Fu and Kane~\cite{FuKaneZ2}. The difference is that, here, $\hat{\cal U}_\Theta=i\sigma_y$ corresponds to the unitary part of the block off-diagonal $\Theta$, obtained in the diagonal basis of the chiral symmetry operator $\Pi$. In this basis, we identify the block off-diagonal part of the Hamiltonian as $\hat{A}_{\bm{k}}$. The transition to this basis is here effected via the transformation $(\Pi+\tau_3)/\sqrt{2}$, which brings the arising chiral symmetry generator $\Pi=\kappa_3\tau_2$ into the form $\Pi=\tau_3$, and leads to:
\bea
\hat{A}_{\bm{k}}&=&-\sum_{s}^{{\rm e,h}}{\cal P}_s\Big[\hat{F}\big(\Delta_{\bm{k}}^s\big)+i\kappa_3\hat{F}\big(h_{\bm{k}}^s\big)\Big]\no\\
&&-\kappa_2\frac{M_\perp\big(\mathds{1}+\eta_1\big)\rho_1\sigma_z+M_{||}\big(\mathds{1}-\eta_1\big)\rho_3\sigma_x}{2}\no\\
&&-\kappa_2\frac{M_\perp\big(\mathds{1}+\zeta_1\big)\lambda_1\sigma_z+M_{||}\big(\mathds{1}-\zeta_1\big)\lambda_3\sigma_y}{2}\,,\quad
\label{eq:bloch_SWC4}
\eea

\noi with ${\cal P}_{\rm e,h}$ ($\hat{F}$) defined once again as in Sec.~\ref{sec:2BM_1D} (App.~\ref{app:Functions}). 

We now move on with the discussion of the various crystalline symmetries, which are identical to the ones dicta\-ting the 1BMs in Sec.~\ref{subsec:SWC4_phase_majorana_dispersive_edge_modes}. Specifically, the antiunitary mirror symmetries $({\rm C_{4v}-C_4}){\cal T}$ belonging to the re\-la\-ted ${\rm G}_{\rm SWC_4}$ point group discussed in Sec.~\ref{subsec:SWC4_phase_majorana_dispersive_edge_modes}, combine with $\Theta$ and give rise to the unitary mirror ope\-ra\-tions ${\cal R}=({\rm C_{4v}- C_4})\mathcal{T}\Theta=\kappa_3({\rm C_{4v}-C_4})$. These lead to a AIII$\oplus$AIII (D$\oplus$D) class in the corresponding HSP when the pairing gap $\Delta_{\bm{k}}$ is even (odd) under the given mirror operation, e.g., for $\Delta_{\bm{k}}\in{\rm B_{1g}}$ the symmetry class is AIII (D) in the $xz$ ($x=y$) and $yz$ ($x=-y$) HSPs. Both AIII and D classes are nontrivial in 1D for a fully-gapped system. Thus, HSPs dictated by the symmetry class AIII$\oplus$AIII [D$\oplus$D] and at the same time exhibit a fully-gapped spectrum, are characterized by a $\mathbb{Z}$ [$\mathbb{Z}_2$] mirror win\-ding number $w_{M,{\rm HSP}}$ [mirror Majorana number ${\cal M}_{M,{\rm HSP}}$].\footnote{${\cal M}_{M,{\rm HSP}}$ is defined as ${\cal M}_{M,{\rm HSP}}={\rm sgn}\prod_{\sigma}{\cal M}_{\sigma,{\rm HSP}}$, where $\sigma=\pm1$ labels the D$\oplus$D blocks. Each ${\cal M}_{\sigma,{\rm HSP}}$ follows from Eq.~\eqref{eq:WeakDZ2}.\label{footnote:mirrorMajorananumber}} In contrast, nodes in HSPs dictated by either AIII$\oplus$AIII or D$\oplus$D are not protected.

On the other hand, nonsymmorphic symmetries can only influence the topological classification at ISPs. Remarkably, the $\Gamma(0,0)$ and ${\rm M}(q,q)$ points are under the simultaneous influence of two TR symmetries which square to $-\mathds{1}$, i.e., $\Theta=\kappa_3{\cal T}$ and $\{{\cal T}\,|\,(\nicefrac{\pi}{Q},\nicefrac{\pi}{Q})\}$, thus observing a fourfold degeneracy. This can be understood in terms of the unitary symmetry $\{\kappa_3\,|\,(\nicefrac{\pi}{Q},\nicefrac{\pi}{Q})\}=e^{i(k_x+k_y)\pi/Q}\kappa_3\lambda_2\rho_2$ which emerges at these two ISPs.\footnote{Note that the above fourfold degeneracy does not lead to hourglass MFs. Following Ref.~\onlinecite{Shiozaki2016May}, we can attribute this to the commutation relation $[\{\kappa_3\,|\,(\nicefrac{\pi}{Q},\nicefrac{\pi}{Q})\},\Pi]=0$ which holds here.}

We now investigate a concrete 2BM, specifically the model defined in Fig.~\ref{fig:Figure18}. Similar to the analysis of the 1BMs in the SWC$_4$ phase, we identify two types of MBZ points, namely the points connected by a single $\bm{Q}$-vector (green dots) and the points connected by both ordering wave vectors (orange dots), cf Fig.~\ref{fig:Figure18}(b). Based on the results of the previous paragraphs, we find that the nodes move on arcs determined by the intersection of the two bands connected by a single $\bm{Q}$-vector, e.g., $\xi_{\bm{k}+3\bm{q}_1-\bm{q}_2}^{\rm e}=\xi^{\rm h}_{\bm{k}-3\bm{q}_1-\bm{q}_2}$. The paths of the nodes upon variations of the magnetic or superconducting gaps are marked by the black dotted lines in the inset of Fig.~\ref{fig:Figure18}(b). Once again, bulk nodes appear strictly for $\Delta^{\rm e}\Delta^{\rm h}>0$, since we here consider an interband-only texture. 

\begin{figure}[t!]
\centering
\includegraphics[width=0.95\columnwidth]{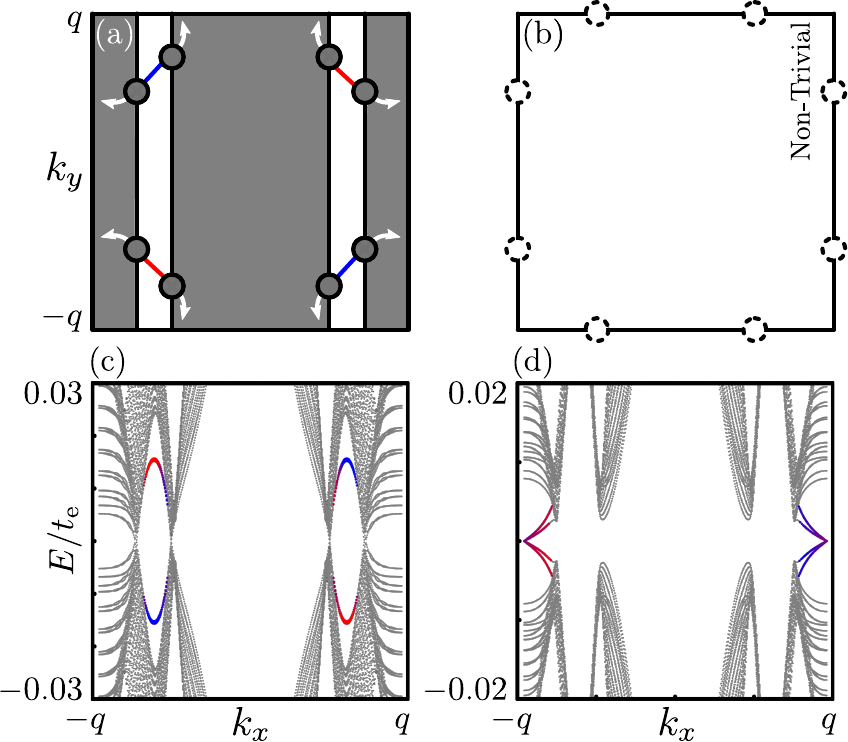}
\caption{Bidirectional and mirror-symmetry-protected quasi-helical Majorana edge modes for the 2BM in the SWC$_4$ phase. (a) The bulk nodes move from the points connected by two $\bm{Q}$-vectors (orange dots in Fig.~\ref{fig:Figure18}(b)), and meet at points connected by a single $\bm{Q}$-vector (green dots in Fig.~\ref{fig:Figure18}(b)), as indicated by the white arrows. (c) The spectrum related to (a), with bidirectional Majorana edge modes ($M^{\rm eh}_{||,\perp} = 0.095\,t_{\rm e}$). Note that here the bidirectional MF modes are not to\-po\-lo\-gi\-cal\-ly protected, due to the similar lack of protection seen by the bulk nodes. (b) Sketch of the MBZ after the nodes have met and annihilated at the points marked by the dotted circles. (d) Resulting quasi-helical edge modes connected to the sketch in (b), where we used $M^{\rm eh}_{||,\perp} = 0.11\,t_{\rm e}$. For clarity, in (d), we show only the modes on a single edge, since the $z$ spin axis electronic spin polarization of the modes on the other edge is exactly opposite. We note that both types of spectra are twofold de\-ge\-ne\-ra\-te for rea\-sons discussed in Footnote~\ref{footnote:deg}. We used open boundary conditions in the $y$ direction, $\Delta^{\rm e}=\Delta^{\rm h}=0.1\,t_{\rm e}$, $M^{\rm e, h}_{z,\perp}=0$ and $L_x=L_y=501$.}
\label{fig:Figure19}
\end{figure}

The presence of bulk nodes goes hand in hand with the emergence of bidirectional MF edge modes, as seen in Fig.~\ref{fig:Figure19}(a) and (c) where we display the nodes and edge spectrum. However, in the present situation, the bulk nodes are not topologically-protected, thus implying the same for the resulting bidirectional edge modes. Both nodes and edge modes are thus removable by considering additional Hamiltonian terms which do not modify the ensuing DIII class. Similar conclusions were drawn for the 1BMs in the SWC$_4$, with the crucial distinction that there the edge modes had a topologically protected cros\-sing at $k_{x,y}=q$. Such protected cros\-sings do not arise for the bidirectional MF modes in Fig.~\ref{fig:Figure19}(c). 

By increasing the energy scale of the magnetic gaps the nodes move on arcs, as indicated by the white arrows in Fig.~\ref{fig:Figure19}(a), and meet up at the green points in the MBZ in Fig.~\ref{fig:Figure18}(b), when the familiar criterion is sa\-ti\-sfied $\Delta^{\rm e}\Delta^{\rm h}=M_{\pm}^2$. In contrast to the 1BMs, here the nodes \textit{do} get lifted when they meet up, since they intersect away from ISPs, as sketched in Fig.~\ref{fig:Figure19}(b). Beyond this point the spectrum is fully-gapped. In the present case, the fourfold degeneracies at ${\rm \Gamma}(0,0)$ and ${\rm M}(q,q)$ addi\-tio\-nal\-ly imply that, there, ${\rm Pf}\big(\hat{\cal W}_{\bm{k}_s}\big)$ features an even number of gap clo\-sings upon sweeping the va\-rious parameters. Hence, the above-mentioned invariants simplify as ${\cal M}^{\rm KP}={\cal M}^{\rm KP}_{k_x=q}{\cal M}^{\rm KP}_{k_y=q}$ where ${\cal M}^{\rm KP}_{k_x=q(k_y=q)}={\rm sgn}[{\rm Pf}(\hat{\cal W}_{\bm{k}_{\rm X(Y)}})]$. In the event of a C$_4$-symmetric energy spectrum, which is actual\-ly the case here, the two invariants are equal. The two weak Majorana Kramers pair (MKP) numbers ge\-ne\-ral\-ly be\-co\-me non\-tri\-vial si\-mul\-ta\-neou\-sly. Nonetheless, here we find that all three inva\-riants remain trivial.

{\color{black}Despite the fact that the DIII} invariants are here all trivial, in Fig.~\ref{fig:Figure19}(d) we indeed find the here-termed quasi-helical edge modes centered at $k_x=q$, which are protected by the mirror symmetry $\sigma_{yz}{\cal T}$. These come in pairs, and their electronic spin-polarization is opposite on opposite edges, similar to what is encountered for their helical counterparts. However, the quasi-helical ones appear only for edges preseving the respective mirror symmetry, in stark contrast to the helical edge modes stemming from the strong DIII invariant in 2D, which emerge for a termination of an arbitrary orientation. Since for the above numerical calculations we have considered $\Delta_{\bm{k}}^{\rm e,h}\sim {\rm A}_{\rm 1g}$, the HSP plane is dictated by the AIII$\oplus$AIII symmetry class. On the other hand, con\-si\-de\-ring a pairing gap $\Delta_{\bm{k}}^{\rm e,h}\sim{\rm B_{2g}}$ imposes the D$\oplus$D symmetry class in the $k_x=q$ HSP, and allows instead for quasi-helical Majorana edge modes protected by a mirror $\mathbb{Z}_2$ inva\-riant. The presence of two possible types of to\-po\-lo\-gi\-cal protection for the touching point, further suggests a different behaviour for the quasi-helical Majorana edge modes in response to external perturbations, e.g., Zeeman fields.

\subsubsection{Magnetic-field-induced TSC phases: Majorana uni/bi-directional, quasi-helical, and chiral edge modes}

We complete the study of the 2BMs in 2D, by con\-si\-de\-ring the effects of an additional Zeeman field on the system discussed in the previous paragraph. In this case, the system undergoes the symmetry class transition DIII\,$\rightarrow$\,D. Therefore, for a fully-gapped bulk energy spectrum, chiral edge modes become accessible. Even more, when the field is aligned with one of the HSPs, mirror-symmetry protected edge modes are also possible. For a magnetic field in the $x$ ($y$) direction, the re\-sul\-ting magnetic point group becomes ${\rm M}_{\rm SWCB_2}=E+\sigma_{xz}\mathcal{T}$ (${\rm M}_{\rm SWCB_2}=E+\sigma_{yz}\mathcal{T}$). HSPs now belong to the BDI symmetry class, since the antiunitary elements of the point group act as TR symmetries. Hence, we can define the mirror winding numbers $\tilde{w}_{M,{\rm HSP}}\in\mathbb{Z}$ similarly to the weak winding numbers $w_{k_{x,y}}$ introduced in Eq.~\eqref{eq:winding_ky}. 

\begin{figure}[t!]
\centering
\includegraphics[width=\columnwidth]{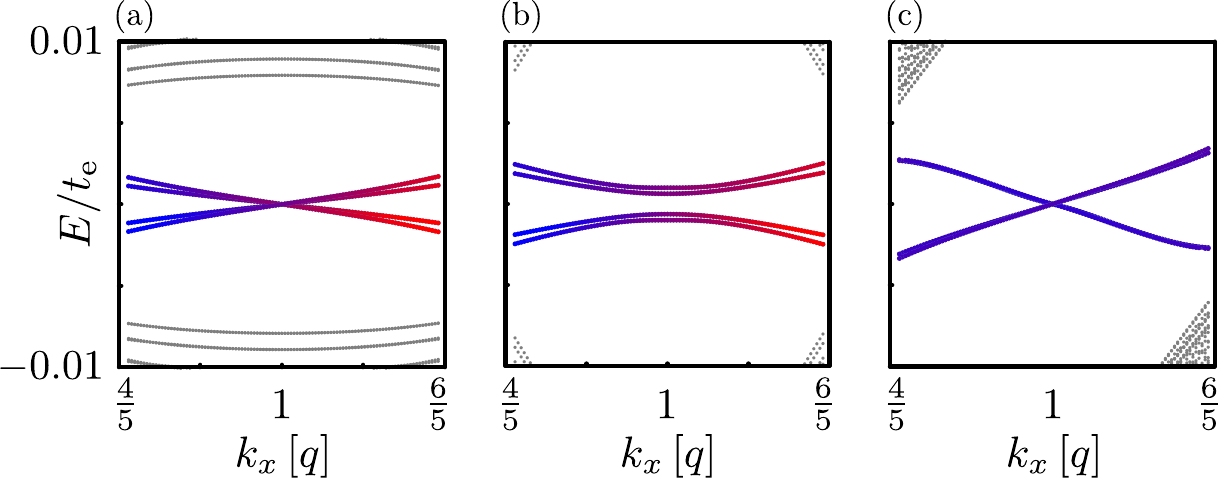}
\caption{Impact of an applied Zeeman field on the quasi-helical Majorana modes of Fig.~\ref{fig:Figure19}(d). Once again, only modes on a single edge are also shown here for clarity. (a) Zoom in the edge mode spectrum of Fig.~\ref{fig:Figure19}(d). (b)[(c)] shows the resul\-ting edge spectrum for an external magnetic field in the $x$[$y$] direction ($B_x=0.04\,t_{\rm e}$) [($B_y=0.05\,t_{\rm e}$)]. The quasi-helical edge modes are protected by the TR symmetry $\sigma_{yz}\mathcal{T}$, except in the case for a magnetic field in the $x$ direction, where the modes get lifted. Hence, the addition of the Zeeman field modifies the symmetry properties at the touching point in such a way, so that the quasi-helical Majorana modes in (a) still are present in (c). The figures were obtained with open boundary conditions along the $y$ direction, $\Delta^{\rm e}=\Delta^{\rm h}=0.1\,t_{\rm e}$, $M^{\rm eh}_{||,\perp}=0.11\,t_{\rm e}$, $M^{\rm e, h}_{z,\perp}=0$ and $L_x=L_y=501$.}
\label{fig:Figure20}
\end{figure}

For concreteness, below we focus on a system with open boundaries along the $y$ direction, and assume that the pairing gap is nonzero in the $xz$ and $yz$ HSPs. For a magnetic field in the $x$ direction, the TR symmetry $\sigma_{xz}\mathcal{T}$ is not preserved by the termination, thus not protecting the quasi-helical modes in the HSP. Evidently the quasi-helical Majorana edge modes in Fig.~\ref{fig:Figure20}(a), become lifted by the term $B_x\sigma_x$, as seen in Fig.~\ref{fig:Figure20}(b). If we instead consider a field perpendicular to the edge, i.e., a nonzero $B_y$ or $B_z$ field, we find that the TR symmetry $\sigma_{yz}\mathcal{T}$ is preserved by the termination, thus allowing for mirror-symmetry protected quasi-helical edge modes shown in Fig.~\ref{fig:Figure20}(c) for a field in the $y$-direction. Finally, we remark that results with a $B_z$ field are not shown, since for the present model the transition to a chiral TSC appears to occur for extremely large values of the magnetic field.

{\color{black}

\section{Experimental Implementation}\label{sec:ExperimentalStuff}

{\color{black}This section discusses the most prominent ca\-te\-go\-ries of systems which exhibit the coveted coexistence of magnetic texture crystals and spin-singlet supercon\-ducti\-vi\-ty, thus being compatible with the topological sce\-na\-rios presented in the previous paragraphs. This section (i) associates the various TSCs discussed in the previous paragraphs with realistic physical systems that can support them, (ii) brings to the attention of the reader a number of already accessible 1D and 2D TSC platforms based on a MHC texture, whose properties neeed to be re\-vi\-si\-ted in the presence of additional magnetic point/space group symmetries (see concluding discussions of Sec.~\ref{sec:SectionII}), and (iii) discusses the conditions for realizing novel TSCs based on the SWC$_4$ texture in itinerant magnets.}

{\color{black}The exi\-sten\-ce of such candidate platforms is here made plausible by relying on ge\-ne\-ral symmetry arguments and results from previous theoretical studies, and does not resort to model specific self-consistent studies. It is important to note that for the qualitative discussion pursued in this section, carrying out self-consistent calculations appears non vital. In this manuscript we have mainly focused on topological phases for which the magnetic ener\-gy scale is required to exceed the pairing gap.} See for instance Fig.~\ref{fig:Figure1} and the topological criterion in Eqs.~\eqref{eq:single_Q_ani_criterion}. Hence, our interest lies in situations with a clear separation between the magnetic and pairing ener\-gy scales, which simplifies the analysis of their interplay. In the following two subsections we elaborate {\color{black}on two di\-stinct limiting situations, in which one of the two above energy scales clearly dominates the other.}

\subsection{Case of dominant magnetic energy scale}

The limit of interest in this subsection describes the scenario where the influence of superconductivity is much weaker than that of magnetism. Below, we discuss five general classes of systems that in certain circumstances can sa\-tisfy this condition and engineer a TSC.

\subsubsection{Hybrid devices with integrated nanomagnets}

The first possibility is to engineer the desired magnetic texture crystal by means of tunable magnets. Such a direction has recently attracted significant attention from both {\color{black}theo\-re\-ti\-cal~\cite{KarstenNoSOC,KlinovajaGraphene,Zutic,Marra,Abiague,FPTA} and experimental~\cite{Kontos,FrolovMagnets} points of view}. The most feasible TSCs that appear to be engineerable in this manner are 1D fully-gapped and 2D gapless TSCs generated by a MHC due to the stray fields of a nanomagnet, which is coupled to a single semiconducting nanowire or an array of semiconducting nanowires in proximity to a conventional superconductor. In these systems, one aims at simultaneously tuning the doping of the semiconductor and the perio\-dicity of the magnetic texture crystal, in order to reduce the threshold magnetic energy scale required for the system to enter the topologically nontrivial phase. 

\subsubsection{Magnet-superconductor interfaces}

Another platform where 2D TSCs with a fully-gapped bulk energy spectrum become accessible are interfaces of a magnetic insulator and a conventional superconductor~\cite{Nakosai,Mendler,MorrTripleQ,MohantaSkyrmion}. {\color{black}The class of so-called} chiral magnets, which are characterized by the absence of inversion symmetry and are dictated by the Dzyaloshinskii-Moriya interaction (DM)~\cite{DM}, were theoretically predicted to harbor SSCs more than two decades ago~\cite{Bogdanov,BogdanovHubert}. Those theoretical predictions fuelled an intense pursuit of such topological magnetic states of matter~\cite{NagaosaTokuraSkX,BogdanovPanagopoulos}, which eventually led to their successful discovery by a number of experimental groups, cf Refs.~\onlinecite{Muelhbauer,Tokura,Heinze,PanagopoulosSkX,Wulfhekel}. Such hybrid platforms have recently drawn renewed attention~\cite{Kubetzka,PanagopoulosSkyrmionVortex}, after the discovery of a triple-Q noncoplanar magnetic texture crystal in Mn/Re (0001)~\cite{TripleQWiesendanger}, and the experimental demonstration of a skyrmion-vortex coupling in such interfaces~\cite{PanagopoulosSkyrmionVortex}. Notably, it has been theoretically shown that the triple-Q magnetic order discussed in the experiment of Ref.~\onlinecite{TripleQWiesendanger} is capable of indu\-cing a 2D TSC with a nonzero Chern number~\cite{MorrTripleQ}, which supports chiral Majorana edge modes. 

To this end, we remark that SSCs ari\-sing in chiral magnets due to lo\-ca\-li\-zed moments typically appear as metastable states which are stabilized by an external magnetic field. In contrast, metallic magnets do not conform to this constraint, and allow for the spontaneous appearance of SSCs~\cite{BogdanovSkMagMet}. Even more, SSCs can also appear even when inversion symmetry is preserved and the DM interaction is absent. This becomes possible as long as the magnet is sufficiently frustrated. The latter pos\-si\-bi\-li\-ty has been shown in Heisenberg models with com\-pe\-ting spin-spin couplings of different ranges~\cite{Kawamura,Mostovoy}.

\subsubsection{Kondo lattice systems}

In the previous section, the magnets in discussion were governed by direct couplings between localized spins which are described in terms of ge\-ne\-ra\-li\-zed Heisenberg models. However, another pos\-si\-bi\-li\-ty is that the magnetic state arises due to indirect spin-spin interactions, which are mediated by conduction electrons. This indirect mechanism is a result of the Ruderman-Kittel-Kasuya-Yosida (RKKY) interac\-tion~\cite{RKKY}, and arises due to the exchange coupling between {\color{black}itinerant electrons and localized spin moments. A rich variety of magnetic phases emerge in these systems, which also inclu\-des} noncoplanar phases such as the SSC texture~\cite{MartinBatista,AzharMostovoy,HayamiEffective,HayamiRev}. Since magnetism {\color{black}in this case stems} from lo\-ca\-li\-zed moments, which are described within a classical picture, the mo\-du\-lus of the moments induced by the magnetic texture crystal is also here spatially constant. Consequently, the systems of present interest are prominent candidates for experimentally realizing TSCs in 1D and 2D which result from the MHC~\cite{Ivar} and SSC textures~\cite{WeiChen}.

\subsubsection{Topological magnetic adatom lattices}

The next general class consisting of systems known to support the coexistence of magnetic texture crystals and superconductivity concerns a conventional superconductor, on top of which, a single chain or a 2D lattice of magnetic adatoms are deposited. See for instance the pro\-po\-sals in Refs.~\cite{Choy,NadgPerge,Nakosai,Selftuned,Pientka,Ojanen,Sedlmayr,Paaske,MorrSkyrmions}. When the exchange coupling of the electrons in the supercon\-ducting substrate is sufficiently strong, low-energy ingap Yu-Shiba-Rusinov states~\cite{YSRS} appear and dictate the topological properties of the system. Such systems appear prominent to harbor fully-gapped TSCs in 1D and 2D, where the magnetic texture crystal is of the MHC or the SSC type, respectively. Note that, within the limit of classical magnetic adatoms, the magnetic moment of the adatoms are equal and fixed, thus giving rise to a magnetization profile whose modulus is spatially constant. 

The type of magnetic order governing these lattices is also here usually controlled by substrate-me\-dia\-ted RKKY spin-spin interactions. The arising magnetic ground state depends on various parameters~\cite{Selftuned,HeimesInter,Paaske,Neuhaus-Steinmetz}. Of primary importance {\color{black}here is how the adatom lattice constant compares} to the lattice constant or the Fermi wavelength of the substrate, since in the present systems these generally differ. Other factors that determine the outcome of the magnetic ground state consist of bandstructure features of the substrate superconductor~\cite{Selftuned}, the strength of the Rashba SOC in the substrate and its interplay with crystal fields~\cite{HeimesInter}, and finally long range RKKY-like spin-spin couplings~\cite{Neuhaus-Steinmetz}.

\subsubsection{Itinerant magnetic superconductors}

In the above four cases, the interplay between magnetism and superconductivity can lead to non-negligible competition-type of effects. For example, in nanowire hybrids the superconducting phase exhibited by the metallic segment which is required to engineer the proximity effect ``melts'' beyond a critical value for the magnetic energy scale. In addition, in topological YSR lattices the pairing term can become substantially suppressed in the vicinity of the adatoms~\cite{BalatskyRMP}, while the RKKY interaction is also rendered shorter-ranged when it is mediated by a conventional superconductor~\cite{Pientka}. Re\-mar\-ka\-bly, the presence of superconductivity in the substrate on top of which a magnetic chain is deposited, can be pivotal for the stabilization of MHC phases~\cite{Paaske}. 

In spite of these notable consequences of the interplay between these two phases of matter, the competition between magnetism and superconductivity is ty\-pi\-cal\-ly weak in the abovementioned systems. The reason is that magnetism and superconductivity originate from different degrees of freedom. Therefore, the above four general categories appear prominent for realizing a number of TSCs discussed in this work, and in particular those possessing a fully-gapped energy spectrum. 

The suppressed competition arising above is however not relevant when discussing itinerant magnetic superconductors, in which case both magnetism and superconductivity arise from the same degrees of freedom. A representative category of such systems, which is currently receiving significant attention from the condensed matter physics community, is the quite broad family of FeSCs. While in these systems the competition between magnetism and superconductivity is generally expected to be substantial, FeSCs have been nonetheless experimentally captured to harbor magnetic superconducting phases~\cite{ni08a,nandi10a,johrendt11,avci11,hassinger,avci14a,bohmer15a,wasser15,allred15a,allred16a,malletta,klauss15,mallettb,wang16a,zheng16a,meier17}. Even more, the here considered MHC and SWC$_4$ have been recently theoretically predicted to appear in doped BaFe$_2$As$_2$ compounds~\cite{Christensen_18}. In fact, these two magnetic texture crystal phases constitute only a part of the possible magnetic textured phases that can appear as stable global minima in an itinerant tetragonal magnet. Among these, one also finds that the SSC$_4$ phase is metastable and only accessible as a local minimum. The above results where obtained in Ref.~\onlinecite{Christensen_18} using a general Landau functional that considers the principal harmonics of magnetic texture crystals.

In the limit of a dominant magnetic energy scale assumed here, magnetic texture crystals which lead to a full gap in the energy spectrum are expected to completely suppress the emergence of superconductivity. At least this picture appears probable to hold in the usual weak coupling limit, where superconductivity originates from Cooper pairing at the FS. This prohibits the ap\-pea\-ran\-ce of TSCs in 1D, since the FS consists of points and any arising FS is due to accidental degeneracies. However, the 2D case does not present such stringent restrictions, since for non-perfectly nested FSs the stabilization of a magnetic texture crystal is expected to gap out only a fraction of the FS, thus leaving behind a reconstructed band structure consisting of Fermi pockets. This remnant FS typically occupies a smaller area than the ori\-gi\-nal FS, and therefore provides a substantially reduced phase space for spin-singlet superconductivity to appear.

\begin{figure}[t!]
\centering
\includegraphics[width=\columnwidth]{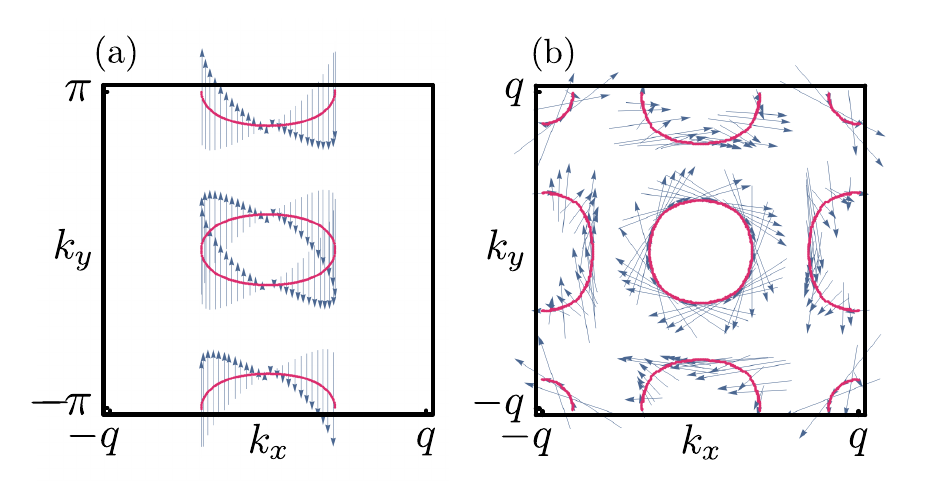}
\caption{{\color{black}(a) and (b) show the reconstructed FS for the 2D 1BM of Fig.~\ref{fig:Figure7} in the presence of a MHC and a SWC$_4$, respectively. The corresponding FSs in the absence of magnetism are shown in Figs.~\ref{fig:Figure7}(b) and~\ref{fig:Figure11}(b). The arrows depict the electronic spin polarization in the $xy$ spin plane, along the reconstructed FS. In the MHC (SWC$_4$) phase the net spin po\-la\-ri\-za\-tion is zero, thus inducing a 1D (2D) Rashba-type of SOC. Hence, the reconstructed FSs consists of Kramers partners which allow for the spontaneous appearance of spin-singlet superconductivity. The figures were obtained for $M_{||,\perp}=2.5\,t$.}}
\label{fig:Figure21}
\end{figure}

Notably, while the shape of the remnant FS may retain inversion symmetry\footnote{{\color{black}See for instance in Table~\ref{table:TableI} that the magnetic point groups of MHC and SWC$_4$ contain the $C_2$ point group element.}}, the FS points connected by inversion do not generally carry opposite spins in the magnetic phase, unless additional symmetries are present. As a matter of fact, this is the case for the MHC and SWC$_4$ phases, because these do not induce a net spin po\-la\-ri\-za\-tion\footnote{{\color{black}This is due to the space group symmetries discussed in Table~\ref{table:TableI}.}}. This is corroborated by panels (a) and (b) of Fig.~\ref{fig:Figure21}, where we depict the spin-polarization of the remnant FS for the model of Fig.~\ref{fig:Figure7}, after a reconstruction is introduced by a MHC and a SWC$_4$ texture. Indeed the presence of the two magnetic texture crystals gives rise to a spin texture in $\bm{k}$ space which is equivalent to a Rashba-type of SOC. Therefore, the emergence of Kramers partners stemming from points connected by inversion allows the spontaneous development of spin-singlet super\-con\-duc\-ti\-vi\-ty. In contrast, the possible appearance of net spin-polarization in the remnant FS implies that spin-singlet supercon\-duc\-ti\-vi\-ty cannot arise for an infinitesimally weak strength of the interaction dri\-ving the Cooper pairing, but instead it needs to reach a threshold value since the pairing su\-scep\-tibility is no longer divergent at zero temperature\footnote{{\color{black}In certain systems the emergence of net spin polarization may significantly enhance the tendency to develop equal spin pairing phases~\cite{SigristUeda}. We assume that this is not the case here.}}.

The structure of the spin-orientation along the FS is also distinct to the one that typically stabilizes finite Cooper-pair momentum phases~\cite{FF,LO}. The latter phases are usually favoured when different FSs are spin-polarized in an antiparallel manner. This is obviously not the case here. Therefore, spin-singlet superconductivity pairing up electrons of opposite wave vectors remains favorable. A natural question to ask is what is the $\bm{k}$-space structure of the spin-singlet gap $\Delta_{\bm{k}}$ which becomes preferred deep in the magnetic phase. To provide an answer, one first needs to re-classify the various pairing terms accor\-ding to the IRs of the magnetic point group. As it has been already discussed in Sec.~\ref{sec:2D_topological_superconductors}, the four 1D IRs of D$_{\rm 4h}$ are bunched together into the two groups $\{{\rm A_{1g},B_{1g}}\}$ and $\{{\rm B_{2g},A_{2g}}\}$ when the MHC sets in, while they all belong to distinct IRs when instead the SWC$_4$ emerges [cf Eqs.~\eqref{eq:GSWC4A1}-\eqref{eq:GSWC4B2}]. Therefore, the emergent $\Delta_{\bm{k}}$ is of the nematic type in the case of MHC, while it is expected to preserve the fourfold-symmetry of the energy spectrum in the case of a SWC$_4$ texture. 

To this end, an important aspect concerning the possible realization of a TSC needs to be discussed. Since the spin-singlet superconductivity emerges to gap out only the helical FS branch which was left ungapped by the magnetic texture crystal, one expects that the system will prefer to develop a highly selective $\bm{k}$-structure, which mainly becomes nonzero in the vicinity of the remnant helical FS. If such a scenario is realized, then a TSC phase does not become accessible, since magnetism and superconductivity do not coexist in the same regions of $\bm{k}$ space. Nevertheless, such a situation appears highly unlikely for systems whose pairing instabilities are driven by interactions which favor only a low number of crystal harmonics. Consequently, nonzero pairing is expected to also appear in the vicinity of the FS gapped out by the magnetic texture crystals and mediate the electron-hole conversion which is required for the charge-neutral Majorana quasiparticles to appear.

\subsection{Case of dominant pairing energy scale}

Concluding this section, we wish to briefly comment on what types of TSCs are still accessible in the antipodal limit than the one discuss above, i.e., in which the strength of the energy scale of the magnetic texture crystal is subdominant to the pairing gap. It is first of all straightforward to infer that, when magnetism and superconductivity coexist in this limit, TSC phases are not accessible when the system is fully gapped. Instead, if nodes appear in the spectrum, then one arrives at si\-tua\-tions similar to the ones discussed in App.~\ref{app:AppendixC6}. There, the nontrivial topology stems from the unconventional cha\-rac\-ter of the pairing term itself, while the influence of the added magnetic texture crystal leads to qualitatively new effects only after its strength exceeds the required thre\-shold to compensate the pairing gap at the nested points of the underlying band structure. 

Besides the above, we remark that an alternative topological scenario becomes accessible when the pairing term contains nodes and the magnetic texture crystal is such, so that it mediates the pairwise scattering of all the nodes. In the event that this process leads to a fully gapped energy spectrum, the bulk system may be classifiable as topologically trivial, but it still allows for the appearance of MZMs. This occurs by trapping the MZMs at the cores of vortex defects introduced in the magnetic texture crystal. Such a possibility is detailed in Ref.~\onlinecite{SteffensenMZM}.
}

\section{Conclusions and Outlook}\label{sec:conclusion_outlook}

We provide a systematic classification of the rich va\-rie\-ty of accessible topological phases and Majorana excitations that appear due to the bulk interplay of spin-singlet superconductivity and representative magnetic texture crystals. This work aims at inspiring new developments in a field which has recently attracted si\-gni\-fi\-cant inte\-rest from both theo\-re\-ti\-cal~\cite{SteffensenMZM,Rex,BalatskySkyrmion,KlinovajaSkyrmion,MikeSkyrmion,Mirlin,Garnier,Mesaros,HeimesInter,Cadez,Choy,KarstenNoSOC,Ivar,KlinovajaGraphene,NadgPerge,Nakosai,KotetesClassi,Selftuned,Pientka,Ojanen,Sedlmayr,WeiChen,XiaoAn,Paaske,Zutic,Marra,Abiague,MorrSkyrmions,MorrTripleQ,FPTA,MohantaSkyrmion,Christensen_18,Mendler} and experimental~\cite{Gerbold,Wiesendanger,Cren,Kontos,Wiebe,Wiesendanger2,Kubetzka,PanagopoulosSkyrmionVortex} sides. Our work is novel in many ways, as it accounts for all possible strong, weak, and crystallline phases ari\-sing in to\-po\-lo\-gi\-cal superconductors (TSCs) induced by {\color{black}a set of particularly relevant magnetic texture crystals, and considers one/two-band systems} which harbor conventional or uncon\-ven\-tio\-nal spin-singlet pai\-ring. As we uncovered here, the concepts of the magnetic and pai\-ring groups play a crucial role in the symmetry classification of these systems, since their interplay controls the topological bulk and boun\-da\-ry properties. {\color{black}Our entire discussion unfolds by further assigning and calculating suitable topological invariants that arise from general classification schemes~\cite{Altland,SchnyderClassi,KitaevClassi,Ryu,TeoKane}. Notably, we show how a number of these abstract invariants emerge in the present context, explicitly provide their construction, and finally clarify their phy\-si\-cal meaning by linking their presence to the quantization imposed on a number of physical quantities, such as the staggered magnetization.}

Our investigation first focuses on 1D systems. This allows bridging our work with previous known results~\cite{BrauneckerSOC,KarstenNoSOC,Ivar,KlinovajaGraphene,NadgPerge,KotetesClassi,Selftuned,Pientka} but also report a {\color{black}long list of} new phe\-no\-me\-na. Even more, it sets the stage for the for\-ma\-lism that we employ in 2D, which relies on a sublattice description, as well as on downfolding to the magnetic Brillouin zone (MBZ). While {\color{black}a rigorous topological classification is extracted by investigating the symmetry of properties of general Hamiltonians within the} sublattice picture, the latter approach exposes transparently the key me\-cha\-nisms which drive the nontrivial topology. In fact, the MBZ description is also computationally advantageous when studying the topological properties in the low-energy sector, since a few number of bands are required for this. 

By following the above approaches, we find a number of new interesting results in 1D. First of all, we construct new crystalline topological invariants which reflect the quantization of the staggered magnetic moment in such systems. In addition, our analysis includes the study of unconventional pairing gaps and discusses how multiple Majorana zero modes appear on a given edge. Ano\-ther important component of this study is the consi\-de\-ra\-tion of two-band models (2BMs). Remarkably, the multiband structure of the magnetization allows interpolating between different symmetry classes, i.e., BDI, AIII and DIII. The former appears when both interband and intraband magnetic scatterings are present. The se\-cond becomes relevant for interband-only scattering, in which case the Majorana edge excitations come in pairs. However, these do not obey a charge-conjugation symmetry and thus each pair should be viewed as a single Andreev zero mode. On the other hand, true Majorana Kramers pairs appear when interband-only scatte\-ring is present and additional TR-symmetry preserving intraband terms are included, e.g., inversion-symmetry-breaking spin-orbit coupling terms. See also  Ref.~\onlinecite{SteffensenMZM}. 

The emergence of Andreev edge modes in topologically nontrivial systems has recently attracted substantial attention~\cite{KlinovajaFF,SticletFF,NagaosaFF,MarraFF,FPTA}. Noteworthy, here we obtain topologically-protected Andreev modes (cf Ref.~\onlinecite{MarraFF}) which are pinned to zero energy in an extended window in parameter space. As a result, these topologically-protected zero modes open perspectives for new quantum computing platforms, since they can constitute the hardware of long-lived Andreev qubits with enhanced protection against decoherence~\cite{Zazunov,HigginbothamParity}. Even more, engineering systems harboring topologically-protected AZMs, opens a new direction in synthesizing topological Andreev bandstructures in synthetic space~\cite{vanHeck14,Yokoyama15,Strambini,Riwar16,Eriksson17,Meyer17,Xie17,MeyerHouzet,LevchenkoTop,Balseiro,PK_PRL,MTM_PRB,Sakurai,Houzet,BelzigABS}. Indeed, such a pursuit in TSCs has so far been una\-voi\-da\-bly restricted to exploiting MZMs in multi-terminal devices~\cite{MeyerHouzet,LevchenkoTop,Balseiro,PK_PRL,MTM_PRB,Sakurai,Houzet}, and theoretical works predict that it gives access to the observation of Weyl points, chiral anomaly, and a number of quantized Josephson transport phenomena. Being in a position to obtain AZMs in multi-terminal platforms such as the one experimentally investigated recently in Ref.~\onlinecite{Pankratova}, provides an alternative and in some cases more robust and less demanding route for such types of synthetic topology.

The 2D investigation brings forward an equally rich list of results. By considering once again one- and two-band models, generally-unconventional multiband pai\-ring, as well as various multiband implementations of representative textures, we uncover an intricate set of flat-band, uni/bi-directional, quasi-helical, helical, and chiral Majorana edge modes, that may be protected by strong, weak, or crystalline topology. Specifically, for a magnetic helix crystal (MHC) we obtain Majorana and Andreev flat bands, which can be viewed as direct extensions of the 1D phenomena. However, new physics appears here due to the interplay of the magnetic and pairing point groups, thus revealing a dichotomy when it comes to the to\-po\-lo\-gi\-cal classification in high-symmetry planes (HSPs). Depending on whether the irreducible representation (IR) of the pairing term belongs to either the $\{{\rm A_{1g},B_{1g}}\}$ or $\{{\rm B_{2g},A_{2g}}\}$ set, we find a different topological classification and, in turn, Majorana edge mode dispersions. 

Considering instead a spin whirl crystal (SWC$_4$) magnetic texture which eliminates the possibility of dispersionless edge modes, since the coexistence of two magnetic helix textures winding in different directions and spin planes, lifts almost all the degeneracies in the MBZ. Strikingly however, mirror and space-group symmetries consistent with the structure of the texture still impose a number of degeneracies in HSPs. These enable the conversion of the flat band edge modes into uni/bi-directional ones. The bidirectional modes constitute dispersive Majorana edge modes with neither a fixed chirality nor helicity, whose spin character depends on the conserved wave number. Notably, besides a few exceptions~\cite{Shiozaki2016May}, this type of excitations has been poorly discussed so far in the literature. 

We find that the emerging topological phases in 2D are mostly of the crystalline or weak type, and become manifest through the appearance of the here-called quasi-helical Majorana edge modes, which present a certain number of similarities with the standard helical edge modes in 2D. Re\-mar\-ka\-bly, strong phases do not become accessible in 2D, because the crystalline symmetries present trivialize the respective strong topological invariants. Specifically, the pre\-sen\-ce of space-group symmetries imposes de\-ge\-ne\-ra\-cies at inversion-symmetric points (ISPs) which, in con\-junction with the fourfold rotational symmetry (C$_4$) present, imposes a nodal bulk spectrum in the one-band models (1BMs), and do not allow for a strong $\mathbb{Z}_2$ inva\-riant in 2BMs despite the fact that a fully-gapped spectrum is accessible there. As we demonstrate, one possible route to unlock genuinely-2D to\-po\-lo\-gi\-cal phases, is by inclu\-ding terms which violate C$_4$ symmetry, while respecting the degeneracies imposed by the magnetic space group. Notably, the violation of C$_4$ symmetry can be either externally imposed via strain engineering, or spontaneously appear in systems with nematic correlations, cf Ref.~\onlinecite{nematic}. 

Another possibility is to consider the additional pre\-sen\-ce of a Zeeman/exchange field, which lifts the degeneracies at ISPs, but still retains a number of crystalline symmetries. As a result, the arising bidirectional modes can be converted into unidirectional depending on the orientation of the field, Majorana chiral edge modes become accessible in 1BMs, and mirror-symmetry-protected quasi-helical edge modes may persist or get gapped out. Despite the fact that space-group symmetries appear to be detrimental for the appearance of genuinely-2D topological phases, we remark that they still constitute a unique pathway to obtain multiply-degenerate Majorana excitations, such as, hourglass Majorana edge modes~\cite{Shiozaki2016May}. While such a possibility did not occur for the models examined, it still constitutes an interesting direction of research. Lastly, we remind that the presence of space-group symmetries are absent for magnetic textures with incommensurate magnetic orde\-ring vectors~\cite{Christensen_18}, but may still be appro\-xi\-ma\-te\-ly preserved in itinerant magnets for low energies.  

At this point, we wish to discuss in more detail prominent candidate physical systems that can host the above-mentioned phenomena. Our framework addresses a single Kramers doublet of the double covering D$_{\rm 4h}$ point group, therefore allowing to describe tetragonal magnets. These systems may for instance correspond to correlated magnetic superconductors, where magnetism and superconductivity coexist microscopically. The desirable scenario is the one where a magnetic texture appears to partially gap out a well-nested Fermi surface~\cite{Christensen_18}, lea\-ving behind re\-con\-struc\-ted pockets, which can be subsequently gapped out by the emergence of supercon\-duc\-ti\-vi\-ty. Si\-mi\-lar to Ref.~\onlinecite{Selftuned}, in this situation, one expects that the resulting magnetic superconductor selftunes into one of the topological phases discussed here. 

Among the possible systems that promise to provide a fertile ground to materialize such a possibility, the fa\-mi\-ly of doped Fe-based superconductors (FeSCs) stands out. Some FeSCs are well known to exhibit a coexistence of magnetism and superconductivity~\cite{ni08a,nandi10a,johrendt11,avci11,hassinger,avci14a,bohmer15a,wasser15,allred15a,allred16a,malletta,klauss15,mallettb,wang16a,zheng16a,meier17}. Re\-fe\-ren\-ce~\onlinecite{Christensen_18} has identified all the possible single- and double-$\bm{Q}$ magnetic ground states that can appear in representative five-orbital mo\-dels of FeSCs, and demonstrated that doping generally enables various magnetic textures, some of which we explore here. The possible subsequent emergence of conventional or unconventional spin-singlet super\-con\-duc\-ti\-vi\-ty can give rise to a number of the topological scenarios discussed here. Moreover, accounting for a weak band dispersion in the third spatial dimension, which may be non-negligible in certain compounds, opens additional perspectives for realizing systems with topologically-protected Weyl and Dirac points~\cite{Burkov}, as well as nodal lines, rings and chains~\cite{NodalLineRev}, thus leading to Majorana/Andreev arc and drumhead surface modes.  

Other physical systems which our results may be applicable to, include hybrid systems~\cite{Wiesendanger,Kontos,Wiebe,KarstenNoSOC,NadgPerge,Nakosai,KotetesClassi,Selftuned,Pientka,Ojanen,Sedlmayr,WeiChen,XiaoAn,Paaske,Zutic,Marra,Abiague,FPTA} such as superconductor-semiconductor nanowire hybrids and topological magnetic lattices. In the former class of systems, it is desirable to impose on the system the desired magnetic texture by external means, i.e., u\-sing nanomagnets~\cite{Kontos,FrolovMagnets}. In this case, the magnetic wave vectors should be tailored to be comparable to the Fermi wave vectors of the underlying hybrid system, which in turn can be controlled by gating the device. On the other hand, the wave vectors describing the magnetic texture appearing in topological chains depends on up to which degree is the electronic spectral weight carried by the superconducting electrons of the substrate~\cite{Brydon,Li,HeimesInter}. In the deep so-called Yu-Shiba-Rusinov limit, the magnetic adatoms can be treated classically, and the modulation of the magnetic texture is determined by a number of factors. These include the spacing of the magnetic adatoms, the size of their moment, the strength of their coupling to the substrate electrons, the possible presence of ISB in the substrate and/or crystal field effects. {\color{black}See Refs.~\onlinecite{HeimesInter,Neuhaus-Steinmetz} for investigations concerning topological magnetic chains.}

We continue with enumerating a number of possible experimental methods to infer the various to\-po\-lo\-gi\-cal phases discus\-sed here, and detect the arising Majorana/Andreev modes. As mentioned already, spin- and angle-resolved photoemission spectroscopy~\cite{Hsieh} can provide information regarding protected degeneracies. Spectroscopic methods are also standard routes to detect Majorana/Andreev excitations~\cite{PALee,JaySpec,Flensberg,JJHe,KotetesSTM}. Here we are particularly interested in spin-resolved scanning tun\-ne\-ling spectroscopy~\cite{JJHe,KotetesSTM} which can probe the spin character of the boundary excitations~\cite{Bena,IsingSpin}. The various MF edge modes lead to a characteristic electronic edge spin polarization. For a TSC induced by a MHC the presence of chiral symmetry confines the electronic spin polarization within a given spin plane, similar to what is encountered in superconductor-semiconductor nanowires~\cite{Bena,JJHe,KotetesSTM}, magnetic chains~\cite{HeimesInter,HeimesSynth}, as well as charged~\cite{MTM_PRB} and neutral~\cite{IsingSpin} p-wave superfluids. TSCs engineered from the SWC$_4$ texture exhibit a wider range of possibilities. As we show in Fig.~\ref{fig:Figure13}, the type of termination is decisive for the spin character of the bidirectional MF modes which possess neither fixed he\-li\-ci\-ty nor chirality. In contrast, unidirectional modes tend to exhibit a higher degree of spin polarization. Chiral and (quasi-)helical MFs have instead a fixed spin character, since they stem from fully-gapped TSCs. Even more, the various dispersionless or dispersive Majorana and Andreev edge modes can be probed in suitably-designed generally-spin-resolved charge and thermal transport experiments~\cite{WangResponse,RyuResponse,FurusakiResponse,BulmashResponse,SigristResponse}. Depen\-ding on whether we have electrically neutral (Majorana) or charged (Andreev) edge excitations one can correspondingly look for cha\-ra\-cte\-ri\-stic features and scaling be\-ha\-viors in thermal and Hall responses~\cite{SigristResponse}.

Finally, we conclude by pointing out that magnetic texture crystals can generally harbor 0D topological defects, such as vortices, {\color{black}which can be further employed to} trap Majorana zero modes. This was brought to light only recently in Ref.~\onlinecite{SteffensenMZM}, where it was also shown that such a mechanism takes place only when the pairing term of the coexisting superconducting order contains nodes in its bulk energy spectrum. By considering also this possibility, spin-singlet superconductors harboring magnetic textures appear to be unique versatile platforms where a multitude of TSC phases can be observed and harnessed for cutting edge applications.

\section*{Acknowledgements}

D.~S. and B.~M.~A. acknowledge support from the Carlsberg Foundation. B.~M.~A. additionally acknow\-ledges support from the Independent Research Fund Denmark grant number DFF-6108-00096. M.~H.~C. acknow\-ledges financial support from U.S. Department of Energy, Office of Scien\-ce, Basic Energy Scien\-ces, Materials Scien\-ce and Engineering Division, under Award No. DE-SC0020045.

\appendix

\section{Sublattice Formulation in 1D}\label{app:AppendixA}

\begin{figure}[t!]
\centering
\includegraphics[width=\columnwidth]{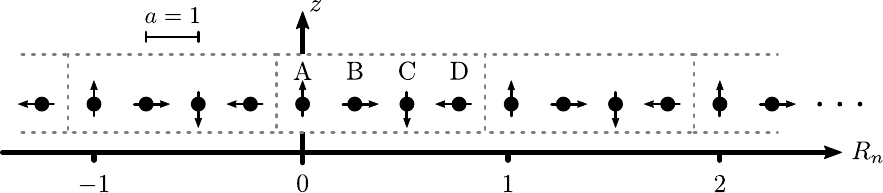}
\caption{Unit cells for a 1D system experiencing a magnetic texture with a 4-site periodicity. The four sites of a unit cell are labelled as \{A,\,B,\,C,\,D\} and the center of inversion is here chosen to be the site A of the $n=0$ unit cell.} 
\label{fig:appFigure1}
\end{figure}

In this appendix, we reformulate our analysis in terms of a four sublattice basis, which is more transparent in regards to the topological classification, since it leads to properly compactified $2\pi$-periodic Hamiltonians. For illustrative purposes, we restrict to the 1D case, since the 2D description is obtainable in a straightforward manner. Within this framework a unit cell consists of 4 sites labelled as \{A,\,B,\,C,\,D\}, cf Fig.~\ref{fig:appFigure1}. We define the spinor:
\bea
\bar{\bm{\psi}}_n^{\dag}=\left(\bm{\psi}_{{\rm A},n}^{\dag},\,\bm{\psi}_{{\rm B},n}^{\dag},\,\bm{\psi}_{{\rm C},n}^{\dag},\,\bm{\psi}_{{\rm D},n}^{\dag}\right)\,,\label{eq:SublatticeSpinor}
\eea

\noi where $n$ now labels a 4-site unit cell. In this basis, a translation $\{\mathds{1}\,|\,a\}$, effects the shift $n\mapsto n+\frac{1}{4}$. Hence, $\{\mathds{1}\,|\,a\}$ and $\{\mathds{1}\,|\,2a\}\equiv \{\mathds{1}\,|\,\nicefrac{\pi}{Q}\}$ [for $Q=\pi/(2a)$] read in wave-number space $k_x\in(-\pi/4,\pi/4]$ (with $a=1$):
\begin{align}
\{\mathds{1}\,|\,a\}=\left(\begin{array}{cccc}
0&0&0&\beta^*\\1&0&0&0\\0&1&0&0\\0&0&1&0
\end{array}\right),
\phd
\{\mathds{1}\,|\,\nicefrac{\pi}{Q}\}=\left(\begin{array}{cccc}
0&0&\beta^*&0\\0&0&0&\beta^*\\1&0&0&0\\0&1&0&0
\end{array}\right),
\end{align}

\noi where we set $\beta=-e^{i4k_x}$ in order for the MBZs of the wave-number shifts and sublattice descriptions to match. Next we identify the action of inversion ${\cal I}$ about the center of inversion which is here set to be the A site of the $n=0$ unit cell. For a Hamiltonian element $\hat{H}_{k_x}$ defined in the respective wave-number spinor of Eq.~\eqref{eq:SublatticeSpinor}, inversion acts as: ${\cal I}\hat{H}_{k_x}=\big[\hat{{\cal I}}^{\dag}\hat{H}_{k_x}\hat{{\cal I}}\big]_{k_x\mapsto-k_x}$ with:
\bea
\hat{{\cal I}}=\left(\begin{array}{cccc}1&0&0&0\\0&0&0&\beta\\0&0&\beta&0\\0&\beta&0&0\end{array}\right)\,.
\eea

\noi We note that the inversion of $k_x$ takes place only after the matrix multiplications. This is because the matrix representation of inversion is $k_x$-dependent in this basis. 
\noi The kinetic energy operator de\-scri\-bing first ($t$), second ($t'$), and third ($t''$) order neighbour hopping is represented as:
\bea
\hat{H}_{{\rm kin},k_x}&=&
-t\left(\begin{array}{cccc}0&1&0&\beta^*\\1&0&1&0\\0&1&0&1\\\beta&0&1&0\end{array}\right)
-t''\left(\begin{array}{cccc}0&\beta^*&0&1\\\beta&0&\beta^*&0\\0&\beta&0&\beta^*\\1&0&\beta&0\end{array}\right)\no\\
&&-t'\left(\begin{array}{cccc}0&0&1+\beta^*&0\\0&0&0&1+\beta^*\\1+\beta&0&0&0\\0&1+\beta&0&0\end{array}\right)\,.
\eea

\noi One verifies that the above kinetic part of the Hamiltonian is invariant under translations and inversion. The Hamiltonian for the MHC texture in Eq.~\eqref{eq:MH}, here reads:
\bea
\hat{H}_{\rm mag}=
\left(\begin{array}{cccc}M_\perp\sigma_z&0&0&0\\0&M_{||}\sigma_x&0&0\\0&0&-M_\perp\sigma_z&0\\0&0&0&-M_{||}\sigma_x\end{array}\right)\,.\quad
\eea

\begin{figure*}[t!]
\centering
\includegraphics[width=\textwidth]{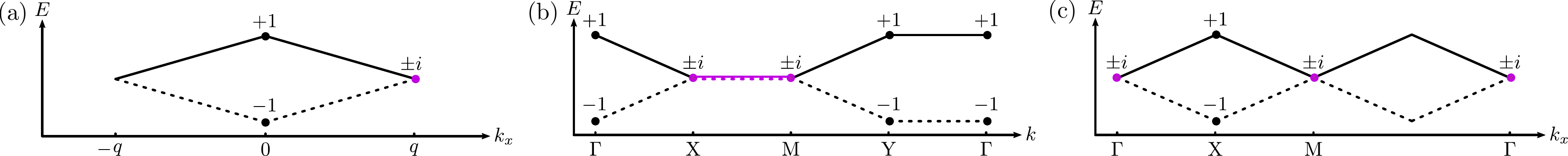}
\caption{Sketches of the entire bulk band structure, or cuts of it, in the case of a MHC in 1D and 2D [panels (a) and (b)], and the SWC$_4$ magnetic phase [(c)]. The labels show the eigenvalues of the space group symmetry which commutes with the Hamiltonian in the given HSP. Degeneracies are colored in purple.}
\label{fig:appFigure2}
\end{figure*}

\begin{figure}[b!]
\centering
\includegraphics[width=\columnwidth]{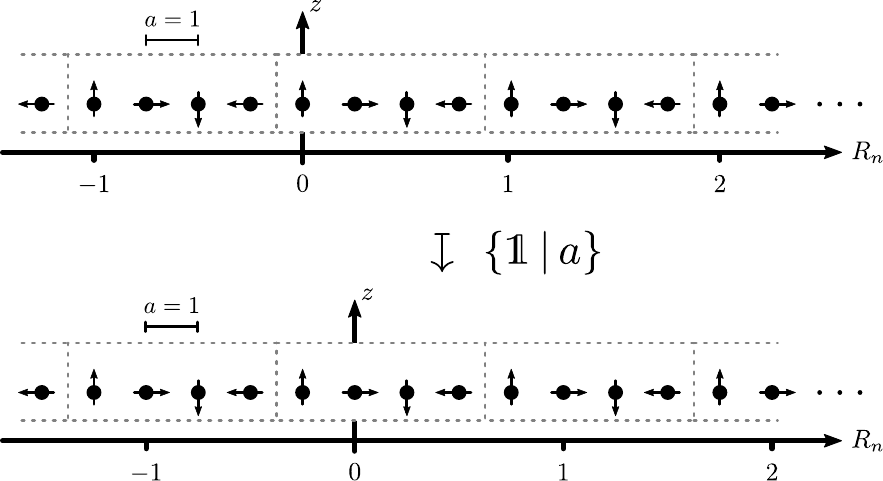}
\caption{Sketch of the magnetic unit cell under the action of $\{\mathds{1}\,|\,a\}$. This translation renders $\{\sigma_{yz}\,|\,\nicefrac{\pi}{Q}\}$ symmorphic, while at the same modifying $\mathcal{R}_{yz}$. See Eq.~\eqref{eq:off_centered_pair}. $\{\sigma_{yz}\,|\,\nicefrac{\pi}{Q}\}$ and $\mathcal{R}_{yz}$ define a pair of offcentered symmetries.}
\label{fig:appFigure3}
\end{figure}

Main target of this appendix is to shed light to the topological classification at ISPs/HSPs. At $k_x=0$, we find that ${\cal I}$ and $\{\mathds{1}\,|\,\nicefrac{\pi}{Q}\}$ possess the twofold degenerate eigenvalues $\pm1$ and $\pm i$, respectively. Instead, at $k_x=\pi/4$, we find that ${\cal I}$ possesses the eigen\-va\-lues $\{1,1,1,-1\}$, while $\{\mathds{1}\,|\,\nicefrac{\pi}{Q}\}$ possesses the twofold de\-ge\-ne\-ra\-te eigenvalues $\pm1$. The emergence of different eigenva\-lues for these symmetry operations at the two ISPs implies that the respective symmetry classes ge\-ne\-ral\-ly differ. In contrast, these symmetry operators are $k_x$-independent in the wave-number-transfer description, thus implying that the various ISPs and HSPs are dictated by the same symmetry and topological properties. The apparent discrepancy is attributed to the fact that the Hamiltonian in the wave-number-transfer description is not compactified. Therefore, caution is needed when performing the classification using this formalism. In fact, we find that the topological classifications coincide for the the ISPs and HSPs where a magnetic gap opens. In contrast, at ISPs/HSPs where a Kramers degeneracy appears and the magnetization becomes ineffective, only the sublattice-based topological classification is valid.

\section{Space Group Symmetry-Protected Degeneracies}\label{app:AppendixB}

Nonsymmorphic space group operations take the form $\{g\,|\mathbf{t}\}$, where $g$ defines a point group operation and $\mathbf{t}$ is a translation by a fraction of a Bravais lattice vector. A space group symmetry is referred to as nonsymmorphic, when no coordinate system can be chosen to remove the translation $\mathbf{t}$ in $\{g\,|\,\mathbf{t}\}$~\cite{Bible,Dresselhaus,Young2015Sep,Shiozaki2016May,Zhao2016Nov}. This is satisfied when $g\mathbf{t}=\mathbf{t}$, i.e., when the involved translation is along a HSP of $g$. If this is not the case, the component of $\mathbf{t}$ which is perpendicular to the HSP of $g$ is obsolete and can be removed. However, such a removal may result in the redefinition of other symmorphic symmetry elements, which in the new basis may involve a translation. The elements that become simultaneously modified in such a process define the so-called set of ``offcentered symmetry elements''. In the main text, we encounter pairs of such offcentered symmetries. As explained below, their pre\-sen\-ce introduces protected de\-ge\-ne\-ra\-cies in the spectrum. See also Refs.~\onlinecite{Fang2015Aug,Yang2017Feb,Zhang2018Jul,Malard2018Oct}, and Figs.~\ref{fig:appFigure2} and~\ref{fig:appFigure3}.

The fact that a genuine nonsymmorphic symmetry requires that the equivalence $g{\bf t}={\bf t}$ should be met, further restricts the systems in which nonsymmorphic symmetries can provide topological protection to boundary modes and thus stabilize crystalline topological phases. Since such a boundary is required to preserve both $g$ and $\bf t$, only 3D systems can exhibit topological crystalline phases induced by nonsymmorphic symmetries. Indeed, edges of 2D systems generally fail to fulfill these criterion, and the presence of nonsymmorphic symmetries can only affect the bulk topological properties of the system. Consider, for example, a 2D system with the nonsymmorphic symmetry $\{\sigma_{xz}\,|\,(1,0)\}$. For this particular case, the edge $(01)$ is invariant under $\sigma_{xz}$ while $(10)$ is preserving $\{\mathds{1}\,|\,(1, 0)\}$, i.e., we can never find an edge which is invariant under the symmetry operation $\{\sigma_{xz}\,|\,(1,0)\}$. Extending the example to 3D systems, we immediately observe that the surface (001) preserves both $\sigma_{xz}$ and $\{\mathds{1}\,|\,(1,0)\}$, and can thus potentially exhibit topological surface states protected by the nonsymmorphic symmetry. Hence we conclude that a nonsymmorphic symmetry cannot induce a crystalline topological phase in 2D systems, except in rare cases where $g\bm{k}=\bm{k}$~\cite{Shiozaki2016May}. 

For the MHC texture in 1D, our system is invariant under a set of symmetries shown in Table~\ref{table:TableI}. Out of these, we find that the symmetry element $\{\sigma_{yz}\,|\,\nicefrac{\pi}{Q}\}$ is rendered symmorphic after translating the magnetic unit cell by $\{\mathds{1}\,|\,a\}$, as shown in Fig.~\ref{fig:appFigure3}. At the same time, the point group element $\mathcal{R}_{yz}=\sigma_{yz}\mathcal{T}\Theta$ is redefined and in the new basis involves a translation. Specifically, the two symmetry elements become redefined as follows:
\bea
\{\sigma_{yz}\,|\,\nicefrac{\pi}{Q}\}\mapsto\sigma_{yz}\,\quad{\rm and}\,\quad{\cal R}_{yz}\mapsto \{{\cal R}_{yz}\,|\,\nicefrac{\pi}{Q}\}\,.\quad
\label{eq:off_centered_pair}
\eea

In fact, it is not possible to choose a coordinate system for which both $\{\sigma_{yz}\,|\,\nicefrac{\pi}{Q}\}$ and ${\cal R}_{yz}$ become regular point group elements. This leads to symmetry-protected degeneracies in the spectrum. To exemplify this, we rely on the relation: $\{\sigma_{yz}\,|\,\nicefrac{\pi}{Q}\}\,{\cal R}_{yz}={\cal R}_{yz}\,\{\sigma_{yz}\,|\,\nicefrac{\pi}{Q}\}e^{ik_x\pi/q}$. By further taking into account that $\Theta^2=+\mathds{1}$, which holds in the here-relevant BDI symmetry class, we obtain $\{\sigma_{yz}\,|\,\nicefrac{\pi}{Q}\} ^ 2=+\mathds{1}e^{ik_{\mathcal{I}}\pi/q}$ for $k_{\mathcal{I}}=0,q$. This leads to the two eigenvalue equations:
\bea
\{\sigma_{yz}\,|\,\nicefrac{\pi}{Q}\}|k_{\mathcal{I}},\pm\rangle&=&\pm e^{ik_{\mathcal{I}}\pi/Q}|k_{\mathcal{I}},\pm\rangle\,,\\
\{\sigma_{yz}\,|\,\nicefrac{\pi}{Q}\}{\cal R}_{yz}|k_{\mathcal{I}},\pm\rangle&=&\pm e^{3ik_{\mathcal{I}}\pi/Q}{\cal R}_{yz}|k_{\mathcal{I}},\pm\rangle\,.
\eea

\noi We thus observe that the two pairs $|k_{\mathcal{I}},\pm\rangle$ and ${\cal R}_{yz}|k_{\mathcal{I}},\pm\rangle$ have the same (opposite) eigenvalues at $k_{\mathcal{I}}=0$ ($k_{\mathcal{I}}=q$). The two states are therefore mutually ``pa\-ral\-lel'' (orthogonal) at $k_{\mathcal{I}}=0$ ($k_{\mathcal{I}}=q$), ultimately leading to a protected degeneracy at $k_x=q$, see Fig.~\ref{fig:appFigure2}(a).

The above degeneracies appear at isolated points of the MBZ. This behavior has to be compared with the consequences of the genuinely nonsymmorphic symmetry $\{\sigma_{xz}\,|\,\nicefrac{\pi}{Q}\}\,k_x=k_x$. The latter can be employed to label the eigenstates of the Hamiltonian $\forall k_x$ in the 1D MBZ. Since the system is effectively spinless, $\Theta ^ 2 = +\mathds{1}$, we find:
\bea
\{\sigma_{xz}\,|\,\nicefrac{\pi}{Q}\}\,|k_x,\pm\rangle=\pm e^{ik_x\pi/Q}|k_x,\pm\rangle.
\eea

\noi From the above eigenvalues, in combination with $\Theta$, we know that the spectrum in the MBZ must follow the sketch in Fig.~\ref{fig:appFigure2}(a), which is compatible with the degeneracies imposed by the pair of offcentered symmetries. The twofold degeneracy at $k_x=q$ is enforced because the eigenvalues $\pm i$ are connected by the antiunitary symmetry $\Theta$. Notably, this degeneracy can be alternatively seen as the result of the emergent TR symmetry effected by $\tilde{\Theta}=\{\sigma_{xz}\,|\,\nicefrac{\pi}{Q}\}\Theta\equiv\{\mathcal{T}\,|\,\pi/Q\}$, with $\tilde{\Theta}^2=\mathds{1}e^{ik_x\pi/q}$.

Similar arguments for the MHC in 2D establish once again that the symmetries $\{\sigma_{yz}\,|\,\nicefrac{\pi}{Q}\}$ and ${\cal R}_{yz}$ constitute a pair of offcentered ones, and impose a twofold de\-ge\-ne\-ra\-cy at $k_x=q$ for all $k_y\in[0,2\pi)$. This gives insight about the key features of the generic band structure which is depicted in Fig.~\ref{fig:appFigure2}(b), and reveals that the pairing gap cannot be compensated by the magnetic gap in this HSP.

We conclude this section with a brief comment on the generic characteristics of the band structure for a 2D system under the influence of a SWC$_4$ texture. In this case, we do not find any offcentered symmetries. Nonetheless, twofold degeneracies still appear as the result of the pre\-sen\-ce of non\-sym\-morphic symmetries. See Fig.~\ref{fig:appFigure2}(c) for a sketch of the general dispersion in the MBZ.

\section{Details on Topological Invariants}\label{app:AppendixC}

\subsection{Winding Number in 1D 1BMs for $\bm{\Delta_{k_x}=\Delta}$}\label{app:AppendixC1}

The calculation is facilitated by noticing the presence of the anti\-uni\-ta\-ry TR symmetry $\tilde{\Theta}=\rho_2\sigma_y{\cal K}$ of the Hamiltonian in Eq.~\eqref{eq:bloch_1BM_ani_helix}. Being $k_x$-independent, $\tilde{\Theta}$ influences the to\-po\-lo\-gi\-cal clas\-sification in the entire MBZ. Note that such a $k_x$-independent symmetry does not appear in the sublattice formulation of the problem. The product involving $\tilde{\Theta}$ and the pre\-existing $\Theta={\cal K}$ symmetry, induces the unitary symmetry $\tilde{{\cal O}}=\tilde{\Theta}\Theta=\rho_2\sigma_y$. In par\-ti\-cu\-lar, this allows us to diagonalize the BdG Hamiltonian into blocks labelled by the eigenvalues of $\tilde{\cal O}$. By per\-for\-ming the unitary transformation induced by the operator $\tilde{{\cal S}}=(\tilde{\cal O}+\sigma_z)/\sqrt{2}$, we obtain the blocks:
\begin{align}
\hat{{\cal H}}_{k_x,\sigma}'=\left[h_{k_x}^{(0)}+h_{k_x}^{(1)}\rho_2+h_{k_x}^{(2)}\eta_3+h_{k_x}^{(3)}\eta_3\rho_2\right]\tau_3-M_\sigma\rho_1\no\\
+M_{-\sigma}\eta_1\rho_1+\left[\Delta_{k_x}^{(0)}+\Delta_{k_x}^{(1)}\rho_2+\Delta_{k_x}^{(2)}\eta_3+\Delta_{k_x}^{(3)}\eta_3\rho_2\right]\tau_1
\label{eq:block_x}
\end{align}

\noi where $M_\sigma=(M_{||}+\sigma M_\perp)/2$, with $\sigma=\pm1$ labeling the eigenvalues of $\sigma_z$ in the new frame. Both blocks reside in BDI class with $\Theta={\cal K}$, $\Xi=\rho_2\tau_2{\cal K}$ and $\Pi=\rho_2\tau_2$. Consequently, the presence of the unitary symmetry effected the symmetry class transition BDI$\rightarrow$BDI$\oplus$BDI, which allows defining a winding number $w_{\sigma}$ for each block. 

One observes that each block leads to a fractional win\-ding number $\pm1/2$. As discussed pre\-viously in Ref.~\onlinecite{HeimesInter}, this peculiarity is due to the choice of the spinor, which, while being con\-ve\-nient, does not gua\-ran\-tee that the Hamiltonian blocks satisfy the compactification criteria required to define the $\mathbb{Z}$ index. As a result, the block winding numbers cannot define two independent to\-po\-lo\-gi\-cal invariants, but they have to be added or subtracted to provide the proper inva\-riant. The correct way to combine them can be inferred based on a well-known li\-mi\-ting case, or, by adding infinitesimal terms which vio\-la\-te the unitary symmetry but preserve $\Theta$, $\Xi$ and $\Pi$. Nevertheless, here it is straightforward to infer how to combine the block invariants, by inve\-sti\-ga\-ting their behavior in the already established result for $M_\perp=M_{||}=M>0$. In this known case, the block win\-ding numbers become:
\bea
w_+^{M_{||,\perp}=M}&=&-\frac{{\rm sgn}\Big\{\big(M^2-\Delta^2\big)\big[1+(\Delta^2-M^2)/(2\mu)^2\big]\Big\}}{2},\no\\
w_-^{M_{||,\perp}=M}&=&\nicefrac{1}{2}\,.
\eea

\noi In order to retrieve the topological invariant of Eq.~\eqref{eq:invariant_k_y_0}, we verify that the winding number should be defined as:
\bea
w=w_--w_+\,.\label{eq:winding_number_1BM_total}
\eea

\subsection{Winding Number in 1D 1BMs for a generic $\bm{\Delta_{k_x}}$}\label{app:AppendixC2}

Here we obtain an expression for the winding number in the case of a generic $\Delta_{k_x}$. To facilitate the derivation of an analytical result, we restrict to the weak coupling limit. We block diagonalize the low-energy BdG Hamiltonian in Eq.~\eqref{eq:Low_Energy_1BM} and find:
\bea
\hat{{\cal H}}_{k_x,\sigma}^{\rm low-en}&=&\left(\xi_{k_x;q}^++\xi_{k_x;q}^-\rho_2\right)\tau_3-M_\sigma\rho_1\no\\
&+&\left(\Delta_{k_x;q}^++\Delta_{k_x;q}^-\rho_2\right)\tau_1\,,
\label{eq:Low_Energy_1BM_Blocks}
\eea

\noi and define
\bea
&&\det\big(\hat{A}_{k_x,\sigma}^{\rm low-en}\big)
=\big(\xi_{k_x;q}^+\big)^2+\big(\Delta_{k_x;q}^+\big)^2-\big(\xi_{k_x;q}^-\big)^2\no\\
&&-\big(\Delta_{k_x;q}^-\big)^2-M_\sigma^2+2i\big(\xi_{k_x;q}^-\Delta_{k_x;q}^+-\xi_{k_x;q}^+\Delta_{k_x;q}^-\big),\quad
\eea

\noi which is of the exact same form as Eq.~(\ref{eq:det_A_1BM}), with the crucial difference that here $k_x\in{\rm MBZ}$, which implies that the contribution of the last term in the topological invariant of Eq.~\eqref{eq:winding_number_1BM_ani} drops out. In addition, when $k_x=0$ constitutes the only wave number where a gap closing takes place, one directly retrieves the result of Eq.~\eqref{eq:winding_number_1BM_Low_Energy} after setting $\Delta_{k_x}=\Delta$.

\subsection{Mirror Invariant in 1D 1BMs}\label{app:AppendixC3}

To evaluate the mirror invariant of Eq.~\eqref{eq:0Dinvariant_k_y_0_ani_sigma_blocks} at $k_x=0$, we restrict to the weak-coupling regime, and block dia\-go\-na\-li\-ze Eq.~\eqref{eq:Low_Energy_1BM} by means of effecting the unitary transformation $({\cal O}_{yz}+\sigma_x)/\sqrt{2}$, which yield the blocks:
\bea
\hat{\cal H}_{k_x=0,\sigma}^{\rm low-en}=\xi_{0;q}^+\tau_3+\Delta_{0;q}^+\tau_1-\sigma \frac{M_\perp+M_{||}\rho_3}{2}\,.
\eea

\noi The above is further block-diagonalizable by introducing the eigenstates of $\rho_3$ labelled by $\rho=\pm1$. Straighforward manipulations following after the definitions of Eq.~\eqref{eq:Augmented_winding_number}, yield the result for each $\hat{\cal H}_{k_x=0,\sigma,\rho}^{\rm low-en}$ block:
\bea
n_{k_x=0,\sigma,\rho}=\sigma\rho\frac{1+{\rm sgn}\left[M_{\rho}^2-(\xi^+_{0;q})^2-(\Delta_{0;q}^+)^2\right]}{2}\,.\no
\eea

\noi Similar to the construction leading to Eq.~\eqref{eq:winding_number_1BM_total}, also here one has to consider combinations of the invariants stemming from the $\rho$ blocks. Specifically, here we need to define $n_{k_x=0,\sigma}=-(n_{k_x=0,\sigma,-}+n_{k_x=0,\sigma,+})/2$.

\subsection{Glide Majorana Parity}\label{app:AppendixC4}

The glide Majorana parity is here defined as a $\mathbb{Z}_2$ invariant for the BDI class in 0D. This is given as the pa\-ri\-ty of the winding number for an interpolation $\hat{\cal H}_{\ell}$ with $\ell\in[0,2\pi)$ connecting the Hamiltonian of interest $\hat{\cal H}_{\pi}$ and a reference Hamiltonian $\hat{\cal H}_{0}$. The winding number reads:
\bea
w_{\rm inter}=\int_0^{2\pi}\frac{{\rm d}\ell}{2\pi i}{\rm Tr}\left(\hat{A}_\ell^{-1}\frac{\rm d}{{\rm d}\ell}\hat{A}_\ell\right)\,.
\eea

\noi By virtue of the charge-conjugation symmetry, we obtain:
\bea
i\pi w_{\rm inter}&=&\ln\big[\det\big(\hat{A}_\pi\big)/\det\big(\hat{A}_0\big)\big]\Rightarrow\no\\
P_G&=&{\rm sgn}\big[(-1)^{w_{\rm inter}}\big]={\rm sgn}\prod_{\ell=0,\pi}\det\big(\hat{A}_\ell\big)\,.\quad
\eea

To obtain the result of Eq.~\eqref{eq:GlideMajoranaCalculation}, we employ the above equation, where each one of the two $\sigma$ blocks of Eq.~\eqref{eq:GlideBlocks1D} is considered as a reference Hamiltonian for the other. In the weak-coupling limit, we project onto the $\eta_3=1$ block and obtain the Hamiltonian blocks:
\bea
\hat{{\cal H}}_{k_x=0,\sigma}^{\rm low-en}=\xi_{0;q}^+\tau_3+\Delta_{0;q}^+\tau_1-M_\sigma\rho_1\,.
\eea

\noi We block-off diagonalize the above blocks via the transformation $(\rho_2\tau_2+\tau_3)/\sqrt{2}$ and find:
\bea
\hat{A}_{k_x=0,\sigma}=-\Delta_{0;q}^+-M_\sigma\rho_3-i\xi_{0;q}^+\rho_2\,.
\eea

\noi By obtaining the determinant of the above upper off-diagonal blocks, we directly find the result of Eq.~\eqref{eq:GlideMajoranaCalculation}. 

\subsection{Winding Number in 1D 2BMs}\label{app:AppendixC5}

To obtain the winding number for an interband-only MHC, we restrict to the low-energy sector of the system and consider the projected spinor: 
\begin{align}
\bm{\chi}_{k_x}^{\rm 2BM}=\frac{\rho_2+\rho_3}{\sqrt{2}}(\bm{\Psi}_{k_x+3q}^{\rm e},\bm{\Psi}_{k_x-3q}^{\rm e},\bm{\Psi}_{k_x+3q}^{\rm h},\bm{\Psi}_{k_x-3q}^{\rm h})^{\intercal},
\end{align}

\noi as well as the corresponding Hamiltonian blocks:
\bea
\hat{{\cal H}}_{k_x}^{\rm low-en}&=&\sum_{s={\rm e,h}}{\cal P}_s\Big\{\Big(\xi_{k_x;-3q}^{s,+}+\xi_{k_x;-3q}^{s,-}\rho_2\Big)\tau_3+\Delta^{s}\tau_1\Big\}\no\\
&&-\frac{\hat{M}_\perp\rho_1\sigma_z+\hat{M}_{||}\rho_3\sigma_x}{2}\,.
\label{eq:Low_Energy_2BM_Blocks}
\eea

\noi By exploiting the $\tilde{\mathcal{O}}=\rho_2\sigma_y$ symmetry, we can block diagonalize the Hamiltonian by means of $\tilde{\mathcal{S}}$ in App.~\ref{app:AppendixC1}:
\bea
\hat{{\cal H}}_{k_x,\sigma}^{\rm low-en}&=&\sum_{s={\rm e,h}}{\cal P}_s\Big\{\Big(\xi_{k_x;-3q}^{s,+}+\xi_{k_x;-3q}^{s,-}\rho_2\Big)\tau_3+\Delta^{s}\tau_1\Big\}\no\\
&&-\hat{M}_{\sigma}\rho_1\,.
\label{eq:Low_Energy_2BM_Blocks}
\eea

\noi By solely considering interband magnetic scattering, i.e. $\hat{M}_{\sigma}=M_{\sigma}\kappa_1$ we find the emergent unitary symmetry $\mathcal{O}=\kappa_3\sigma_y$ in the original basis, which allows for yet another block diagonalization via $(\tilde{\mathcal{S}}^{\dagger}\mathcal{O}\tilde{\mathcal{S}}+\rho_1)/\sqrt{2}$:
\bea
\hat{{\cal H}}_{k_x,\rho}^{\rm low-en}&=&\sum_{s={\rm e,h}}{\cal P}_s\Big\{\Big(\xi_{k_x;-3q}^{s,+}+\rho\xi_{k_x;-3q}^{s,-}\kappa_3\Big)\tau_3+\Delta^{s}\tau_1\Big\}\no\\
&&-\rho M_{\sigma}\kappa_1\,,
\label{eq:Low_Energy_2BM_Blocks_Again}
\eea

\noi where $\rho=\pm1$ label the eigenvalues of the matrix $\rho_1$. Hence, we block off-diagonalize the Hamiltonian via the unitary transformation $(\Pi+\tau_3)/\sqrt{2}$, and obtain:
\bea
\det\big(\hat{A}_{k_x,\sigma,\rho}^{\rm low-en}\big)&=&\xi_{k_x+\rho 3q}^{\rm e}\xi_{k_x-\rho 3q}^{\rm h}+\Delta^{\rm e}\Delta^{\rm h}-M_\sigma^2\no\\
&+&i\left(\xi_{k_x+\rho 3 q}^{\rm e}\Delta^{\rm h}-\xi_{k_x-\rho3q}^{\rm h}\Delta^{\rm e}\right)\,.\label{eq:2BM_1D_MH_detA}
\eea

A nonzero intraband magnetization components restores the BDI$\oplus$BDI class found in 1BMs and yields:
\bea
\hat{A}^{\rm low-en}_{k_x,\sigma}=-\sum_{s={\rm e,h}}{\cal P}_s\Big\{i\left(\xi_{k_x;-3q}^{s,+}\rho_2+\xi_{k_x;-3q}^{s,-}\right)\no\\
+\Delta^s+M_{\sigma}^s\rho_3\Big\}-\kappa_1M_{\sigma}^{\rm eh}\rho_3\,.\qquad\label{eq:2BM_single_Q_low_ham}
\eea

\subsection{Invariants for 2D 1BMs in the MHC phase\\ and with $\bm{\Delta_{\bm{k}}\sim\{{\rm B_{2g},A_{2g}}\}}$}\label{app:AppendixC6}

For a gap function transforming as the $\rm B_{2g}$ or $\rm A_{2g}$ IR, the resulting point group becomes $\rm G_{\rm MHC}$ accompanied by the space group symmetries $\{\mathcal{T},\sigma_{xz,yz}^{{\cal Q}}\,|\,(\nicefrac{\pi}{Q},0)\}$. As discussed in Sec.~\ref{sec:1BM_2D_MH}, these modified symmetries lead to a BDI$\oplus$BDI class in the $k_x=0$ HSP and a AI$\oplus$AI class for $k_y=\{0,\pi\}$, since, for the latter, the gap function va\-ni\-shes. Specifically, for these HSPs, denoted by the wave-vectors $\bm{k}_{\mathcal{R}_{xz}}$, the Hamiltonian takes the simple form in the weak-coupling regime:
\begin{align}
\hat{{\cal H}}_{\bm{k}_{\mathcal{R}_{xz}},\sigma,\tau}=\tau\left[\xi^+_{\bm{k}_{\mathcal{R}_{xz}};\bm{q}_1}+\xi^-_{\bm{k}_{\mathcal{R}_{xz}};\bm{q}_1}\rho_2\right]-M_\sigma\rho_1,
\end{align}

\noi where $\tau=\pm1$ labels the two AI blocks. This class only supports a strong mirror $\mathbb{Z}$ invariant for a nodal spectrum, with the associated invariant defined similar to Eq.~\eqref{eq:vorticity_mirror}. Specifically we find the normalized complex function entering in the invariant:
\begin{align}
&Z_{\epsilon,k_x,\sigma,\tau}=
\\
&\frac{\xi_{\bm{k}_{\mathcal{R}_{xz}-\bm{q}_1}}\xi_{\bm{k}_{\mathcal{R}_{xz}+\bm{q}_1}}+\epsilon^2+M_{\sigma}^2+2i\tau\epsilon\,\xi^+_{\bm{k}_{\mathcal{R}_{xz}};\bm{q}_1}}{\sqrt{(\xi_{\bm{k}_{\mathcal{R}_{xz}-\bm{q}_1}}\xi_{\bm{k}_{\mathcal{R}_{xz}+\bm{q}_1}}+\epsilon^2+M_{\sigma}^2)^2+4\epsilon^2(\xi^+_{\bm{k}_{\mathcal{R}_{xz}};\bm{q}_1})^2}}\,.\no
\end{align}

Instead for the $k_x=0$ HSP, the BdG Hamiltonian in the weak-coupling limit becomes:
\bea
\hat{\mathcal{H}}_{\bm{k}=(0,k_y)}&=&\xi_{\bm{k}=(q,k_y)}\tau_3-\Delta_{\bm{k}=(q,k_y)}\rho_2\tau_1\no\\
&-&(M_{\perp}\rho_1\sigma_z+M_{||}\rho_3\sigma_x)/2\,.
\eea

\begin{figure}[t!]
\centering
\includegraphics[width=0.95\columnwidth]{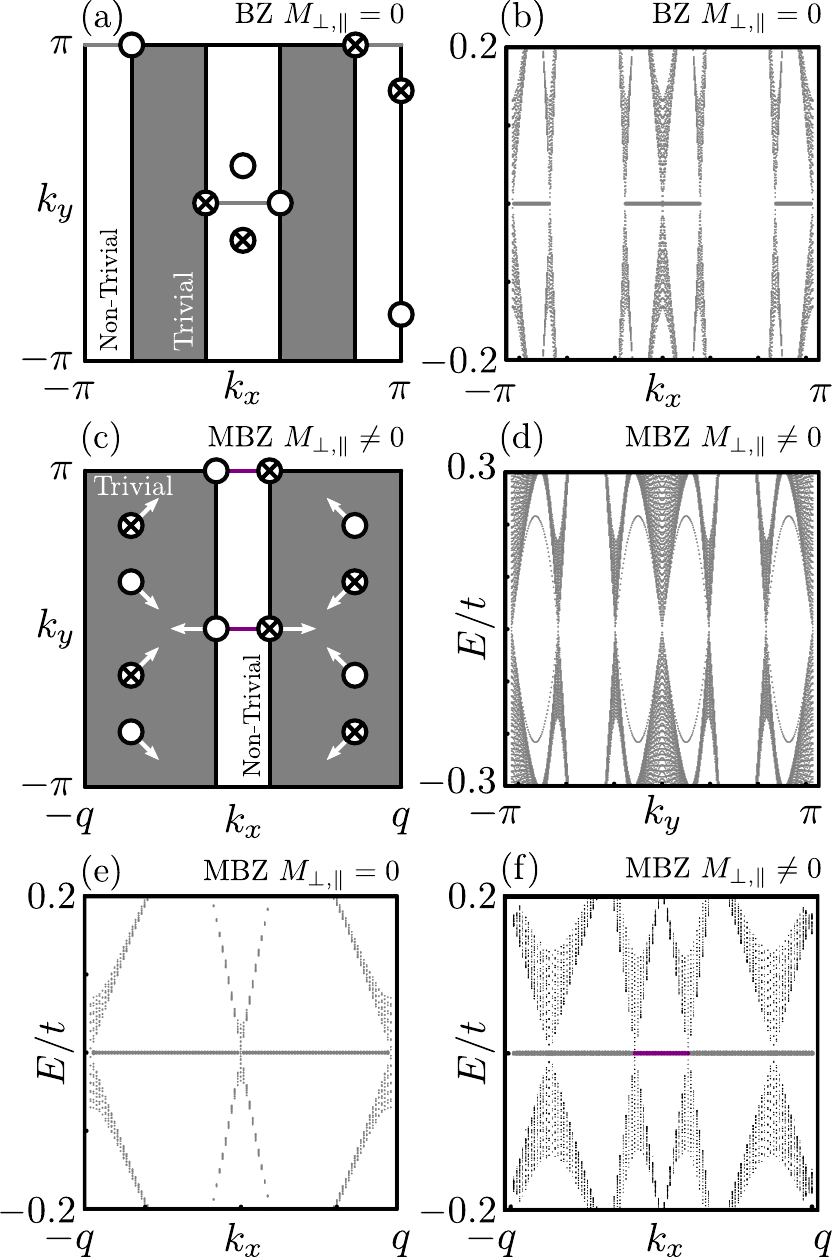}
\caption{Nodal spectrum for the 1BM in Fig.~\ref{fig:Figure7} in the MHC phase with $\Delta_{\bm{k}}=\Delta\sin k_x \sin k_y\sim{\rm B_{2g}}$. (a) displays the nodes in the first BZ in the absence of magnetism. These symmetry enforced nodes give rise to MFBs, as seen in (b). (c) illustrates the same as in (a), after we include magnetism and downfold to the MBZ. We see that the initial nodes in (a) split and move on straight lines. This ultimately lifts the MFBs on edges perpendicular to the magnetic ordering wave-vector $\bm{Q}$, see (d), since the vorticity of the nodes cancel when projected onto this edge. In contrast, we see in (f) that new MFBs (marked in purple) are established on edges parallel to $\bm{Q}$ in addition to the MFBs inherited from the case of $M_{||}=M_{\perp}=0$ (marked in gray), cf. (e). We used $L_x=L_y=401$, $\Delta=1\,t$ throughout, and $M_{\perp,||}=0.8\,t$ in (d) and (f).}
\label{fig:appFigure4}
\end{figure}

\noi For this case we can only define a strong invariant for a fully gapped spectrum, namely the glide $w_{G,k_x=0}$ and mirror $w_{M,k_x=0}$ winding numbers, see Eqs.~\eqref{eq:glide_winding_number_def} and~\eqref{eq:mirror_winding_number_def}, after replacing $\epsilon\mapsto k_y$ in the latter. For the symmetry $\{\mathcal{T}\,|\,(\nicefrac{\pi}{Q},0)\}\Theta\equiv\tilde{\cal O}$ we block diagonalize the Hamiltonian via $\tilde{\cal S}$ and find:
\bea
\hat{\mathcal{H}}_{\bm{k}=(0,k_y),\sigma}&=&\xi_{\bm{k}=(q,k_y)}\tau_3-\Delta_{\bm{k}=(q,k_y)}\rho_2\tau_1-M_{\sigma}\rho_1\,.\quad\quad\,\,\,\,
\eea

\noi After block off-diagonalizing the above Hamiltonian by means of the unitary operator $(\rho_2\tau_2+\tau_3)/\sqrt{2}$, we obtain:
\bea
\det (\hat{A}_{k_y,\sigma})&=&\xi_{\bm{k}=(q,k_y)}^2-\Delta_{\bm{k}=(q,k_y)}^2\no\\
&-&M_{\sigma}^2+2i\xi_{\bm{k}=(q,k_y)}\Delta_{\bm{k}=(q,k_y)}\,.\no
\eea

For the remaining two symmetries $\{\sigma_{yz}^{{\cal Q}}\,|\,(\nicefrac{\pi}{Q},0)\}$ and ${\cal R}_{yz}^{{\cal Q}}$, which in fact commutes with $\tilde{\cal O}$, we find that their matrix representations coincide in each $\sigma$ block. This leads to the additional block diagonalization by means of the unitary operator $(\rho_1\tau_3+\rho_2)/\sqrt{2}$: 
\bea
\hat{{\cal H}}_{\bm{k}=(0,k_y),\sigma,\rho}=\xi_{\bm{k}=(q,k_y)}\tau_3-\rho\Delta_{\bm{k}=(q,k_y)}\tau_1-\rho M_{\sigma}\tau_3\,,\no\\
\eea

\noi which we further block off-diagonalize with by means of a unitary transformation with operator $(\tau_2+\tau_3)/\sqrt{2}$, and find $\hat{A}_{\bm{k}=(0,k_y),\sigma,\rho}=-i\big[\xi_{\bm{k}=(q,k_y)}-\rho M_{\sigma}\big]+\rho\Delta_{\bm{k}=(q,k_y)}$.

In order to exemplify the above, we consider in the following the 1BM from Fig.~\ref{fig:Figure7} in the MHC phase, with a pairing function transforming as the $\rm B_{2g}$ IR, specifically $\Delta_{\bm{k}}=\Delta\sin k_x\sin k_y$. Through relation Eq.~\eqref{eq:Condition}, we find the gap closing points $\bm{k}_{\rm c}:\big\{k_y=0,\,\pi,\,{\rm or}\,k_x=\pm\arccos\big(-2t\cos k_y \cos q/\mu\big)\big\}$. For the values of $q$ and $\mu$ used in Fig.~\ref{fig:Figure7}, we obtain the simple relation $\bm{k}_{\rm c}:\{k_y=0,\,\pi,\,{\rm or}\,k_x=\pm k_y\}$. For the gap closing points $\bm{k}_{\rm c}=(k_x,0)$ and $(k_x,\pi)$ we straightforwardly find the gap clo\-sing cri\-te\-rion $\xi_{\bm{k}_{\mathcal{R}_{xz}}-\bm{q}_1}\xi_{\bm{k}_{\mathcal{R}_{xz}}+\bm{q}_1}=M_{\sigma}^2$. When the magnetic wave-vector $\bm{Q}_1$ coincides with the nesting vector $\bm{Q}_{\rm N}$, we observe that the spectrum is nodal even for $M_{\perp}=M_{||}=0$ with $\bm{k}_{\rm c}=(0,0)$ and $(0,\pi)$. For a nonzero $M_{\perp,||}$ the nodes move along the $xz$ HSPs. The remaining set of nodes, i.e., the ones mo\-ving along the lines $k_x=\pm k_y$ appear for $M^2_{\sigma}=\sqrt{\xi_{\bm{k}-\bm{q}_1}\xi_{\bm{k}+\bm{q}_1}+\Delta_{\bm{k}-\bm{q}_1}\Delta_{\bm{k}+\bm{q}_1}}$. Similar to the nodes mo\-ving along the HSPs $k_y=\{0\,\pi\}$, also here we find a nodal spectrum for $M_{\perp}=M_{||}=0$ for $k_x=\pm k_y=q$ and $q + \pi$. Upon increasing $M_{\perp,||}$ the nodes move along the lines $k_x=\pm k_y$.

In Fig.~\ref{fig:appFigure4} we sketch the path of the nodes, the vorti\-ci\-ties and resulting edge spectra. As seen in (a), already in the absence of the MHC, the spectrum contains zeros which are enforced by the symmetry of $\Delta_{\bm{k}}$. These give rise to topologically-protected MFBs. This is exemplified in (b) using open boun\-dary conditions in the $y$ axis. Since in this case $M_{\perp,||}=0$, we show the ori\-gi\-nal BZ. After switching on magnetism and transferring to the MBZ, we observe in (c) that the nodes split and start to move on straight lines as we vary the pairing and magnetic gaps. This splitting lifts the preexisting MFBs on edges perpendicular to the $\bm{Q}$ vector, as a consequence of the cancelling vorticities. This becomes more transpa\-rent by comparing (b) and (d). Instead, for edges parallel to the $\bm{Q}$-vector, we recover the MFBs in (b), but we also obtain newly-established MFBs stemming from the split nodes. This is illustrated in (e), where we show the same as in (b) but now in the MBZ, and in (f) with $M_{\perp,||}\neq0$.\\

\section{Functions for the representation of the BdG Hamiltonian in 2D}\label{app:Functions}

The matrix function $\hat{F}(f_{\bm{k}})$ has the form:
\bea
\hat{F}\big(f_{\bm{k}}\big)&=&f^{(0, 0)}_{\bm{k}}+f_{\bm{k}}^{(0, 1)}\rho_2+f_{\bm{k}}^{(0, 2)}\eta_3+f_{\bm{k}}^{(0, 3)}\eta_3\rho_2\no\\
&+&\lambda_2\big[f^{(1, 0)}_{\bm{k}}+f^{(1, 1)}_{\bm{k}}\rho_2+f^{(1, 2)}_{\bm{k}}\eta_3+f^{(1, 3)}_{\bm{k}}\eta_3\rho_2\big]\no\\
&+&\zeta_3\big[f^{(2, 0)}_{\bm{k}}+f_{\bm{k}}^{(2, 1)}\rho_2+f_{\bm{k}}^{(2, 2)}\eta_3+f_{\bm{k}}^{(2, 3)}\eta_3\rho_2\big]\no\\
&+&\zeta_3\lambda_2\big[f_{\bm{k}}^{(3, 0)}+f_{\bm{k}}^{(3, 1)}\rho_2+f_{\bm{k}}^{(3, 2)}\eta_3+f_{\bm{k}}^{(3,3)}\eta_3\rho_2\big],\no\label{eq:F}
\eea

\noi with the functions $f_{\bm{k}}^{(t,s)}=(-1)^{s+t}f_{-\bm{k}}$ defined as:
\bea
f_{\bm{k}}^{(0,s)}&=&\big[f^{(s)}_{\bm{k}-\bm{q}_2}+f^{(s)}_{\bm{k}+\bm{q}_2}+f^{(s)}_{\bm{k}+3\bm{q}_2}+f^{(s)}_{\bm{k}-3\bm{q}_2}\big]/4\,,\no\\
f_{\bm{k}}^{(1,s)}&=&\big[f^{(s)}_{\bm{k}-\bm{q}_2}-f^{(s)}_{\bm{k}+\bm{q}_2}+f^{(s)}_{\bm{k}+3\bm{q}_2}-f^{(s)}_{\bm{k}-3\bm{q}_2}\big]/4\,,\no\\
f_{\bm{k}}^{(2,s)}&=&\big[f^{(s)}_{\bm{k}-\bm{q}_2}+f^{(s)}_{\bm{k}+\bm{q}_2}-f^{(s)}_{\bm{k}+3\bm{q}_2}-f^{(s)}_{\bm{k}-3\bm{q}_2}\big]/4\,,\no\\
f_{\bm{k}}^{(3,s)}&=&\big[f^{(s)}_{\bm{k}-\bm{q}_2}-f^{(s)}_{\bm{k}+\bm{q}_2}-f^{(s)}_{\bm{k}+3\bm{q}_2}+f^{(s)}_{\bm{k}-3\bm{q}_2}\big]/4\,.\no
\label{eq:DefsShiftedVars_2D}
\eea

\end{document}